\documentclass[aps, pra, onecolumn, tightenlines, letterpaper, amsmath, amssymb, preprintnumbers, superscriptaddress, floatfix, longbibliography, nofootinbib]{revtex4-2}

\usepackage{dsfont}
\usepackage{amsmath}
\usepackage{amssymb}
\usepackage{physics}
\usepackage{tabu}
\usepackage{tabularx}
\usepackage{bm}
\usepackage{tikz}
\usetikzlibrary{quantikz}
\usepackage{textgreek}
\usepackage{multirow}
\usepackage{makecell}

\makeatletter
\newcommand{\fmarki}{*}
\newcommand{\fmarkii}{\ensuremath{\dagger}}
\newcommand{\fmarkiii}{\ensuremath{\ddagger}}
\newcommand{\fmarkiv}{\ensuremath{\mathsection}}
\newcommand{\fmarkv}{\ensuremath{\mathparagraph}}
\newcommand{\fmarkvi}{\ensuremath{\|}}
                
\def\@fnsymbol#1{{\ifcase#1\or \fmarki\or \fmarkii\or \fmarkiii\or \fmarkiv\or \fmarkv\or \fmarkvi \else\@ctrerr\fi}}
\makeatother

\renewcommand{\fmarkvi}{\$}

\usepackage[hypertexnames=false]{hyperref}
\hypersetup{
    colorlinks=true,       
    linkcolor=blue,          
    citecolor=blue,        
    filecolor=blue,      
    urlcolor=blue           
}

\newcolumntype{Y}{>{\centering\arraybackslash}X}

\makeatletter
\pretocmd\frontmatter@thefootnote{\color{black}}{}{}
\makeatother

\usepackage{orcidlink}

\begin{document}

\begin{figure}
  \vskip -1.cm
  \leftline{\includegraphics[width=0.15\textwidth]{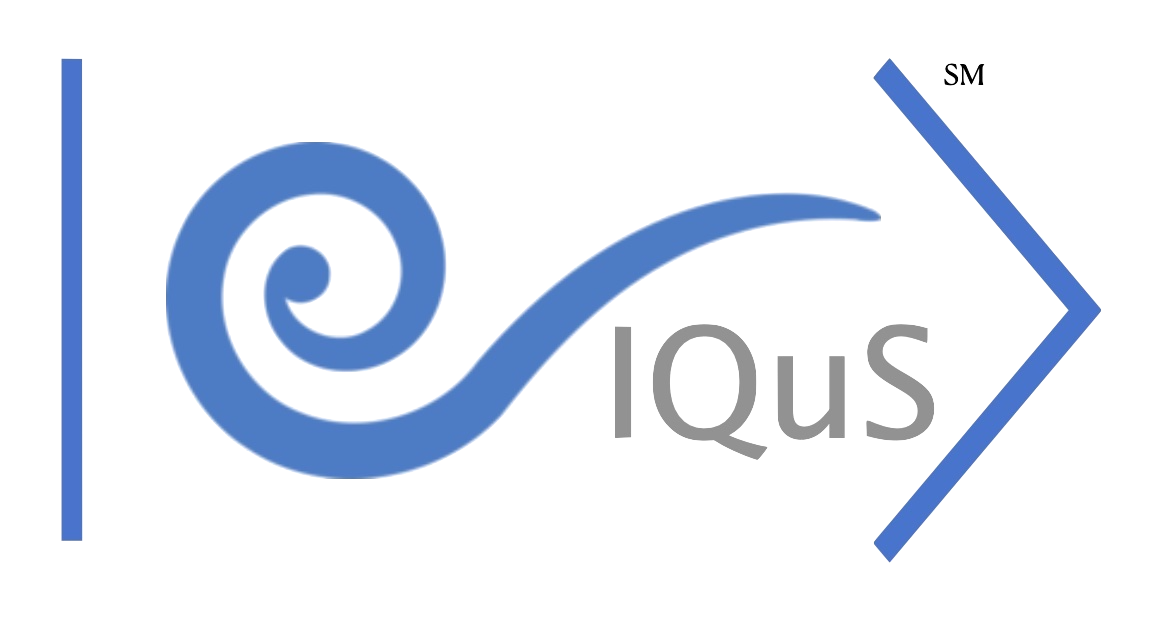}}
\end{figure}

\title{Preparations for Quantum Simulations of Quantum Chromodynamics in \texorpdfstring{\boldmath$1+1$}{1+1} Dimensions:
(I) Axial Gauge}

\author{Roland C.~Farrell\,\orcidlink{0000-0001-7189-0424
}}
\email[Corresponding author, ]{rolanf2@uw.edu}
\affiliation{InQubator for Quantum Simulation (IQuS), Department of Physics, University of Washington, Seattle, WA 98195, USA.}
\author{Ivan A.~Chernyshev\,\orcidlink{0000-0001-8289-1991}}
\email{ivanc3@uw.edu}
\affiliation{InQubator for Quantum Simulation (IQuS), Department of Physics, University of Washington, Seattle, WA 98195, USA.}
\author{Sarah J.~M.~Powell\,\orcidlink{0000-0002-5228-8291}}
\email{spow9@uw.edu}
\affiliation{Department of Physics and Astronomy, York University, Toronto, ON M3J 1P3, Canada.}
\author{Nikita A.~Zemlevskiy\,\orcidlink{0000-0002-0794-2389}}\email{zemlni@uw.edu}
\affiliation{InQubator for Quantum Simulation (IQuS), Department of Physics, University of Washington, Seattle, WA 98195, USA.}
\author{Marc Illa\,\orcidlink{0000-0003-3570-2849}}
\email{marcilla@uw.edu}
\affiliation{InQubator for Quantum Simulation (IQuS), Department of Physics, University of Washington, Seattle, WA 98195, USA.}
\author{Martin J.~Savage\,\orcidlink{0000-0001-6502-7106}}
\email{mjs5@uw.edu}
\affiliation{InQubator for Quantum Simulation (IQuS), Department of Physics, University of Washington, Seattle, WA 98195, USA.}

\preprint{IQuS@UW-21-027, NT@UW-22-05}
\date{\today}

\begin{abstract}
\noindent
Tools necessary for quantum simulations of $1+1$ dimensional quantum chromodynamics are developed.
When formulated in axial gauge and with two flavors of quarks, this system requires 12 qubits per spatial site with the gauge fields included via non-local interactions.
Classical computations and D-Wave's quantum annealer {\tt Advantage} are used to determine the hadronic spectrum, enabling a decomposition of the masses and a study of quark entanglement.
Color ``edge” states confined within a screening length of the end of the lattice are found. 
IBM's 7-qubit quantum computers, {\tt ibmq\_jakarta} and {\tt ibm\_perth}, are used to compute dynamics from the trivial vacuum in one-flavor QCD with one spatial site. 
More generally, the Hamiltonian and quantum circuits for time evolution of $1+1$ dimensional $SU(N_c)$ gauge theory with $N_f$ flavors of quarks are developed, and the resource requirements for large-scale quantum simulations are estimated.
\end{abstract}

\maketitle

\newpage{}

{\hypersetup{linkcolor=black}
\tableofcontents
}
\newpage{}

\section{Introduction}
\noindent
Simulations of the real-time dynamics of out-of-equilibrium, finite density quantum systems is a major goal of Standard Model (SM)~\cite{Glashow:1961tr,Higgs:1964pj,Weinberg:1967tq,Salam:1968rm,Politzer:1973fx,Gross:1973id} physics research and is expected to be computed efficiently~\cite{Lloyd1073} with ideal quantum computers~\cite{5392446,5391327,Benioff1980,Manin1980,Feynman1982,Fredkin1982,Feynman1986,doi:10.1063/1.881299,williamsNASAconference}.  
For recent reviews, see Refs.~\cite{Preskill:2021apy,Klco:2021lap,Bauer:2022hpo}.
Developing such capabilities would enable precision predictions of particle production and fragmentation in beam-beam collisions at the LHC and RHIC, of the matter-antimatter asymmetry production in the early universe, and of the structure and dynamics of dense matter in supernova and the neutrino flavor dynamics therein.
They would also play a role in better understanding protons and nuclei, particularly their entanglement structures and dynamics, and in exploring exotic strong-interaction phenomena such as color transparency. First steps are being taken toward simulating quantum field theories
(QFTs) using currently available, NISQ-era (Noisy Intermediate Scale Quantum) quantum devices~\cite{Preskill2018quantumcomputingin}, by studying low-dimensional and truncated many-body systems (see for example, Refs.~\cite{Hauke:2013jga,Banuls:2013jaa,Zohar:2016iic,Muschik:2016tws,Martinez:2016yna,Buyens:2016hhu,Banuls:2016lkq,Gonzalez-Cuadra:2017lvz,Dumitrescu:2018njn,PhysRevA.98.032331,Lu:2018pjk,Kaplan:2018vnj,Stryker:2018efp,Yeter-Aydeniz:2018mix,PhysRevD.101.074512,Avkhadiev:2019niu,Bauer:2019qxa,Klco:2019xro,Klco:2019yrb,Banuls:2019bmf,Luo:2019vmi,Funcke:2019zna,Davoudi:2019bhy,Magnifico:2019kyj,Mishra:2019xbh,Shehab:2019gfn,Yang_2020,Kharzeev:2020kgc,Shaw2020quantumalgorithms,PhysRevD.101.074512,PhysRevD.103.094501,Halimeh:2020ecg,Halimeh:2020djb,VanDamme:2020rur,Haase:2020kaj,Yeter-Aydeniz:2020jte,Davoudi:2021ney,ARahman:2021ktn,PhysRevLett.122.050403,Stryker:2021asy,aidelsburger2021cold,Bauer:2021gup,VanDamme:2021njp,Halimeh:2021lnv,Knaute:2021xna,Halimeh:2021vzf,Thompson:2021eze,Yeter-Aydeniz:2021mol,Yeter-Aydeniz:2021olz,Funcke:2021aps,Zhang:2021bjq,Rahman:2022rlg,Deliyannis:2022uyh,Bauer:2021gek,Illa:2022jqb,Mildenberger:2022jqr,Milsted:2020jmf}). 
These studies are permitting first quantum resource estimates to be made for more realistic simulations.

There has already been a number of quantum simulations of latticized $1+1$D quantum electrodynamics (QED, the lattice Schwinger model), starting with the pioneering work of Martinez {\it et al.}~\cite{Martinez:2016yna}.
The Schwinger model shares important features with quantum chromodynamics (QCD), such as charge screening, a non-zero fermion condensate, nontrivial topological charge sectors and a $\theta$-term.
Quantum simulations of the Schwinger model have been performed using quantum computers~\cite{Martinez:2016yna,Klco:2018kyo,Lu:2018pjk,Kokail:2018eiw,Nguyen:2021hyk,Thompson:2021eze}, and 
there is significant effort being made to extend this progress to higher dimensional QED~\cite{Zohar:2011cw,Zohar:2012ay,Tagliacozzo:2012vg,Zohar:2012ts,Wiese:2013uua,Marcos:2014lda,Kuno:2014npa,Bazavov:2015kka,Kasper:2015cca,Brennen:2015pgn,Kuno:2016xbf,Zohar:2016iic,Kasper:2016mzj,Gonzalez-Cuadra:2017lvz,Ott:2020ycj,Paulson:2020zjd,Kan:2021nyu,aidelsburger2021cold,Bauer:2021gek}.
These, of course, build upon far more extensive and detailed classical simulations of this model and analytic solutions of the continuum theory. 
There is also a rich portfolio of classical and analytic studies of $1+1$D $SU(N_c)$ gauge theories~\cite{Wilson:1994fk,Heinzl:1995jn,PhysRevD.31.2020,LIGTERINK2000983c,LIGTERINK2000215}, with some seminal papers preparing for quantum simulations~\cite{PhysRevD.11.395,PhysRevA.73.022328,Zohar:2012xf,PhysRevLett.110.125303,Tagliacozzo:2012vg,Banuls:2017ena,PhysRevD.101.074512,PhysRevD.103.094501,Paulson:2020zjd}, with the recent appearance of quantum simulations of a 1-flavor ($N_f=1$)  $1+1$D $SU(2)$ lattice gauge theory~\cite{Atas:2021ext}.
An attribute that makes such calculations attractive for early quantum simulations is that the gauge field(s) are uniquely constrained by Gauss's law at each lattice site.  
However, this is also a limitation for understanding higher dimensional theories where the gauge field is dynamical.  After pioneering theoretical works developing the formalism and also end-to-end  simulation protocols nearly a decade ago, it is only recently that first quantum simulations of the dynamics of a few plaquettes of gauge fields  have been performed~\cite{PhysRevD.101.074512,PhysRevD.103.094501,Atas:2021ext,Rahman:2022rlg}.

Due to its essential features, 
quantum simulations of the Schwinger model provide benchmarks for QFTs and quantum devices for the foreseeable future.  
Moving toward simulations of QCD requires including non-Abelian local gauge symmetry and multiple flavors of dynamical quarks. Low-energy, static and near-static observables in the continuum theory in $1+1$D
are well explored analytically and numerically, with remarkable results demonstrated, particularly in the 't Hooft model of large-$N_c$~\cite{tHooft:1973alw,tHooft:1974pnl} 
where the Bethe-Salpeter equation becomes exact. For a detailed discussion of $1+1$D $U(1)$ and $SU(N_c)$ gauge theories, see Refs.~\cite{Frishman:2010tc,Frishman2014book}.
Extending such calculations to inelastic scattering to predict, for instance, exclusive processes in high-energy hadronic collisions 
is a decadal challenge.

In $3+1$D QCD, the last 50 years have seen remarkable progress in using classical high-performance computing to provide robust numerical results using lattice QCD, e.g., Refs.~\cite{Joo:2019byq,Aoki:2021kgd}, 
where the quark and gluon fields are discretized in spacetime. Lattice QCD is providing complementary and synergistic results to those obtained in experimental facilities, moving beyond what is possible with analytic techniques alone. 
However, the scope of classical computations, even with beyond-exascale computing platforms~\cite{osti_1369223,Habib:2016sce,Joo:2019byq}, is limited by the use of a less fundamental theory (classical) to simulate a more fundamental theory (quantum).

Building upon theoretical progress in identifying 
candidate theories for early exploration (e.g., Ref.~\cite{Sala:2018dui}),
quantum simulations of $1+1$D non-Abelian gauge theories including matter were recently performed~\cite{Atas:2021ext} for a $N_c=2$ local gauge symmetry with one flavor of quark, $N_f=1$.
The Jordan-Wigner (JW) mapping~\cite{Jordan:1928wi} was used to define the lattice theory, and 
Variational Quantum Eigensolver (VQE)~\cite{Peruzzo_2014} quantum circuits were developed and used on IBM's quantum devices~\cite{IBMQ} 
to determine the vacuum energy, along with meson and baryon masses.  
Further, there have been previous quantum simulations of 
1- and 2-plaquette systems in $N_c=2,3$ Yang-Mills lattice gauge theories~\cite{Klco:2019evd,Ciavarella:2021nmj,Ciavarella:2021lel,ARahman:2021ktn,Illa:2022jqb} that did not include quarks.
Simulations of such systems are developing rapidly~\cite{Ciavarella:2021lel,Rahman:2022rlg} due to algorithmic and hardware advances. In addition, distinct mappings of these theories are being pursued~\cite{Brower:1997ha,Banerjee:2012xg,Tagliacozzo:2012df,Alexandru:2019nsa,Ji:2020kjk,Wiese:2021djl,Caspar:2022llo}.

This work focuses on the quantum simulation of $1+1$D $SU(N_c)$ lattice gauge theory for arbitrary $N_c$ and $N_f$.
Calculations are primarily done in $A^{(a)}_x=0$ axial (Arnowitt-Fickler) gauge,\footnote{For a discussion of Yang-Mills in axial gauge, see, for example, Ref.~\cite{Reinhardt:1996dy}.}
which leads to non-local interactions in order to define the chromo-electric field contributions to the energy density via Gauss's law.
This is in contrast to Weyl gauge, $A_t^{(a)}=0$, where contributions remain local.
The resource estimates for asymptotic quantum simulations of the Schwinger model in Weyl gauge have been recently performed~\cite{Shaw:2020udc}, and also for Yang-Mills gauge theory based upon the Byrnes-Yamamoto mapping~\cite{Kan:2021xfc}.
Here, the focus is on near-term, and hence non-asymptotic, quantum simulations to better assess the resource requirements for quantum simulations of non-Abelian gauge theories with multiple flavors of quarks. 
For concreteness, $N_f=2$ QCD is studied in detail, including the mass decomposition of the low-lying hadrons (the $\sigma$- and $\pi$-meson, the single baryon and the two-baryon bound state), color edge-states, entanglement structures within the hadrons and quantum circuits for time evolution.
Further, results are presented for the quantum simulation of a $N_f=1$, single-site system, using IBM's quantum computers~\cite{IBMQ}.
Such quantum simulations will play a critical role in evolving the functionality, protocols and workflows to be used in $3+1$D simulations of QCD, including the preparation of scattering states, time evolution and subsequent particle detection.
As a step in this direction, in a companion to the present paper, the results of this work have been applied to the quantum simulation of $\beta$-decay of a single baryon in $1+1$D QCD~\cite{PhysRevD.107.054513}.
Motivated by the recent successes in co-designing efficient multi-qubit operations in trapped-ion systems~\cite{Andrade:2021pil,Katz:2022czu}, 
additional multi-qubit or qudit operations are identified, 
specific to lattice gauge theories,
that would benefit from being native operations on quantum devices.

\section{QCD with Three Colors and Two flavors in \texorpdfstring{\boldmath$1+1$}{1+1}D}
\label{sec:Nc3Nf2}
\noindent
In $3+1$D, 
the low-lying spectrum of $N_f=2$ QCD is remarkably rich. 
The lightest hadrons are the $\pi$s, which are identified as the pseudo-Goldstone bosons associated with the spontaneous breaking of the approximate global $SU(2)_L\otimes SU(2)_R$ chiral symmetry, which becomes exact in the chiral limit where the $\pi$s are massless. At slightly higher mass are the broad $I=0$ spinless 
resonance, $\sigma$, and the narrow $I=0$, $\omega$, and $I=1$, $\rho$, vector resonances as well as the multi-meson continuum. 
The proton and neutron, which are degenerate in the isospin limit and the absence of electromagnetism,
are the lightest baryons, forming 
an $I=J=1/2$ iso-doublet.  
The next lightest baryons, which become degenerate with the nucleons in the large-$N_c$ limit (as part of a 
large-$N_c$ tower), are the four $I=J=3/2$ $\Delta$ resonances.
The nucleons bind together to form the periodic table of nuclei, the lightest being the deuteron, an $I=0$, $J=1$ neutron-proton bound state with a binding energy of $\sim 2.2~{\rm MeV}$, which is to be compared to the mass of the nucleon $M_N\sim 940~{\rm MeV}$.  
In nature, the low-energy two-nucleon systems have S-wave scattering lengths that are much larger than the range of their interactions, rendering them unnatural. Surprisingly, this unnaturalness persists for a sizable range of light-quark 
masses, e.g., Refs.~\cite{Beane:2002xf,Epelbaum:2012iu,Berengut:2013nh,Wagman:2017tmp,NPLQCD:2020lxg}. 
In addition, this unnaturalness, and the nearby renormalization-group fixed point~\cite{Kaplan:1998tg,Kaplan:1998we}, provides the starting point for a systematic effective field theory expansion about unitarity~\cite{Kaplan:1998tg,Kaplan:1998we,vanKolck:1998bw,Chen:1999tn}.
Much of this complexity is absent in a theory with only one flavor of quark.

As a first step toward $3+1$D QCD simulations of real-time dynamics of nucleons and nuclei, we will focus on preparing to carry out quantum simulations of $1+1$D QCD with $N_f=2$ flavors of quarks. While the isospin structure of the theory is the same as in $3+1$D, the lack of spin and orbital angular momentum significantly reduces the richness of the hadronic spectrum and S-matrix.
However, many of the relevant features and processes of $3+1$D QCD that are to be addressed by quantum simulation in the future are present in $1+1$D QCD.
Therefore, quantum simulations in $1+1$D are expected to provide inputs to the development of quantum simulations of QCD.

\subsection{Mapping \texorpdfstring{\boldmath$1+1$}{1+1}D QCD onto Qubits}
\noindent
The Hamiltonian describing non-Abelian lattice gauge field theories in arbitrary numbers of spatial dimensions was first given by Kogut and Susskind (KS) in the 1970s~\cite{Kogut:1974ag,Banks:1975gq}. For $1+1$D QCD with $N_f = 2$ discretized onto $L$ spatial lattice sites, which are mapped to 2L $q$, $\overline{q}$ sites to separately accommodate quarks and antiquarks,  the KS lattice Hamiltonian is
\begin{align}
    H_{\rm{KS}} 
     = & 
    \sum_{f=u,d}\left[
        \frac{1}{2 a} \sum_{n=0}^{2L-2} \left ( \phi_n^{(f)\dagger} U_n \phi_{n+1}^{(f)}
        \ +\ {\rm h.c.} \right ) 
    \: + \: 
    m_f \sum_{n=0}^{2L-1} (-1)^{n} \phi_n^{(f)\dagger} \phi_n^{(f)} 
    \right]
    \: + \: 
    \frac{a g^2}{2} 
    \sum_{n=0}^{2L-2} 
    \sum_{a=1}^{8}
    | {\bf E}^{(a)}_n|^2
    \nonumber\\
     & - \: \frac{\mu_B}{3} \sum_{f=u,d} \sum_{n=0}^{2L-1} \phi_n^{(f)\dagger} \phi^{(f)}_n 
    \ - \: 
    \frac{\mu_{I}}{2} \sum_{n=0}^{2L-1}\left(\phi_n^{(u)\dagger} \phi^{(u)}_n \ - \
    \phi_n^{(d)\dagger} \phi^{(d)}_n  \right)
    \ .
    \label{eq:KSHam}
\end{align}
The masses of the $u$- and $d$-quarks are $m_{u,d}$,
$g$ is the strong coupling constant at the spatial lattice spacing $a$,
$U_n$ is the spatial link operator in Weyl gauge
$A_t^{(a)}=0$,
$\phi^{(u,d)}_n$ are the $u$- and $d$-quark field operators which transform in the fundamental representation of $SU(3)$
and 
${\bf E}^{(a)}_n$ is the chromo-electric field associated with the $SU(3)$ generator,
$T^a$.
For convention, we write, for example, $\phi^{(u)}_n=(u_{n,r}, u_{n,g}, u_{n,b})^T$ to denote the $u$-quark field(s) at the $n^{\rm th}$ site in terms of 3 colors $r,g,b$.
With an eye toward simulations of dense matter systems, chemical potentials for baryon number, $\mu_B$, and the third component of isospin, $\mu_I$, are included.
For most of the results presented in
this work, the chemical potentials will be set to zero, $\mu_B=\mu_I = 0$,
and there will be exact isospin symmetry, $m_u=m_d \equiv m$.
In Weyl gauge and using the chromo-electric basis of the link operator $|{\bf R},\alpha,\beta\rangle_n$,
the contribution from the energy in the chromo-electric field from each basis state is proportional to the Casimir of the irrep ${\bf R}$.\footnote{
For an irrep, ${\bf R}$, represented by a tensor with $p$ upper indices and $q$ lower indices, $T^{a_1 \cdots a_p}_{b_1 \cdots b_q}$,
the Casimir provides
\begin{equation}
    \sum_{b=1}^{8}| {\bf E}^{(a)}_n|^2\ |{\bf R},\alpha,\beta\rangle_n = \frac{1}{3}\left( p^2+q^2+p q + 3 p + 3 q \right)\  |{\bf R},\alpha,\beta\rangle_n \ .
\label{eq:Casi}
\end{equation}
The indices $\alpha$ and $\beta$ specify the color state in the left (L) and right (R) link Hilbert spaces respectively. 
States of a color irrep {\bf R} are labelled by their total color isospin $T$, third component of color isospin $T^z$ and color hypercharge $Y$, i.e., $\alpha = (T_L, T^z_L, Y_L)$ and $\beta = (T_R, T^z_R, Y_R)$.
\label{foot:irrep}}
The fields have been latticized
such that the quarks reside on even-numbered sites, $n=0,2,4,6,\ldots$, and antiquarks reside on odd-numbered sites, $n=1,3,5,\ldots$.
Open boundary conditions (OBCs) are employed in the spatial direction, 
with a vanishing background chromo-electric field.
For simplicity, 
the lattice spacing will be set equal to $1$.

The KS Hamiltonian in Eq.~(\ref{eq:KSHam}) 
is constructed in Weyl gauge.
A unitary transformation can be performed on Eq.~(\ref{eq:KSHam}) to eliminate the gauge links~\cite{Sala:2018dui}, with Gauss's Law 
uniquely providing the energy in the chromo-electric field in terms of a non-local sum of products of charges, i.e., the Coulomb energy. 
This is equivalent to formulating the system in axial gauge~\cite{PhysRev.127.1821,weinberg1995quantum}, $A^{(a)}_x = 0$, from the outset.
The Hamiltonian in Eq.~(\ref{eq:KSHam}), when formulated with $A^{(a)}_x = 0$, becomes
\begin{align}
    H 
    = & 
    \sum_{f=u,d}\left[ 
        \frac{1}{2} \sum_{n=0}^{2L-2} \left ( \phi_n^{(f)\dagger} \phi_{n+1}^{(f)}
        \ +\ {\rm h.c.} \right ) 
    \: + \: 
    m_f \sum_{n=0}^{2L-1} (-1)^{n} \phi_n^{(f)\dagger} \phi_n^{(f)} 
    \right]
    \: + \: 
    \frac{g^2}{2}
    \sum_{n=0}^{2L-2} 
    \sum_{a=1}^{8}
    \left ( \sum_{m \leq n} Q^{(a)}_m \right ) ^2    \nonumber\\
     & - \: \frac{\mu_B}{3} \sum_{f=u,d} \sum_{n=0}^{2L-1} \phi_n^{(f)\dagger} \phi^{(f)}_n  
    \ - \:
    \frac{\mu_{I}}{2} \sum_{n=0}^{2L-1}\left(\phi_n^{(u)\dagger} \phi^{(u)}_n \ - \
    \phi_n^{(d)\dagger} \phi^{(d)}_n  \right)
    \ ,
    \label{eq:GFHam}
\end{align}
where the color charge operators on a given lattice site are the sum of contributions from the $u$- and $d$-quarks,
\begin{equation}
    Q^{(a)}_m \ =\ 
    \phi^{(u) \dagger}_m T^a \phi_m^{(u)}\ +\ 
    \phi^{(d) \dagger}_m T^a \phi_m^{(d)}
    \ .
    \label{eq:SU3charges}
\end{equation}
To define the fields, 
boundary conditions with $A_0^{(a)}(x)=0$ at spatial infinity and zero background chromo-electric fields are used, with Gauss's law sufficient to determine them at all other points on the lattice,
\begin{equation}
   {\bf E}^{(a)}_n  = \sum_{m\leq n} Q^{(a)}_m \ .
\end{equation}
In this construction, a state is completely specified by the fermionic occupation at each site. This is to be contrasted with the Weyl
gauge construction where both fermionic occupation and the $SU(3)$ multiplet defining the chromo-electric field are required.

\begin{figure}[!ht]
    \centering
    \includegraphics[width=15cm]{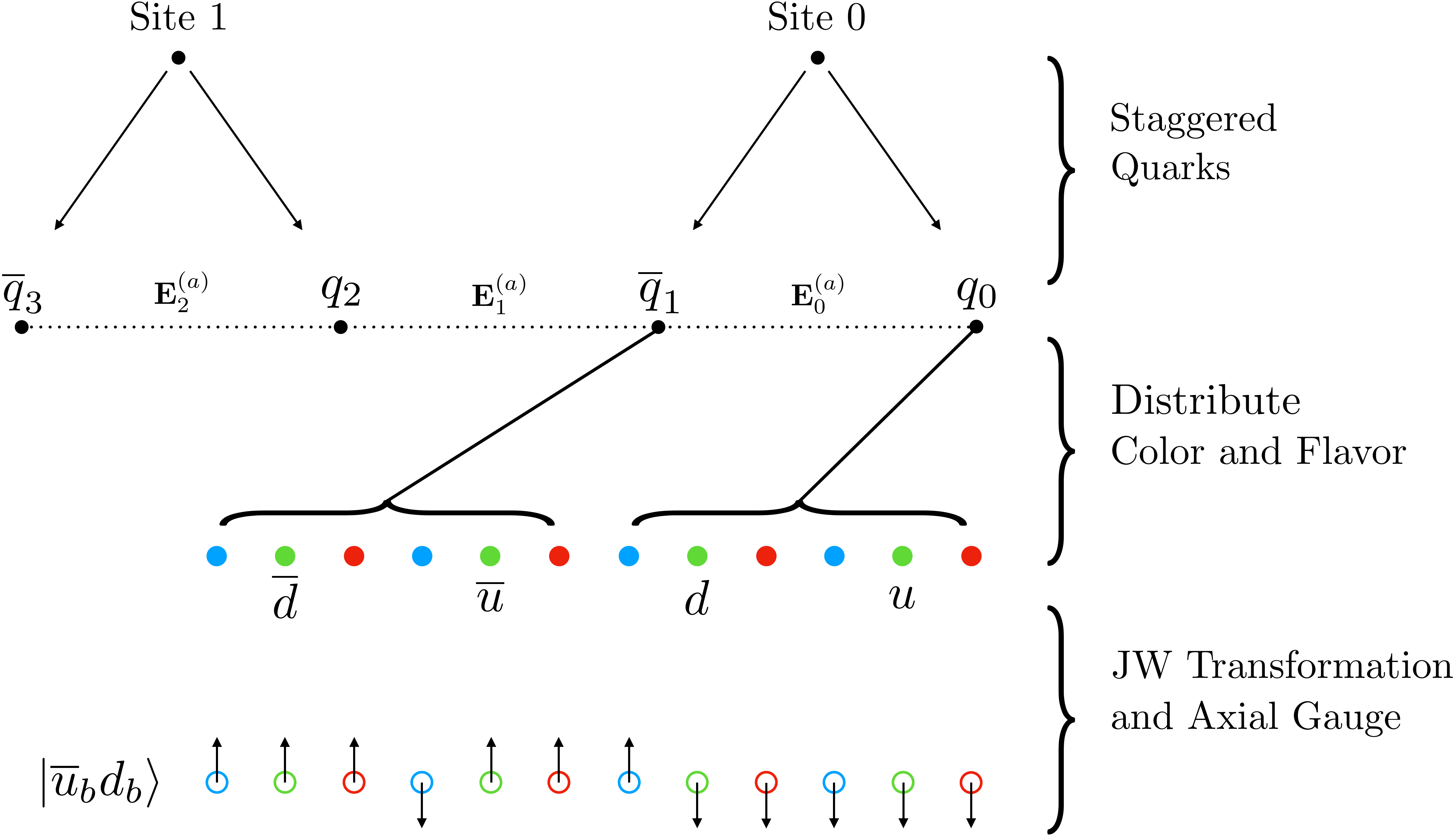}
    \caption{
    The encoding of $N_f=2$ QCD onto a lattice of spins describing $L=2$ spatial sites. 
    Staggering is used to discretize the quark fields, which doubles the number of lattice sites, with (anti)quarks on (odd) even sites.
    The chromo-electric field resides on the links between quarks and antiquarks. 
    Color and flavor degrees of freedom of each quark and antiquark site are distributed over six qubits with a JW mapping, and axial gauge along with Gauss's law are used to remove the chromo-electric fields. 
    A quark (antiquark) site is occupied if it is spin up (down), and the example spin configuration corresponds to the state $\ket{\overline{u}_b \, d_b}$.}
    \label{fig:2flavLayout}
\end{figure}

There are a number of ways that this system,
with the Hamiltonian given in Eq.~(\ref{eq:GFHam}), could be mapped
onto the register of a quantum computer.
In this work, both a staggered discretization and a JW transformation~\cite{1928ZPhy...47..631J} are chosen to map the $N_c=3$ and $N_f=2$
quarks to 6 qubits, with ordering $d_b, d_g, d_r, u_b, u_g, u_r$,
and the antiquarks associated with the same spatial site adjacent with ordering 
$\overline{d}_b, \overline{d}_g, \overline{d}_r, \overline{u}_b, \overline{u}_g, \overline{u}_r$.
This is illustrated in Fig.~\ref{fig:2flavLayout} and
requires a total of 12 qubits per spatial lattice site (see App.~\ref{app:hamConst} for more details). 
The resulting JW-mapped Hamiltonian is the sum of the following five terms:
\begin{subequations}
    \label{eq:H2flav}
    \begin{align}
    H = & \ H_{kin}\ +\ H_m\ +\ H_{el} \ +\ 
    H_{\mu_B}\ +\ H_{\mu_I} \ ,\\[4pt]
    H_{kin} = & \ -\frac{1}{2} \sum_{n=0}^{2L-2} \sum_{f=0}^{1} \sum_{c=0}^{2} \left[ \sigma^+_{6n+3f+c} \left ( \bigotimes_{i=1}^{5}\sigma^z_{6n+3f+c+i} \right )\sigma^-_{6(n+1)+3f+c} +\rm{h.c.} \right]\ ,
        \label{eq:Hkin2flav}\\[4pt]
    H_m = & \ \frac{1}{2} \sum_{n=0}^{2L-1} \sum_{f=0}^{1} \sum_{c=0}^{2} m_f\left[ (-1)^{n} \sigma_{6n + 3f + c}^z + 1\right]\ ,
        \label{eq:Hm2flav}\\[4pt]
    H_{el} = & \ \frac{g^2}{2} \sum_{n=0}^{2L-2}(2L-1-n)\left( \sum_{f=0}^{1} Q_{n,f}^{(a)} \, Q_{n,f}^{(a)} \ + \
        2 Q_{n,0}^{(a)} \, Q_{n,1}^{(a)}
         \right)   \nonumber \\[4pt]
        & + g^2 \sum_{n=0}^{2L-3} \sum_{m=n+1}^{2L-2}(2L-1-m) \sum_{f=0}^1 \sum_{f'=0}^1 Q_{n,f}^{(a)} \, Q_{m,f'}^{(a)} \ ,
         \label{eq:Hel2flav}\\[4pt]
    H_{\mu_B}  = & \ -\frac{\mu_B}{6} \sum_{n=0}^{2L-1} \sum_{f=0}^{1} \sum_{c=0}^{2}  \sigma_{6n + 3f + c}^z \ ,
        \label{eq:HmuB2flav}\\[4pt]
    H_{\mu_I} = & \ -\frac{\mu_I}{4} \sum_{n=0}^{2L-1} \sum_{f=0}^{1} \sum_{c=0}^{2} (-1)^{f} \sigma_{6n + 3f + c}^z  \ ,
        \label{eq:HmuI2flav}
    \end{align}
\end{subequations}
where now repeated adjoint color indices, $(a)$, are summed over,
the flavor indices, $f=0,1$, correspond to $u$- and $d$-quark flavors and $\sigma^\pm = (\sigma^x \pm i \sigma^y)/2$.
Products of charges are given in terms of spin operators as
\begin{align}
    Q_{n,f}^{(a)} \, Q_{n,f}^{(a)} =& \frac{1}{3}(3 - \sigma^z_{6n+3f} \sigma^z_{6n+3f+1} - \sigma^z_{6n+3f} \sigma^z_{6n+3f+2} - \sigma^z_{6n+3f+1} \sigma^z_{6n+3f+2}) \ ,  \nonumber \\[4pt]
    Q_{n,f}^{(a)} \, Q_{m,f'}^{(a)} =& \frac{1}{4}\bigg [2\big (\sigma^+_{6n+3f}\sigma^-_{6n+3f+1}\sigma^-_{6m+3f'}\sigma^+_{6m+3f'+1} + \sigma^+_{6n+3f}\sigma^z_{6n+3f+1}\sigma^-_{6n+3f+2}\sigma^-_{6m+3f'}\sigma^z_{6m+3f'+1}\sigma^+_{6m+3f'+2} \nonumber \\[4pt]
    &+\sigma^+_{6n+3f+1}\sigma^-_{6n+3f+2}\sigma^-_{6m+3f'+1}\sigma^+_{6m+3f'+2} + { \rm h.c.}\big ) + \frac{1}{6}\sum_{c=0}^{2} \sum_{c'=0}^2( 3 \delta_{c c'} - 1 ) \sigma^z_{6n+3f+c}\sigma^z_{6m+3f'+c'} \bigg ] \ .
    \label{eq:QnfQmfp}
\end{align}
A constant has been added to $H_m$ to ensure that all basis states contribute positive mass. The Hamiltonian for $SU(N_c)$ gauge theory with $N_f$ flavors in the fundamental representation is presented in Sec.~\ref{sec:NcNf}. 
Note that choosing $A^{(a)}_x = 0$ gauge and enforcing Gauss's law has resulted in all-to-all interactions, the double lattice sum in $H_{el}$.

For any finite lattice system, there are color non-singlet states in the spectrum, which are unphysical and have infinite energy in the continuum and infinite-volume limits.
For a large but finite system, OBCs can also support finite-energy color non-singlet states which are localized to the end of the lattice (color edge-states).\footnote{Low-energy edge-states that have global charge in a confining theory can also be found in the simpler setting of the Schwinger model. 
Through exact and approximate tensor methods, we have verified that these states exist on lattices up to length $L=13$, and they are expected to persist for larger $L$.} 
The existence of such states in the spectrum is independent of the choice of gauge or fermion mapping.
The naive ways to systematically examine basis states and preclude such configurations is found to be impractical due to the non-Abelian nature of the gauge charges
and the resulting entanglement between states required for color neutrality.
A practical way to deal with this problem is to add a term to the Hamiltonian that 
raises the energy of color non-singlet states.
This can be accomplished by including the energy density in the chromo-electric field beyond the end of the lattice with a large coefficient $h$.  
This effectively adds the energy density in a finite chromo-electric field over a large spatial extent beyond the end of the lattice.
In the limit $h\rightarrow\infty$, only states with a vanishing chromo-electric field beyond the end of the lattice remain at finite energy, rendering the system within the lattice to be a color singlet.
This new term in the Hamiltonian is
\begin{equation}
    H_{\bf 1} = \frac{h^2}{2} \sum_{n=0}^{2L-1}
    \left( \sum_{f=0}^{1} Q_{n,f}^{(a)} \, Q_{n,f}^{(a)} \ + \  
    2 Q_{n,0}^{(a)} \, Q_{n,1}^{(a)}
     \right) \ + \  h^2 \sum_{n=0}^{2L-2} \sum_{m=n+1}^{2L-1}\sum_{f=0}^1 \sum_{f'=0}^1 Q_{n,f}^{(a)} \, Q_{m,f'}^{(a)} \ ,
     \label{eq:Hpen}
\end{equation}
which makes a vanishing contribution when the sum of charges over the whole lattice is zero; otherwise, it makes a contribution $\sim h^2$.

\subsection{Spectra for \texorpdfstring{\boldmath$L=1, 2$}{L=1,2} Spatial Sites}
\label{sec:Exact}
\noindent
The spectra and wavefunctions of systems with a small number of lattice sites can be determined by diagonalization of the Hamiltonian.
In terms of spin operators, the $N_f=2$ Hamiltonian in Eq.~(\ref{eq:H2flav}) decomposes into sums of tensor products of Pauli matrices. The tensor product factorization can be exploited to perform an exact diagonalization relatively efficiently. 
This is accomplished by first constructing a basis
by projecting onto states with specific quantum numbers, and then building the Hamiltonian in that subspace. 
There are four mutually commuting symmetry generators that allow states to be labelled by $(r,g,b,I_3)$: redness, greenness, blueness and the third component of isospin.
In the computational (occupation) basis, states are represented by bit strings of $0$s and $1$s. For example, the $L=1$ state with no occupation is $\ket{000000111111}$.\footnote{Qubits are read from right to left, e.g., $\ket{q_{11} \, q_{10}\, \ldots \, q_{1}\, q_{0}}$. Spin up is $\ket{0}$ and spin down is $\ket{1}$.} Projecting onto eigenstates of $(r,g,b,I_3)$ amounts to fixing the total number of $1$s in a substring of a state.
The Hamiltonian is formed by evaluating matrix elements of Pauli strings between states in the basis, and only involves $2\times 2$ matrix multiplication.
The Hamiltonian matrix is found to be sparse, as expected, and the low energy eigenvalues and eigenstates can be found straightforwardly.
As the dimension of the Hamiltonian grows exponentially with the spatial extent of the lattice, this method becomes intractable for large system sizes, as is well known.

\subsubsection{Exact Diagonalizations, Color Edge-States and Mass Decompositions of the Hadrons}
\noindent
For small enough systems, an exact diagonalization of the Hamiltonian matrix in the previously described basis can be performed.
Without chiral symmetry and its spontaneous breaking, the energy spectrum in $1+1$D does not contain a massless isovector state (corresponding to the QCD pion) in the limit of vanishing quark masses.
In the absence of 
chemical potentials for baryon number, $\mu_B=0$, or isospin, $\mu_I=0$,
the vacuum, $\ket{\Omega}$, has $B=0$
(baryon number zero) and $I=0$ (zero total isospin). 
The $I=0$
$\sigma$-meson is the lightest meson,
while the $I=1$ $\pi$-meson is the next lightest.
The lowest-lying eigenstates in the 
$B=0$ spectra for $L=1,2$
(obtained from
exact diagonalization of the Hamiltonian) 
are given in Table~\ref{tab:specB0}. 
The masses are defined by their energy gap to the vacuum, 
and all results in this section are for $m_u=m_d=m=1$.
\begin{table}[!ht]
\renewcommand{\arraystretch}{1.2}
\begin{tabularx}{0.3\textwidth}{||c | Y | Y | Y ||} 
\hline
\multicolumn{4}{||c||}{$L=1 $} \\
 \hline
 $g^2$ & $E_{\Omega}$ & $M_{\sigma}$ & $M_{\pi}$ \\
 \hline\hline
 8 & -0.205 & 5.73 & 5.82 \\ 
 \hline
 4 & -0.321 & 4.37 & 4.47\\
 \hline
 2 & -0.445 & 3.26 & 3.30 \\
 \hline
 1 & -0.549 & 2.73 & 2.74\\
 \hline
 1/2 & -0.619 & 2.48 & 2.48\\
 \hline
 1/4 & -0.661 & 2.35 & 2.36 \\
 \hline
 1/8 & -0.684 &  2.29 & 2.30 \\
 \hline
\end{tabularx}
\renewcommand{\arraystretch}{1}
\qquad\qquad\qquad
\renewcommand{\arraystretch}{1.2}
\begin{tabularx}{0.3\textwidth}{||c | Y | Y | Y ||} 
\hline
\multicolumn{4}{||c||}{$L=2 $} \\
 \hline
 $g^2$ & $E_{\Omega}$ & $M_{\sigma}$ & $M_{\pi}$\\
 \hline\hline
 8 & -0.611 & 5.82 & 5.92 \\ 
 \hline
 4 & -0.949 & 4.41 & 4.49 \\
 \hline
 2 & -1.30 & 3.27 & 3.31 \\
 \hline
 1 & -1.58 & 2.72 & 2.74 \\
 \hline
 1/2 & -1.77 & 2.45 & 2.46 \\
 \hline
 1/4 & -1.88 & 2.30 & 2.31 \\
 \hline
 1/8 & -1.94 & 2.22 & 2.22 \\
 \hline
\end{tabularx}
\renewcommand{\arraystretch}{1}
\caption{
The vacuum energy and the masses of the $\sigma$- and $\pi$-mesons for $1+1$D QCD with $N_f=2$ for systems with $L=1,2$
spatial sites. These results are insensitive to $h$ as they are color singlets.}
\label{tab:specB0}
\end{table}
By examining the vacuum energy density 
$E_{\Omega}/L$, it is clear that, as expected, this number of lattice sites is insufficient to fully contain hadronic correlation lengths.  
While Table~\ref{tab:specB0} shows the energies of color-singlet states, there are also non-singlet states in the spectra with similar masses,
which become increasingly localized near the end of the lattice, as discussed in the previous section.

\begin{figure}[!ht]
    \centering
    \includegraphics[width=14cm]{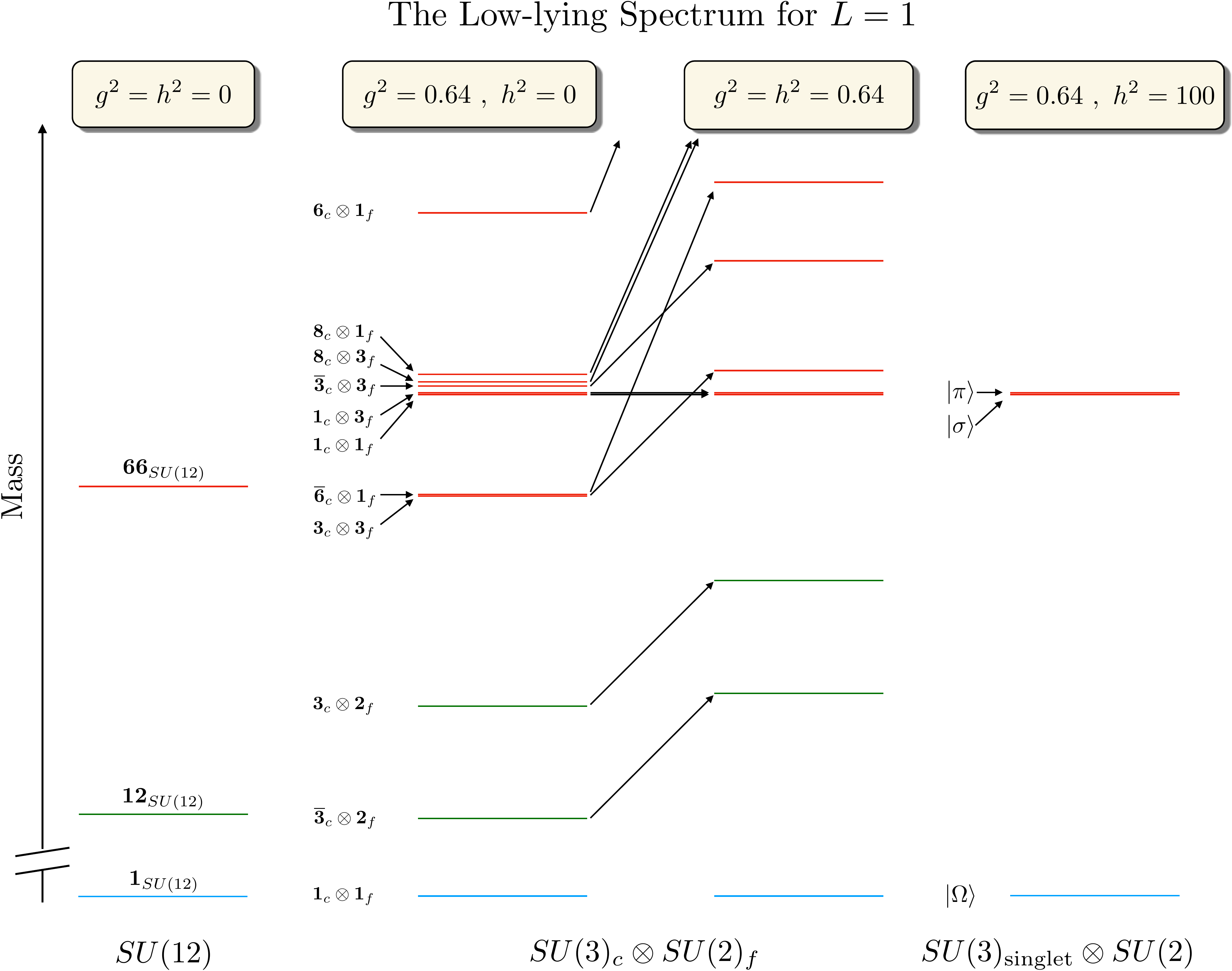}
    \caption{
    The spectrum of the Hamiltonian as the couplings $g$ and $h$ increase. 
    For $g=h=0$ there is an exact $SU(12)$ symmetry and the color-singlet 
    $\sigma$- and $\pi$-mesons are a part of the antisymmetric ${\bf 66}$ irrep. 
    When $g>0$ and $h=0$, the spectrum splits into irreps of global $SU(3)_c \otimes SU(2)_f$ with 
    color non-singlet states among the low-lying states.
    Increasing $h>0$ pushes non-singlet color states out of the low-lying spectrum. 
    Notice that the $\sigma$ and $\pi$ masses are insensitive to $h$, as expected.}
    \label{fig:specDegenh}
\end{figure}

It is informative to examine the spectrum of the $L=1$ system as both $g$ and $h$ are slowly increased and, in particular, take note of the relevant symmetries. For $g=h=0$,
with contributions from only the hopping and mass terms,
the system exhibits a global $SU(12)$ symmetry 
where the spectrum is that of free quasi-particles; see App.~\ref{app:freeSym}.
The enhanced global symmetry at this special point restricts the structure of the spectrum to the ${\bf 1}$ and ${\bf 12}$ of $SU(12)$ as well as the antisymmetric combinations of fundamental irreps, ${\bf 66}, {\bf 220}, \ldots$.  
For $g>0$, these $SU(12)$ irreps split into irreps of color $SU(3)_c$ and flavor $SU(2)_f$. 
The ${\bf 12}$ corresponds to single quark ($q$) or antiquark ($\overline{q}$) excitations
(with fractional baryon number), and splits into ${\bf 3}_c\otimes {\bf 2}_f$ for quarks and $\overline{\bf 3}_c\otimes
{\bf 2}_f$ for antiquarks.  In the absence of OBCs, these states would remain degenerate, but the boundary condition of vanishing background
chromo-electric field is not invariant under 
$q\leftrightarrow \overline{q}$ and the quarks get pushed to higher mass. As there is no chromo-electric energy associated with exciting an
antiquark at the end of the lattice in this mapping, the $\overline{\bf 3}_c\otimes {\bf 2}_f$ states remains low in the spectrum until $h\gg0$.
The ${\bf 66}$ corresponds to two-particle excitations, and contains all combinations of $qq$, $\overline{q}q$ and 
$\overline{q} \overline{q}$ excitations. 
The mixed color symmetry (i.e., neither symmetric or antisymmetric) of $\overline{q}q$ excitations allows for states with
${\bf 1}_c\otimes {\bf 1}_f
\oplus
{\bf 1}_c\otimes {\bf 3}_f
\oplus
{\bf 8}_c\otimes {\bf 1}_f
\oplus
{\bf 8}_c\otimes {\bf 3}_f
$,
while the $qq$ excitations with definite color symmetry allow for 
${\bf 6}_c\otimes {\bf 1}_f
\oplus \overline{\bf 3}_c\otimes {\bf 3}_f
$
and
$\overline{q}\overline{q}$ excitations allow for
$\overline{\bf 6}_c\otimes {\bf 1}_f
\oplus {\bf 3}_c\otimes {\bf 3}_f$,
saturating the $66$ states in the multiplet.
When $g>0$, these different configurations split in energy, and when
$h\gg0$, only color-singlet states are left in the low-lying spectrum. Figure~\ref{fig:specDegenh} shows the evolution of the spectrum as 
$g$ and $h$ increase.
The increase in mass of non-singlet color states  with $h$ is proportional to the Casimir of the $SU(3)_c$ representation which is evident in Fig.~\ref{fig:specDegenh} where, for example, the increase in the mass of the ${\bf 3}_c$s and $\overline{{\bf 3}}_c$s between $h^2 = 0$ and $h^2=0.64$ are the same.

\begin{figure}[!ht]
    \centering
    \includegraphics[width=12cm]{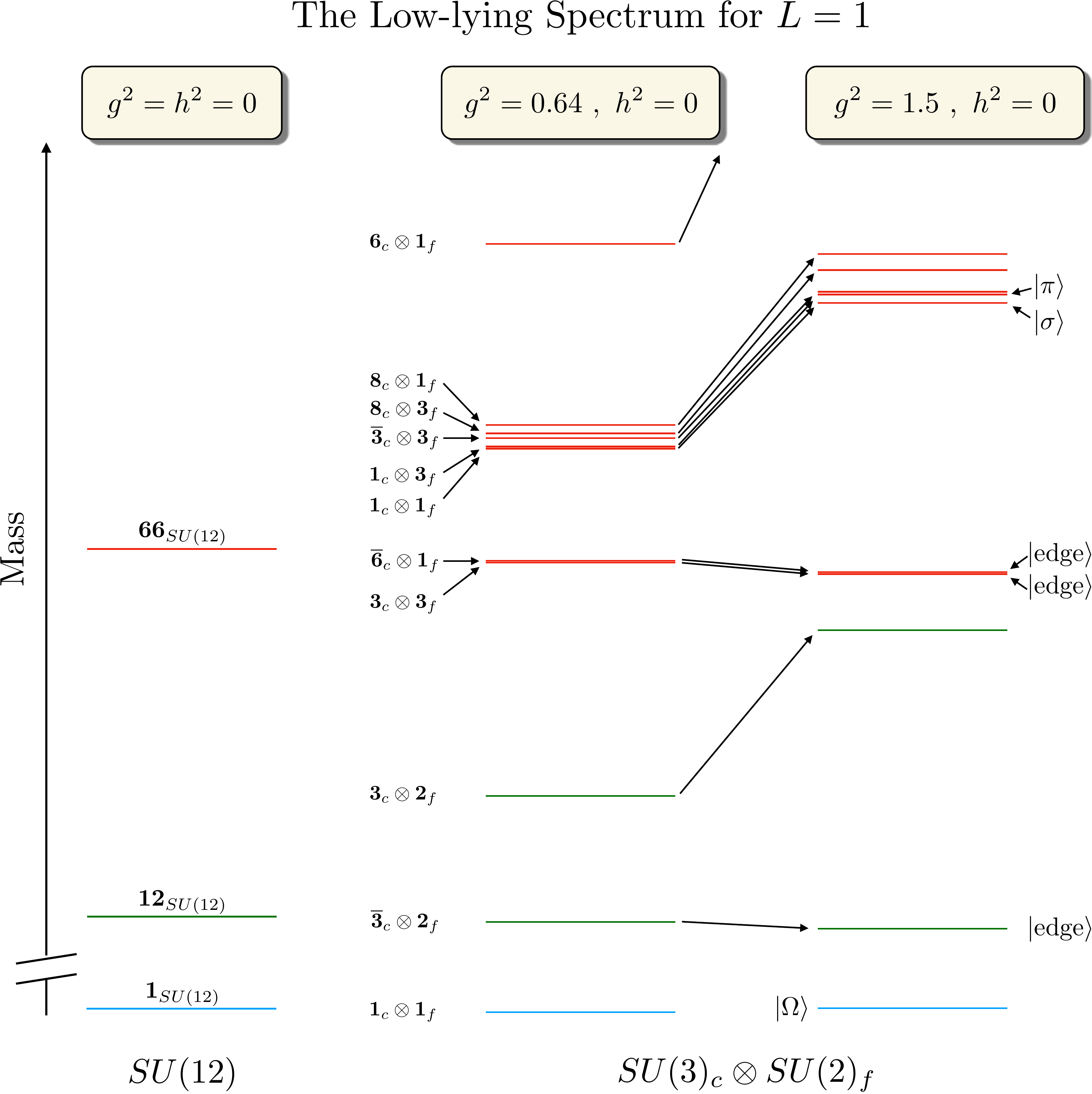}
    \caption{
    The spectrum of the Hamiltonian as $g$ increases for $h=0$. 
    When $g=h=0$ there is an exact $SU(12)$ symmetry and the $\sigma$- and $\pi$-mesons are a part of the antisymmetric ${\bf 66}$ irrep. 
    When $g>0$  but $h=0$ the spectrum splits into irreps of global $SU(3)_c \otimes SU(2)_f$, and non-singlet color states remain in the low-lying spectrum. 
    Increasing $g$ shifts all but the antiquark $\ket{\text{edge}}$ (states) to higher mass.
    }
    \label{fig:specDegeng}
\end{figure}

The antiquark states are particularly interesting as they correspond to edge states that are not ``penalized" in energy by the chromo-electric field when $h=0$. 
These states have an approximate
$SU(6)$ symmetry where the $6$ antiquarks transform in the fundamental.  
This is evident in the spectrum shown in 
Fig.~\ref{fig:specDegeng}
by the presence of a $\overline{{\bf 3}}_c \otimes {\bf 2}_f$ 
and nearly degenerate 
$\overline{\bf 6}_c\otimes {\bf 1}_f$ 
and 
${\bf 3}_c\otimes {\bf 3}_f$
which are identified as states of a 
${\bf 15}$ 
(an antisymmetric irrep of $SU(6)$)
that do not increase in mass as $g$ increases.
This edge-state $SU(6)$ symmetry is not exact 
due to interactions from the hopping term that couple the edge $\overline{q}$s to the rest of the lattice. 
These colored edge states are artifacts of OBCs and will persist in the low-lying spectrum for larger lattices.

Figures~\ref{fig:specDegenh} and \ref{fig:specDegeng}
reveal the near-degeneracy of the  $\sigma$- and $\pi$-mesons throughout the range of couplings $g$ and $h$, suggesting another approximate symmetry, which
can be understood in the small and large $g$ limits.
For small $g^2$, the effect of $H_{el} = \frac{g^2}{2}(Q_{0,u}^{(a)} + Q_{0,d}^{(a)})^2$ on the the $SU(12)$-symmetric spectrum can be obtained through perturbation theory.
To first order in $g^2$, the shift in the energy of any state is equal to the expectation value of $H_{el}$.
The $\sigma$- and $\pi$-meson states are both quark-antiquark states in the {\bf 66} irrep of $SU(12)$, and therefore, both have a ${\bf 3}_c$ color charge on the quark site and receive the same mass shift.\footnote{This also explains why
there are three other states nearly degenerate with the mesons, as seen in Fig.~\ref{fig:specDegenh}. 
Each of these states carry a ${\bf 3}_c$ or $\overline{{\bf 3}}_c$ color charge on the quark site and consequently have the same energy at first order in perturbation theory.
}
For large $g^2$, the only finite-energy excitations of the trivial vacuum (all sites unoccupied) are bare baryons and antibaryons,
and the spectrum is one of non-interacting color-singlet baryons.
Each quark (antiquark) site hosts $4$ distinct baryons (antibaryons) in correspondence with the multiplicity of the $I=3/2$ irrep.
As a result, the $\sigma$, $\pi$, $I=2,3$ mesons, deuteron and antideuteron are all degenerate.

\begin{figure}[!ht]
    \centering
    \includegraphics[width=14cm]{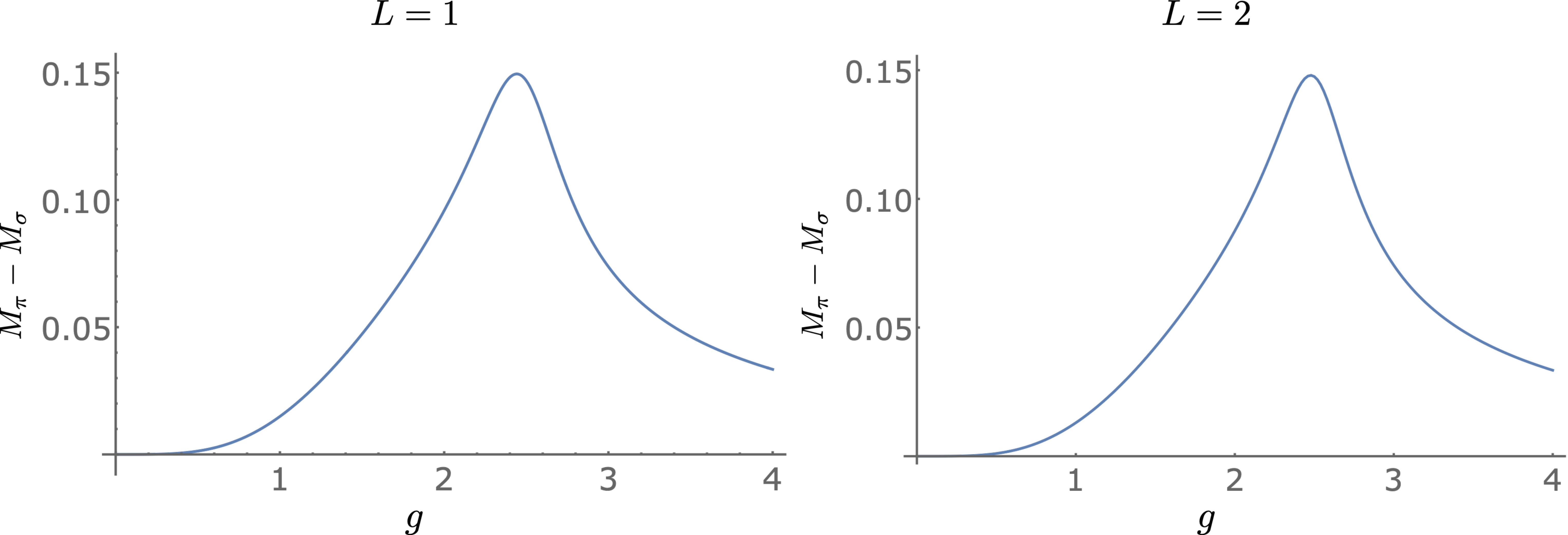}
    \caption{
    The mass splitting between the $\sigma$- and $\pi$-mesons for $L=1$ (left panel) and $L=2$ (right panel).}
    \label{fig:PiSigSplit}
\end{figure}

The $\sigma$- and $\pi$-meson mass splitting is shown in Fig.~\ref{fig:PiSigSplit} and has a clear maxima for $g \sim 2.4$. 
Intriguingly, this corresponds to the maximum of the linear entropy between quark and antiquarks (as discussed in Sec.~\ref{sec:eigenent}),
and suggests a connection between symmetry, via degeneracies in the spectrum, and entanglement.
This shares similarities with the correspondence between Wigner's $SU(4)$ spin-flavor
symmetry~\cite{PhysRev.51.106,PhysRev.51.947,PhysRev.56.519},
which becomes manifest in low-energy nuclear forces in the large-$N_c$ limit of QCD~\cite{Kaplan:1995yg,Kaplan:1996rk},
and entanglement suppression in
nucleon-nucleon scattering found in Ref.~\cite{Beane:2018oxh} (see also Refs.~\cite{Beane:2021zvo,Low:2021ufv,Beane:2020wjl}).

Color singlet baryons are also present in this system, formed by contracting the color indices of three quarks with a Levi-Civita tensor (and antibaryons are formed from three antiquarks).
A baryon is composed of three $I=1/2$ quarks in the (symmetric) $I=3/2$ configuration and in a (antisymmetric) color singlet.
It will be referred to as the $\Delta$, highlighting its similarity to the $\Delta$-resonance in $3+1$D QCD. 
Interestingly, there is an isoscalar $\Delta\Delta$ bound state, which will be referred to as the deuteron. 
The existence of a deuteron makes this system valuable from the standpoint of quantum simulations of the formation of nuclei in a model of reduced complexity.
The mass of the $\Delta$, $M_{\Delta}$, and the binding energy of the deuteron, $B_{\Delta \Delta} = 2 M_{\Delta} - M_{\Delta \Delta}$, are shown in Table~\ref{tab:specBneq0} for a range of strong couplings.
\begin{table}[!ht]
\renewcommand{\arraystretch}{1.2}
\begin{tabularx}{0.3\textwidth}{||c | Y | Y ||} 
\hline
\multicolumn{3}{||c||}{$L=1 $} \\
 \hline
 $g^2$ & $M_{\Delta}$ & $B_{\Delta \Delta}$\\
 \hline\hline
 8 & 3.10 & $2.61\times 10^{-4}$\\ 
 \hline
 4 & 3.16 & $5.48\times 10^{-4}$\\
 \hline
 2 & 3.22 & $6.12\times 10^{-4}$\\
 \hline
 1 & 3.27 & $3.84\times 10^{-4}$\\
 \hline
 1/2 & 3.31 & $1.61\times 10^{-4}$\\
 \hline
 1/4 & 3.33 & $5.27\times 10^{-5}$\\
 \hline
 1/8 & 3.34 & $1.52\times 10^{-5}$\\
 \hline
\end{tabularx}
\renewcommand{\arraystretch}{1}
\qquad\qquad\qquad
\renewcommand{\arraystretch}{1.2}
\begin{tabularx}{0.3\textwidth}{|| c | Y | Y ||} 
\hline
\multicolumn{3}{||c||}{$L=2 $} \\
 \hline
 $g^2$ & $M_{\Delta}$ & $B_{\Delta \Delta}$\\
 \hline\hline
 8  & 3.10 & $2.50\times 10^{-4}$\\ 
 \hline
 4  & 3.16 & $4.95\times 10^{-4}$\\
 \hline
 2  & 3.21 & $5.07\times 10^{-4}$\\
 \hline
 1  & 3.24 & $4.60\times 10^{-4}$\\
 \hline
 1/2  & 3.25 & $1.53\times 10^{-3}$\\
 \hline
 1/4 & 3.23 & $3.91\times 10^{-3}$\\
 \hline
 1/8 & 3.20 & $3.35\times 10^{-3}$\\
 \hline
\end{tabularx}
\renewcommand{\arraystretch}{1}
\caption{
The mass of the $\Delta$ and the binding energy of the deuteron 
in $1+1$D QCD with $N_f=2$ for systems with $L=1,2$ spatial sites.}
\label{tab:specBneq0}
\end{table}

\begin{figure}[!ht]
    \centering
    \includegraphics[width=\columnwidth]{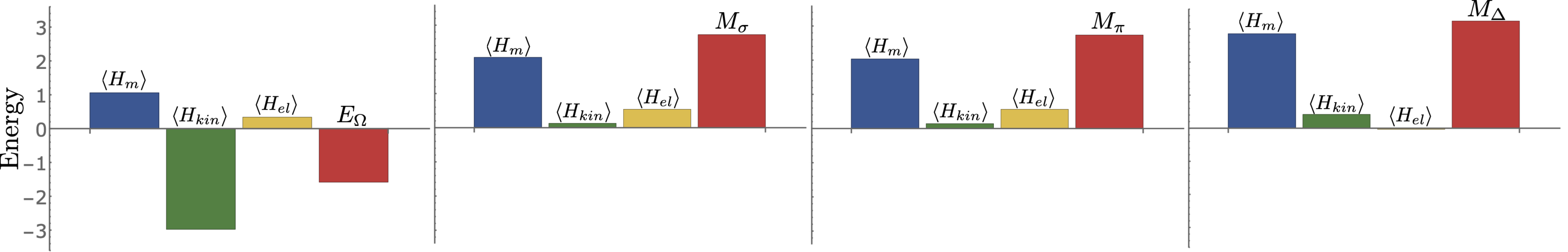}
    \caption{
    The decomposition of vacuum energy ($E_{\Omega}$) and the masses of the lightest hadrons ($M_{\sigma}$, $M_{\pi}$ and $M_{\Delta}$) 
    into contributions from the mass, the kinetic and the chromo-electric field terms in the Hamiltonian, defined in axial gauge, for $1+1$D QCD with $N_f=L=2$ and $m=g=1$.}
    \label{fig:massdeco}
\end{figure}

Understanding and quantifying the structure of the lowest-lying hadrons is a priority for nuclear physics research~\cite{LongRangePlan}.
Great progress has been made, experimentally, analytically and computationally,
in dissecting the mass and angular momentum of the proton (see, for example, Refs.~\cite{deFlorian:2009vb,Nocera:2014gqa,COMPASS:2015mhb,Yang:2018nqn,Alexandrou:2020sml,Ji:2021mtz,Wang:2021vqy,Lorce:2021xku}). 
This provides, in part, the foundation for anticipated precision studies at the future electron-ion collider (EIC)~\cite{Boer:2011fh,Accardi:2012qut} at Brookhaven National Laboratory.
Decompositions of the vacuum energy and the masses of the $\sigma$, $\pi$ and $\Delta$ are shown in Fig.~\ref{fig:massdeco} where, for example, the chromo-electric contribution to the
$\sigma$ is $\langle H_{el} \rangle = \bra{\sigma}  H_{el} \ket{\sigma} - \bra{\Omega}  H_{el} \ket{\Omega}$.
These calculations demonstrate the potential of future quantum simulations in being able to quantify decompositions of properties of the nucleon,
including in dense matter.
For the baryon states, it is $H_{el}$ that is responsible for the system coalescing into localized color singlets in order to minimize the energy in the chromo-electric field (between spatial sites).

\begin{figure}[!ht]
    \centering
    \includegraphics[width=14cm]{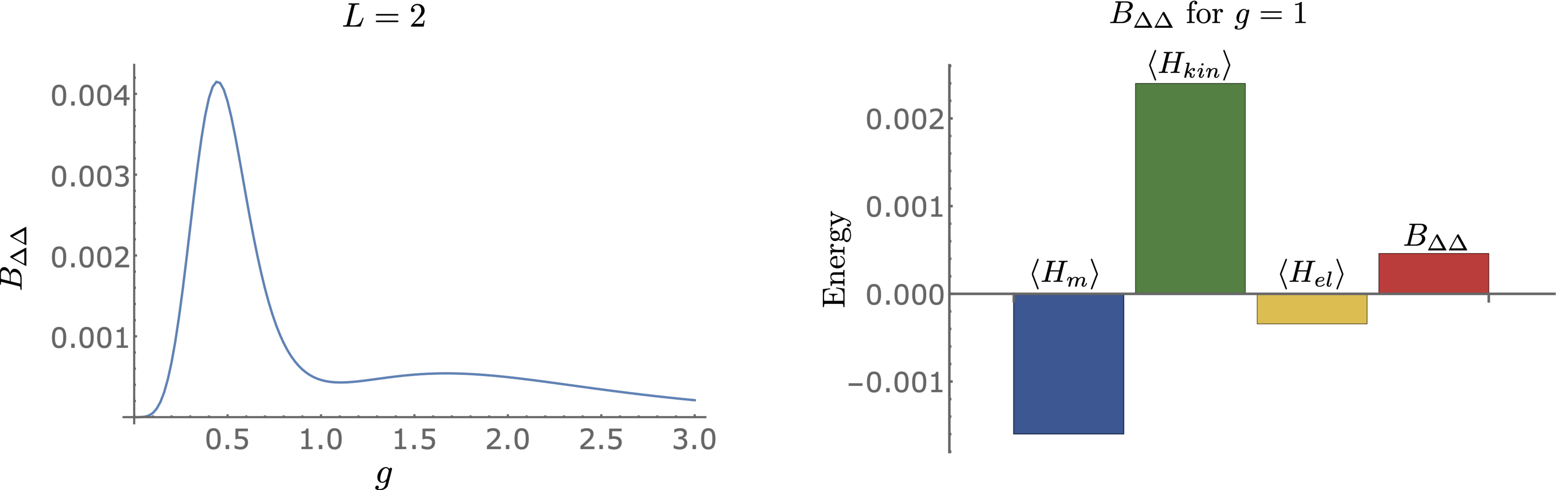}
    \caption{The left panel shows the deuteron binding energy, $B_{\Delta \Delta}$, for $m=1$ and $L=2$. The right panel shows the decomposition of $B_{\Delta \Delta}$ into contributions from the Hamiltonian for $g=1$.}
    \label{fig:deutBE}
\end{figure}

The deuteron binding energy is shown in the left panel of Fig.~\ref{fig:deutBE} as a function of $g$.
While the deuteron is unbound at $g=0$ for obvious reasons, it is also unbound at large $g$ because the spectrum is that of non-interacting color-singlet (anti)baryons.
Therefore, the non-trivial aspects of deuteron binding for these systems is for intermediate 
values of $g$. The decomposition of $B_{\Delta \Delta}$ is shown in the right panel of Fig.~\ref{fig:deutBE}, where, for example, the chromo-electric contribution is
\begin{equation}
    \langle H_{el} \rangle = 2 \big ( \bra{\Delta}  H_{el} \ket{\Delta} - \bra{\Omega}  H_{el} \ket{\Omega} \big ) - \big (\bra{\Delta \Delta}  H_{el} \ket{\Delta \Delta} - \bra{\Omega}  H_{el} \ket{\Omega} \big ) \ .
    \label{eq:deutBEdec}
\end{equation}
The largest contribution to the binding energy is $\langle H_{kin} \rangle$, which is the term responsible for creating $q \overline{q}$ pairs.
This suggests that meson-exchange may play a significant role in the attraction between baryons,
as is the case in $3+1$D QCD, but larger systems will need to be studied before 
definitive conclusions can be drawn.
One consequence of the lightest baryon 
being $I=3/2$ is that, for $L=1$, 
the $I_3=+3/2$ state completely occupies the up-quark sites.
Thus the system factorizes into an inert up-quark sector and a dynamic down-quark sector, and the absolute energy of the lowest-lying baryon state can be written as 
$E_{\Delta} =
M_{\Delta} + E_{\Omega}^{2 f} = 
3m + E_{\Omega}^{1 f}$,
where
$E_{\Omega}^{1,2 f}$ is the
vacuum energy of the 
$N_f=1,2$ flavor systems.
Analogously, the deuteron absolute energy is 
$E_{\Delta \Delta}=6m$, and therefore the 
deuteron binding energy can be written as
$B_{\Delta \Delta}= 2(3m+E_{\Omega}^{1 f}-E_{\Omega}^{2 f}) - (6m-E_{\Omega}^{2 f})
= 2E_{\Omega}^{1 f}-E_{\Omega}^{2 f}$.
This is quite a remarkable result because, in this system, the deuteron binding energy depends only on the difference between
the $N_f=1$ and $N_f=2$ vacuum energies, being bound when $2 E_{\Omega}^{1 f} - E_{\Omega}^{2 f} > 0$.
As has been discussed previously, it is the 
$q\overline{q}$ contribution from this difference that dominates the binding.
%

\subsubsection{The Low-Lying Spectrum Using D-Wave's Quantum Annealers}
\label{sec:dwave_spectrum}
\noindent
The low-lying spectrum of this system can also be determined through annealing by using
D-Wave's quantum annealer (QA) {\tt Advantage}~\cite{DwaveLeap}, 
a device with 5627 superconducting flux qubits, with a 15-way qubit connectivity via Josephson junctions rf-SQUID couplers~\cite{PhysRevB.80.052506}. 
Not only did this enable the determination of the energies of low-lying states, but it also assessed the ability of this quantum device to isolate nearly degenerate states.
The time-dependent Hamiltonian
of the device, which our systems are to be mapped,
are of the form of an Ising model, with the freedom to specify the single- and two-qubit coefficients. Alternatively, the Ising model can be rewritten in a quadratic unconstrained binary
optimization (QUBO) form, $f_Q(x)=\sum_{ij} Q_{ij}x_i x_j$, 
where $x_i$ are binary variables 
and $Q_{ij}$ is a QUBO matrix, which contains the coefficients of single-qubit ($i=j$) and two-qubit ($i\neq j$) terms. 
The QUBO matrix is the input that is submitted to 
{\tt Advantage}, with the output being a bit-string that minimizes $f_Q$.
Due to the qubit connectivity of {\tt Advantage},
multiple physical qubits are chained together to recover the required connectivity, limiting the system size that can be annealed.

The QA {\tt Advantage} was used to
determine the lowest three states in the $B=0$ sector of the $L=1$ system, with $m=g=1$ and $h=2$, following techniques presented in Ref.~\cite{Illa:2022jqb}. 
In that work, the objective function to be minimized is defined as $F=\langle \Psi \rvert \tilde{H} \lvert \Psi \rangle -\eta \langle \Psi| \Psi \rangle$~\cite{doi:10.1021/acs.jctc.9b00402}, where $\eta$ is a parameter that is included to avoid the null solution, and its optimal value 
can be iteratively tuned to be as close to the ground-state energy as possible. 
The wavefunction is expanded in a finite dimensional orthonormal basis $\psi_{\alpha}$, $\lvert \Psi \rangle =\sum^{n_s}_{\alpha} a_\alpha |\psi_{\alpha}\rangle$, which in this case reduces the dimensionality of $H$ to $88$, defining $\tilde{H}$, thus making it feasible to study with {\tt Advantage}. 
The procedure to write the objective function in a QUBO form can be found in Ref.~\cite{Illa:2022jqb} (and briefly described in App.~\ref{app:dwave}), where the coefficients $a_{\alpha}$ are
digitized using $K$ binary variables~\cite{doi:10.1021/acs.jctc.9b00402}, and the adaptive QA eigenvalue solver is implemented by using the zooming method~\cite{Chang:2019,ARahman:2021ktn}. To reduce the uncertainty in the resulting energy and wavefunction, due to the noisy nature of this QA, the iterative procedure
described in Ref.~\cite{Illa:2022jqb} was used, where the (low-precision) solution obtained from the machine after several zooming steps constituted the starting point of a new anneal. This led to a reduction of the uncertainty by an order of magnitude (while effectively only doubling the resources used).

\begin{figure}[!ht]
    \centering
    \includegraphics[width=\columnwidth]{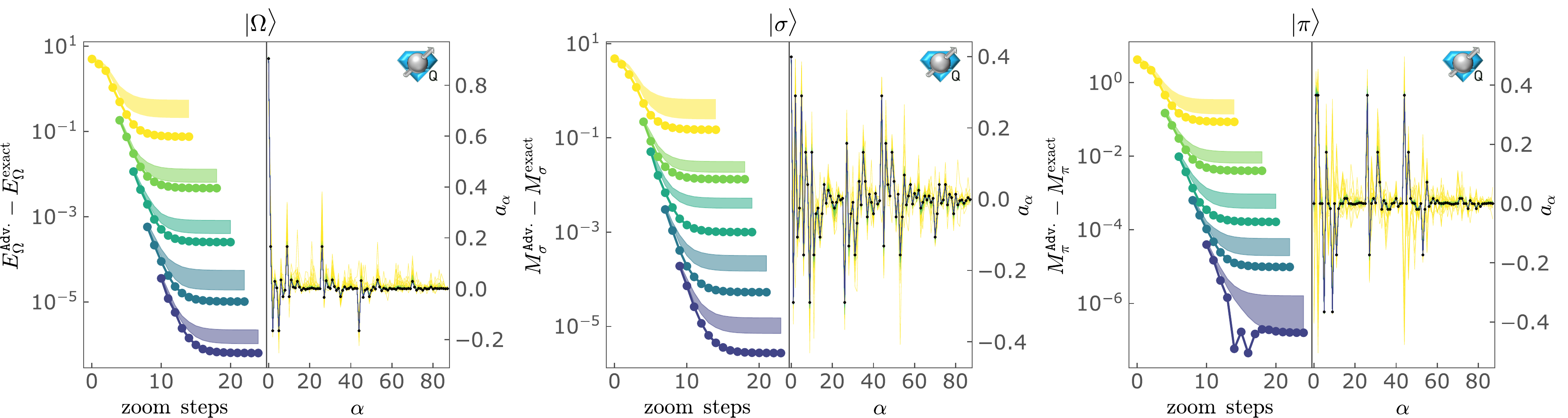}
    \caption{
    Iterative convergence of the energy, masses and wavefunctions for the three lowest-lying states in the $B=0$ sector of $1+1$D QCD with $N_f=2$ and $m=g=L=1$: vacuum (left), $\sigma$-meson (center) and $\pi$-meson (right). 
    The different colors correspond to different steps of the iterative procedure that is described in the main text. 
    The oscillatory behavior seen in the right panel around the 15th zoom step is discussed in App.~\ref{app:dwave}.
    The blue icons in the upper right indicate that this calculation was done on a quantum device~\cite{Klco:2019xro}.}
    \label{fig:QAresults}
\end{figure}

Results obtained using {\tt Advantage} are shown in Fig.~\ref{fig:QAresults}, where the three panels show the convergence of the energy of the 
vacuum state (left), the mass of the $\sigma$-meson (center) and the mass of the $\pi$-meson (right) as a function of zoom steps, as well as comparisons to the exact wavefunctions. The bands in the plot correspond to 68\% confidence intervals determined from 20 independent runs of the annealing workflow, where each corresponds to $10^3$ anneals with an annealing time of $t_A=20$ $\mu$s, and the points correspond to the lowest energy found by the QA. The parameter $K$ in the digitization of $a_{\alpha}$ is set to $K=2$. The parameter $\eta$ is first set close enough to the corresponding energy (e.g., $\eta=0$ for the ground-state), and for the subsequent iterative steps it is set to the lowest energy found in the previous step. The first two excited states are nearly degenerate, and after projecting out the ground state, {\tt Advantage} finds both states in the first step of the iterative procedure (as shown by the yellow lines in the $\pi$ wavefunction of Fig.~\ref{fig:QAresults}). 
However, after one iterative step, the QA converges to one of the two excited states. 
It first finds the second excited state (the $\pi$-meson), and once this state is known with sufficient precision, it can be projected out to study the other excited state. 
The converged values for the energies and masses of these states are shown in Table~\ref{tab:QAresults},
along with the exact results. The uncertainties in these values should be understood as uncertainties on an upper bound of the energy (as they result from a variational calculation). For more details see App.~\ref{app:dwave}.
\begin{table}[!ht]
\renewcommand{\arraystretch}{1.2}
\begin{tabularx}{0.56\textwidth}{|| c | c | Y | Y | Y ||} 
\hline
\multicolumn{5}{||c||}{$L=1$} \\
 \hline
 \multicolumn{2}{||c|}{} & $\ket{\Omega}$ & $\ket{\sigma}$ & $\ket{\pi}$ \\
 \hline\hline
 \multirow{2}{*}{Exact} & Energy & $-0.5491067$ & $2.177749$ & $2.1926786$ \\ 
 \cline{2-5}
 & Mass & - & $2.726855$ & $2.7417853$\\
 \hline
 \multirow{2}{*}{\tt Advantage} &Energy & $-0.5491051(6)$ & $2.177760(4)$ & $2.1926809(7)$ \\ 
 \cline{2-5}
 & Mass & - & $2.726865(4)$ & $2.7417860(9)$\\
 \hline
\end{tabularx}
\renewcommand{\arraystretch}{1}
\caption{Energies and masses of the three lowest-lying states in the $B=0$ sector of $1+1$D QCD with $N_f=2$ and $m=g=L=1$. Shown are the exact results from diagonalization of the Hamiltonian matrix and those obtained from D-Wave's {\tt Advantage}.}
\label{tab:QAresults}
\end{table}
%

\subsubsection{Quark-Antiquark Entanglement in the Spectra via Exact Diagonalization}
\label{sec:eigenent}
\begin{figure}[!ht]
    \centering
    \includegraphics[width=14cm]{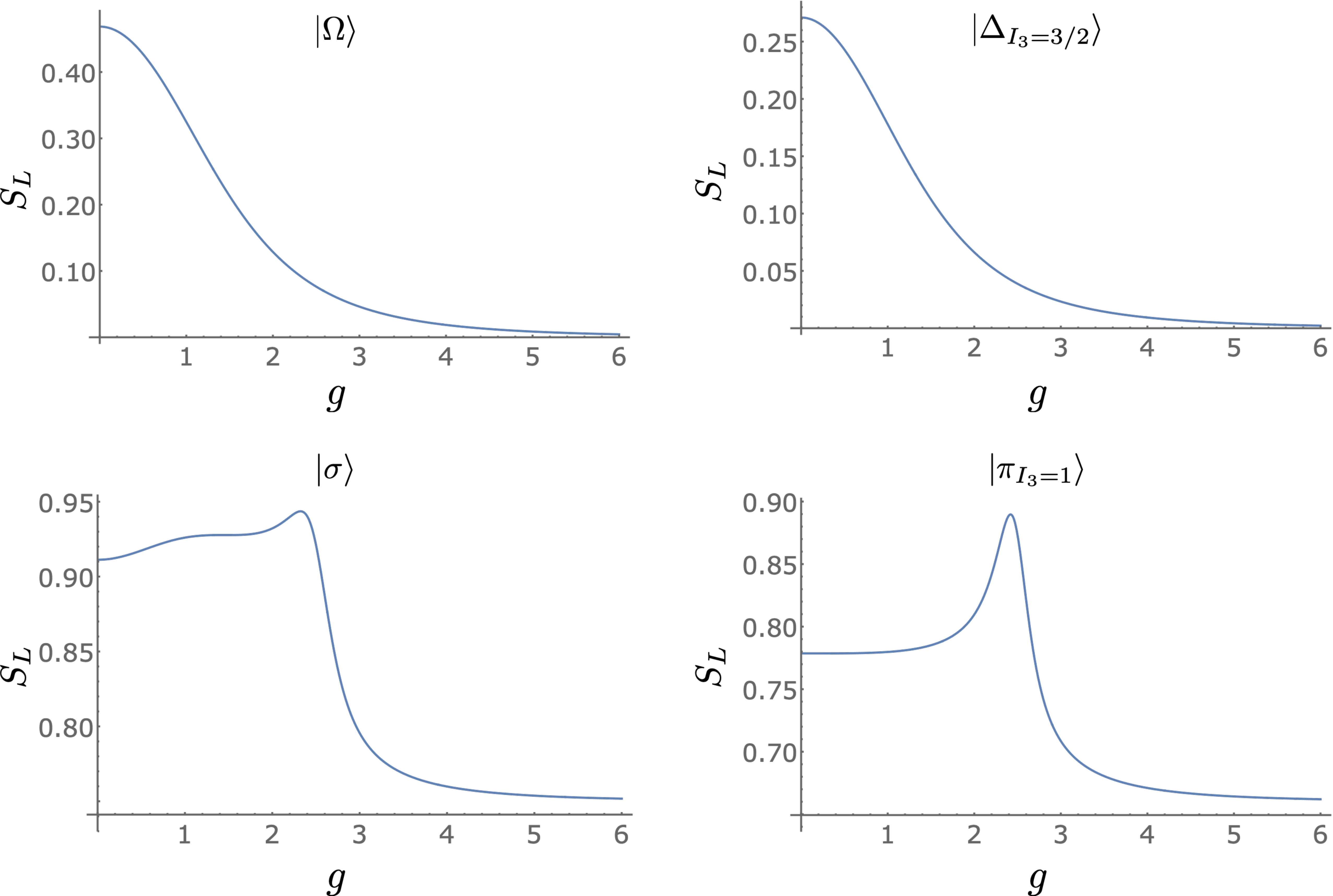}
    \caption{The linear entropy between quarks and antiquarks in $\ket{\Omega}$, $\ket{\Delta_{I_3=3/2}}$, $\ket{\sigma}$ and $\ket{\pi_{I_3=1}}$ for $m=L=1$.}
    \label{fig:m1N2Linent}
\end{figure}

\noindent
With $h \gg g$, the eigenstates of the Hamiltonian are color singlets and irreps of isospin. 
As these are global quantum numbers (summed over the lattice) the eigenstates are generically entangled among the color and isospin components at each lattice site.  With the hope of gaining insight into $3+1$D QCD, aspects of the entanglement structure of the $L=1$ wavefunctions are explored via exact methods.
An interesting measure of entanglement for these systems is the linear entropy between quarks and antiquarks, defined as
\begin{equation}
    S_L = 1 - \Tr [\rho_q^2]
    \ ,
\end{equation}
where $\rho_q =  \Tr_{\overline{q}} [\rho]$ and $\rho$ is a density matrix of the system. Shown in Fig.~\ref{fig:m1N2Linent} is the linear entropy between quarks and antiquarks in
$\ket{\Omega}$, $\ket{\sigma}$, $\ket{\pi_{I_3=1}}$ and $\ket{\Delta_{I_3=3/2}}$ as a function of $g$.
The deuteron is not shown as there is only one basis state contributing for $L=1$.

\begin{figure}
    \centering
    \includegraphics[width=14cm]{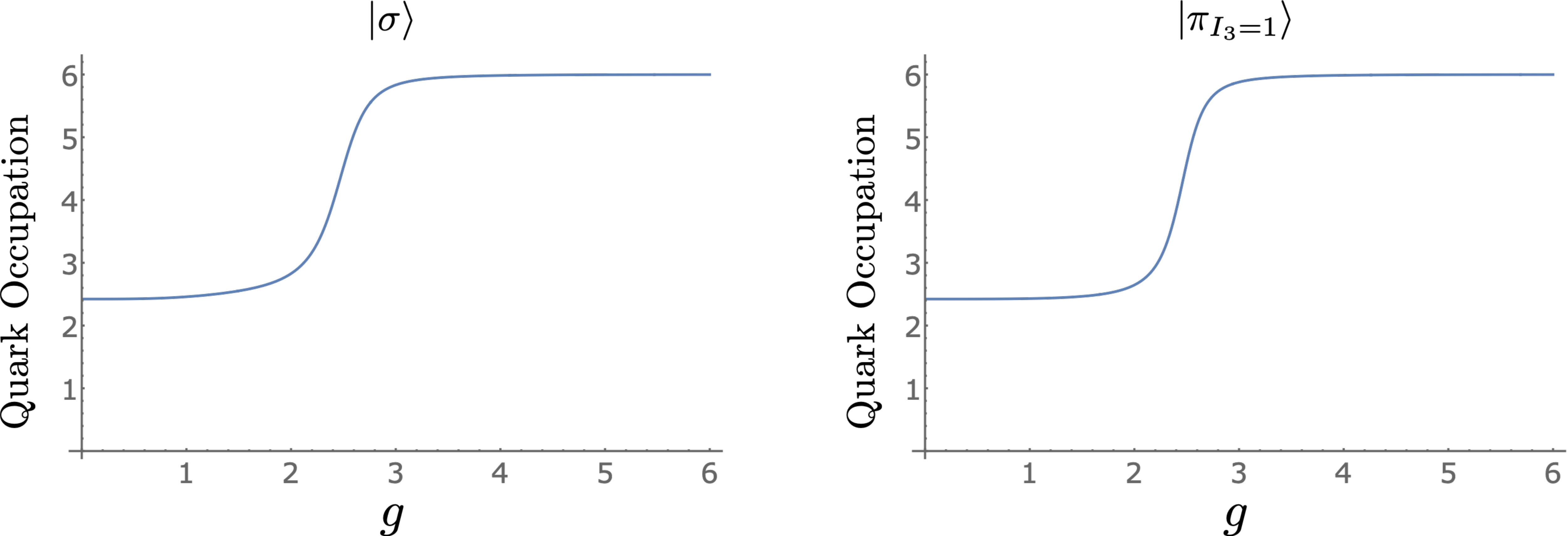}
    \caption{The expectation value of quark occupation in the $\ket{\sigma}$ and $\ket{\pi_{I_3 = 1}}$ for $m=L=1$.}
    \label{fig:m1N2Occ}
\end{figure}

The scaling of the linear entropy in the vacuum and baryon with $g$ can be understood as follows. 
As $g$ increases, color singlets on each site have the least energy density.
The vacuum becomes dominated by the unoccupied state and the $\Delta$ becomes dominated by the ``bare" $\Delta$ with all three quarks located on one site in a color singlet. 
As the entropy generically scales with the number of available states, 
the vacuum and baryon have decreasing entropy for increasing $g$.
The situation for the $\pi$ and $\sigma$ is somewhat more interesting. 
For small $g$, their wavefunctions are dominated by $q \overline{q}$ excitations on top of the trivial vacuum,  which minimizes the contributions from the mass term. 
However, color singlets are preferred as $g$ increases, 
and the mesons become primarily composed of baryon-antibaryon ($B \overline{B}$) excitations.
There are more $q \overline{q}$ states than 
$B \overline{B}$ states with a given $I_3$, 
and therefore there is more entropy at small $g$ than large $g$. 
The peak at intermediate $g$ occurs at the crossover between these two regimes where the meson has a sizable contribution from both $q \overline{q}$ and $B \overline{B}$ excitations. 
To illustrate this, 
the expectation value of total 
quark occupation (number of quarks plus the number of antiquarks) is shown in Fig.~\ref{fig:m1N2Occ}. 
For small $g$, the occupation is near $2$ since the state is mostly composed of $q \overline{q}$,
while for large $g$ it approaches $6$ as the state mostly consists of $B \overline{B}$. 
This is a transition from the excitations being 
``color-flux tubes" between quark and antiquark of the same color to bound states of color-singlet baryons and antibaryons.
%

\subsection{Digital Quantum Circuits}
\label{sec:Circuits}
\noindent
The Hamiltonian for $1+1$D QCD with arbitrary $N_c$ and $N_f$, when written in terms of spin operators, can be naturally mapped onto a quantum device with qubit registers. In this section the time evolution for systems with $N_c = 3$ and $N_f=2$ are developed.

\subsubsection{Time Evolution}
\begin{figure}[!ht]
    \centering
    \includegraphics[height=0.3\textheight]{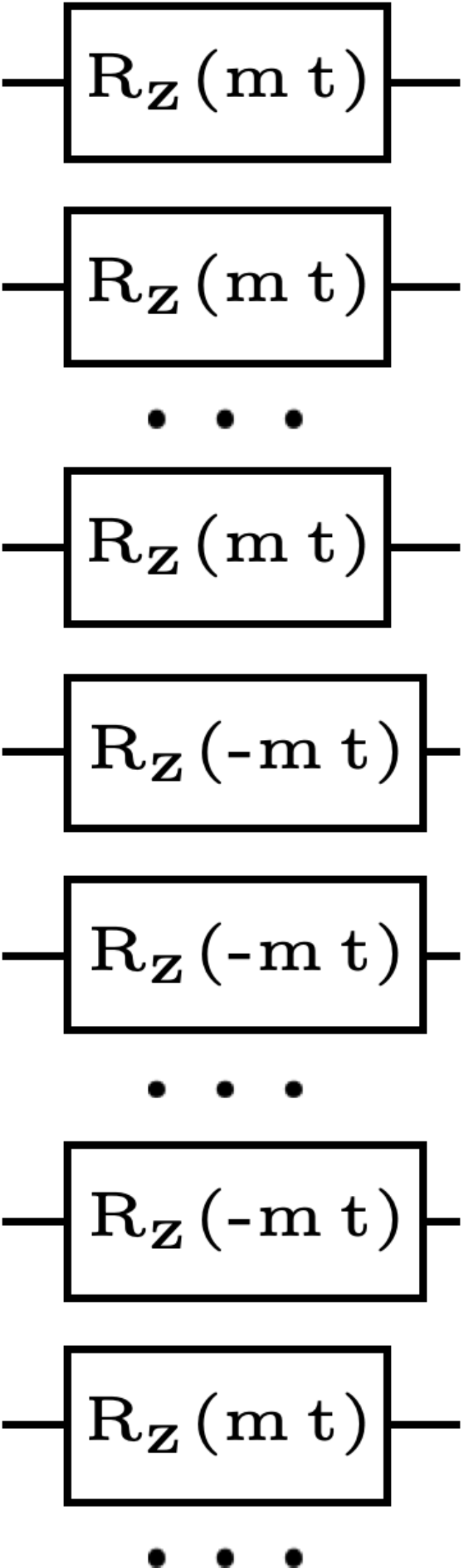}
    \caption{The quantum circuit that implements
    time evolution by the mass term,
    $U_m(t) = \exp(- i H_m t)$.}
    \label{circ:Um}
\end{figure}

\noindent
To perform time evolution on a quantum computer, the operator $U(t) = \exp(-i H t)$ is reproduced by a sequence of gates applied to the qubit register.
Generally, a Hamiltonian cannot be directly mapped to such a sequence efficiently, but each of the elements in a Trotter decomposition can, with systematically reducible errors.
Typically, the Hamiltonian is divided into Pauli strings whose unitary evolution can be implemented with quantum circuits that are readily constructed.
For a Trotter step of size $t$, the circuit that implements the time evolution from the mass term, $U_m(t) = \exp(- i H_m t)$, is shown in Fig.~\ref{circ:Um}.
The staggered mass leads to quarks being rotated by a positive angle and antiquarks being rotated by a negative angle. 
Only single qubit rotations about the z-axis are required for its implementation, with 
$R_Z(\theta) = \exp(-i \theta Z/2)$.
The circuit that implements
the evolution from the baryon chemical potential, $\mu_B$,
$U_{\mu_B}(t) = \exp(- i H_{\mu_B} t)$, 
is similar to $U_m(t)$  with 
$m \to \mu_B/3$, and with both quarks and antiquarks rotated by the same angle. 
Similarly, the circuit that implements the evolution from
the isospin chemical potential, $\mu_I$,
$U_{\mu_I}(t) = \exp(- i H_{\mu_I} t)$, 
is similar to $U_m(t)$ with $m \to
\mu_I/2$ and up (down) quarks rotated by a negative (positive) angle.

\begin{figure}[!ht]
    \centering
    \includegraphics[width=17cm]{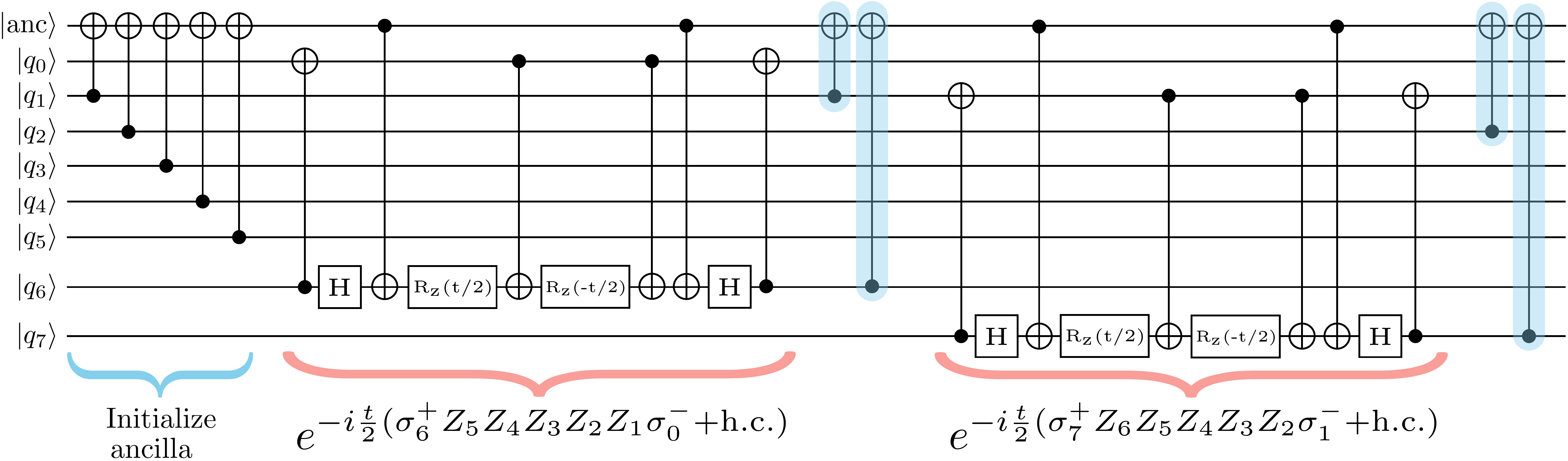}
    \caption{
    A circuit that implements the time evolution from two sequential hopping terms.
    Implementing $\exp(-i H_{kin} t)$ in Eq.~(\ref{eq:Hkin2flav}) is a straightforward extension of this circuit.}
    \label{circ:UkinAnc}
\end{figure}

The kinetic piece of the Hamiltonian, Eq.~(\ref{eq:Hkin2flav}), is composed of hopping terms of the form
\begin{equation}
H_{kin} \ \sim \ 
    \sigma^+ ZZZZZ \sigma^- + \rm{h.c.} \ .
    \label{eq:hop}
\end{equation}
The $\sigma^+$ and $\sigma^-$ operators enable quarks and antiquarks to move between sites with the same color and flavor 
(create $\overline{q}^\alpha_i q_\alpha^i$ pairs)
and the string of $Z$ operators 
incorporates the signs from Pauli statistics.
The circuits for Trotterizing these terms are 
based on circuits in Ref.~\cite{Stetina:2020abi}. We introduce an ancilla to 
accumulate the parity of the JW string of $Z$s.
This provides a mechanism for the 
different hopping terms to re-use 
previously computed
(partial-)parity.\footnote{An ancilla was used similarly in Ref.~\cite{Qchem2014}.}
The circuit for the first two hopping terms is shown in Fig.~\ref{circ:UkinAnc}.
The first circuit operations initialize 
the ancilla to store the parity of the string of $Z$s between the first and last qubit of the string. Next, the system is evolved
by the exponential of the hopping term. After the exponential of each hopping term, the ancilla is modified for the parity of the subsequent hopping term 
(the CNOTs highlighted in blue).
Note that the hopping of quarks, or antiquarks, of different flavors and colors commute, and the Trotter decomposition is exact (without Trotterization errors) over a single spatial site.

\begin{figure}[!ht]
    \centering
    \includegraphics[width=10cm]{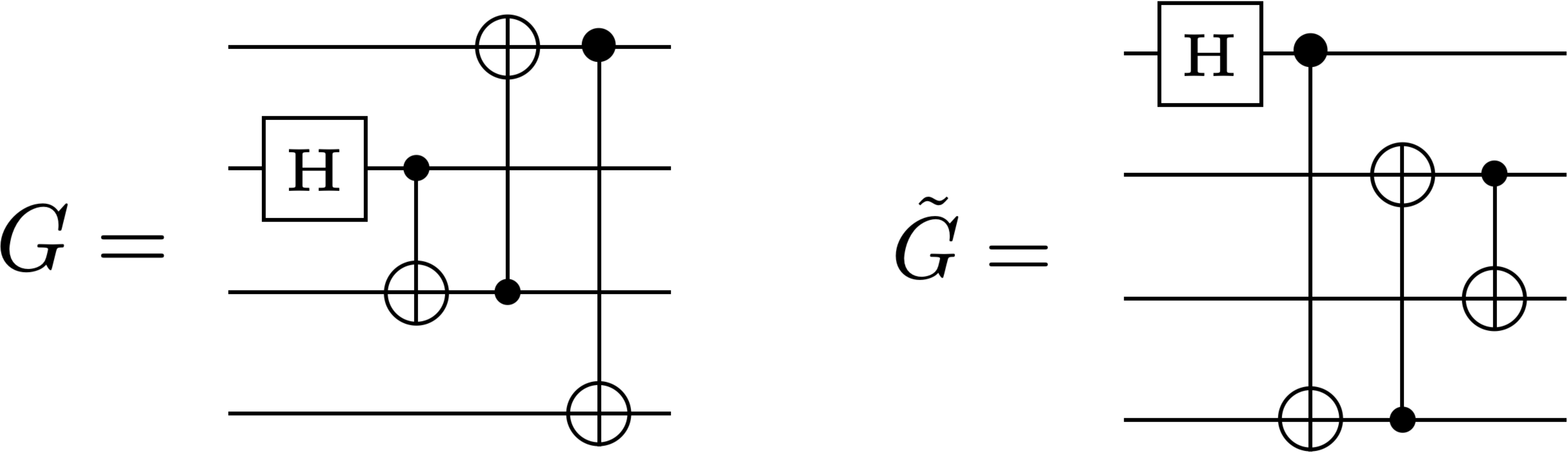}
    \caption{Two GHZ state-preparation circuits.}
    \label{circ:GHZ}
\end{figure}

\begin{figure}[!ht]
    \centering
    \includegraphics[width=15cm]{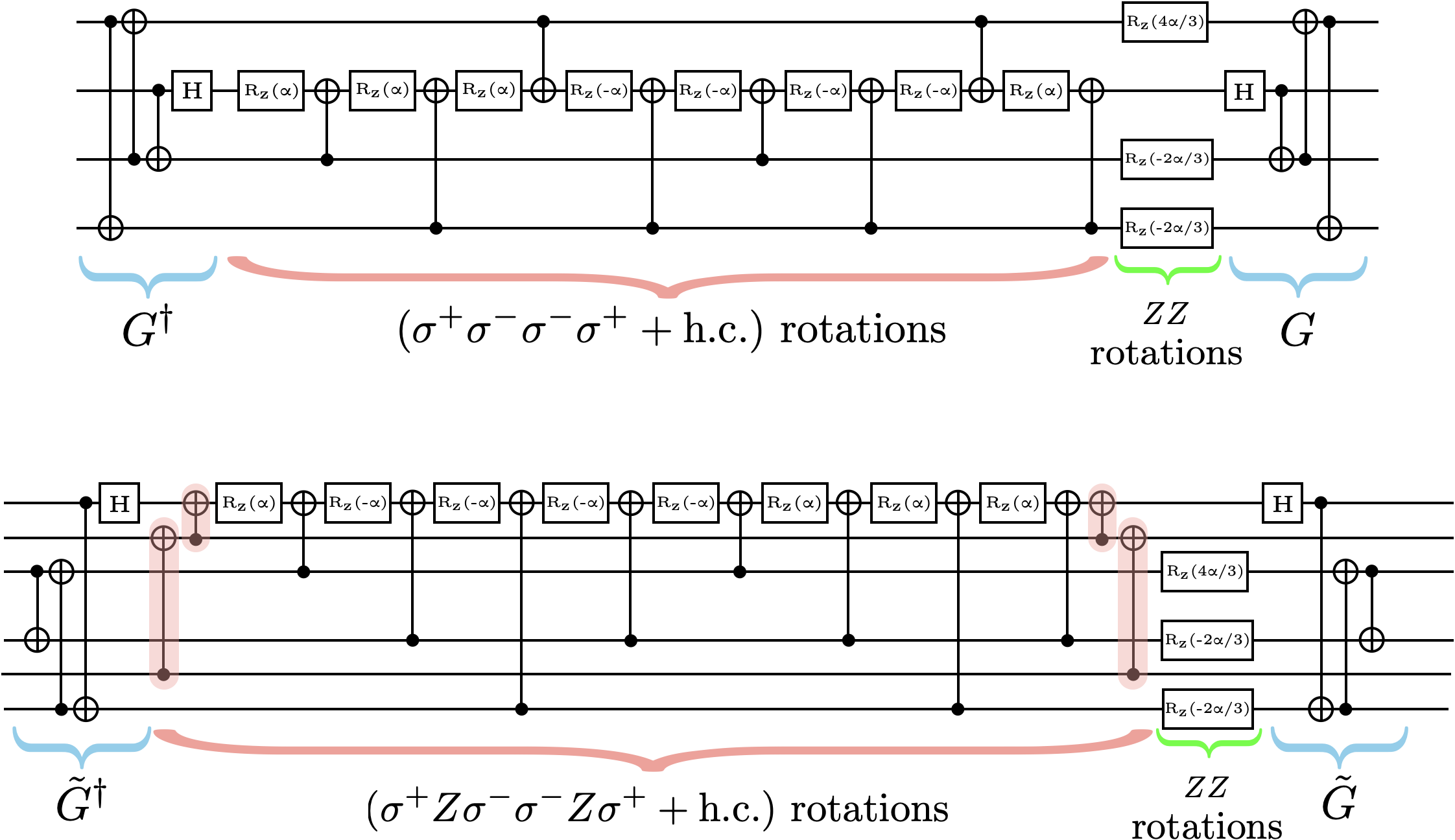}
    \caption{The circuits that implement the time evolution of ${\exp}(-8 i \alpha Q_{n,f}^{(a)} \, Q_{m,f'}^{(a)})$. 
    Specifically, the upper circuit implements
    $\exp{-i 4 \alpha [ (\sigma^+\sigma^-\sigma^-\sigma^+ + {\rm h.c.}) +  \frac{1}{12}(2 IZIZ -IZZI - ZIIZ) ]}$, 
    while the lower circuit implements
    $\exp{-i 4 \alpha[ (\sigma^+ Z \sigma^-\sigma^- Z \sigma^- + {\rm h.c.}) +  \frac{1}{12}(2 ZIIZII -IIZZII - ZIIIIZ) ]}$.
    The CNOTs highlighted in red account for the $Z$s in $\sigma^+ Z \sigma^- \sigma^- Z \sigma^+$.
    For 
    $SU(3)$ with $N_f=2$ and 
    $L=1$, the required evolution operators have $\alpha = t g^2 /8$.}
    \label{circ:UpmmpZZ}
\end{figure}

Implementation of the time-evolution 
induced by the energy density in the 
chromo-electric field, $H_{el}$, 
given in Eq.~(\ref{eq:QnfQmfp}),
is the most challenging due to its 
inherent non-locality in axial gauge.
There are two distinct types of contributions: One is from same-site interactions and the other from interactions between different sites.
For the same-site interactions, the operator is the product of charges 
$Q_{n,f}^{(a)} \, Q_{n,f}^{(a)}$, which contains only $ZZ$ operators, and is digitized with the standard two CNOT circuit.\footnote{Using the native $ZX$ gate on IBM's devices allows this to be done with a single two-qubit entangling gate~\cite{Kim2021ScalableEM}.}
The $Q_{n,f}^{(a)} \, Q_{m,f'}^{(a)}$ operators contain 4-qubit interactions of the form 
$(\sigma^+ \sigma^- \sigma^- \sigma^+ + \ {\rm h.c.})$
and 
6-qubit interactions of the form 
$(\sigma^+ Z \sigma^- \sigma^- Z \sigma^+ + \ {\rm h.c.})$,
in addition to $ZZ$ contributions.
The manipulations required to implement the 6-qubit operators parallel those required for the 4-qubit operators, and here only the latter is discussed in detail.
These operators can be decomposed into eight mutually commuting terms,
\begin{equation}
\sigma^+ \sigma^- \sigma^- \sigma^+ + {\rm h.c.} = \frac{1}{8}(XXXX + YYXX + YXYX - YXXY - XYYX + XYXY + XXYY + YYYY) \ .
\label{eq:pmmp}
\end{equation}
The strategy for identifying the corresponding time evolution circuit is to first apply a unitary that diagonalizes every term, apply the diagonal rotations, and finally, act with the inverse unitary to return to the computational basis. 
By only applying diagonal rotations, 
many of the CNOTs can be arranged to cancel.
Each of the eight Pauli strings
in Eq.~(\ref{eq:pmmp})
takes a state in the computational basis to the corresponding bit-flipped state (up to a phase). 
This suggests that the desired eigenbasis
pairs together states with their bit-flipped counterpart, which is an inherent property of the GHZ basis~\cite{Stetina:2020abi}. 
In fact, any permutation of the GHZ state-preparation circuit diagonalizes the interaction. 
The two that will be used,
denoted by $G$ and $\tilde G$,
are shown in Fig.~\ref{circ:GHZ}.
In the diagonal bases, the Pauli strings
in Eq.~(\ref{eq:pmmp}) become
\begin{align}
    G^{\dagger}\ (\sigma^+ \sigma^- \sigma^- \sigma^+ + {\rm h.c.})\ G =&\ \frac{1}{8} \left (IIZI - ZIZZ - ZZZZ + ZIZI + IZZI - IIZZ - IZZZ + ZZZI \right ) \ , \nonumber \\
    \tilde{G}^{\dagger}\ (\sigma^+ \sigma^- \sigma^- \sigma^+ + {\rm h.c.})\ \tilde{G} =&\ \frac{1}{8} \left (IIIZ - IZZZ - IIZZ + ZIIZ + IZIZ - ZZZZ - ZIZZ + ZZIZ\right ) \ .
    \label{eq:pmmpdiag}
\end{align}
Another simplification comes from the fact that 
$ZZ$ in the computational basis becomes  
a single $Z$ in a GHZ basis if the GHZ state-preparation circuit has a CNOT connecting the two $Z$s. 
For the case at hand, this implies
\begin{align}
    G^{\dagger}\ (IZZI + IZIZ + ZIIZ)\ G =& \ IZII + IIIZ + ZIII \ , \nonumber \\[4pt]
    \tilde{G}^{\dagger}\ (ZIZI + IZZI + ZIIZ)\ \tilde{G} =& \ IIZI + IZII + ZIII  \ .
    \label{eq:ZZGHZ}
\end{align}
As a consequence, all nine $ZZ$ terms in $Q_{n,f}^{(a)} \, Q_{m,f'}^{(a)}$ 
become single $Z$s in a GHZ basis, thus requiring no additional CNOT gates to implement. 
Central elements of the circuits 
required to implement time evolution of the chromo-electric energy density
are shown in Fig.~\ref{circ:UpmmpZZ},
which extends the circuit presented in Fig.~4 of Ref.~\cite{Stetina:2020abi} to non-Abelian gauge theories.
More details on these circuits can be found in App.~\ref{app:circ}.

\subsubsection{Trotterization, Color Symmetry and Color Twirling}
\label{sec:colorBreak}
\noindent
After fixing the gauge, the Hamiltonian is no longer manifestly invariant under local  $SU(3)$ gauge transformations. 
However, as is well known, observables of the theory are correctly computed from such a gauge-fixed Hamiltonian, which possesses a remnant global $SU(3)$ symmetry.
This section addresses the extent to which this symmetry is preserved by Trotterization of the time-evolution operator. 
The focus will be on 
the $N_f=1$ theory as including additional flavors
does not introduce new complications.

Trotterization of the mass and kinetic parts of the Hamiltonian,
while having  non-zero commutators between some terms, preserves the global $SU(3)$ symmetry.
The time evolution of $Q_{n}^{(a)} \, Q_{n}^{(a)}$
can be implemented in a unitary operator without Trotter errors, and, therefore, does not break $SU(3)$. 
On the other hand,
the time evolution induced by 
$Q_{n}^{(a)} \, Q_{m}^{(a)}$
is implemented by the operator being divided into 
four terms:
$(Q^{(1)}_n \, Q^{(1)}_{m} + Q^{(2)}_n \, Q^{(2)}_m)$,  $(Q^{(4)}_n \, Q^{(4)}_m + Q^{(5)}_n \, Q^{(5)}_m)$, $(Q^{(6)}_n \, Q^{(6)}_m + Q^{(7)}_n \, Q^{(7)}_m)$ and $(Q^{(3)}_n \, Q^{(3)}_m + Q^{(8)}_n \, Q^{(8)}_m)$. In order for global $SU(3)$ to be unbroken, 
the sum over the entire lattice
of each of the 8 gauge charges must be unchanged under time evolution. 
Therefore, 
the object of interest is the commutator
\begin{equation}
\mathcal{C} = \left [ \sum_{n=0}^{2L-1}Q^{(a)}_{n} \ , \ Q^{(\tilde{b})}_{m}\cdot Q^{(\tilde{b})}_{l} \right ] \ ,
\label{eq:Qcomm}
\end{equation}
where $\tilde{b}$ is summed over the elements of one of the pairs in $\{(1,2),\,(4,5),\,(6,7),\,(3,8)\}$. 
It is found that this commutator only vanishes if $a=3$ or $a=8$, or if $\tilde{b}$ is summed over all $8$ values (as is the case for the exact time evolution operator). 
Therefore, Trotter time evolution does not preserve the global off-diagonal $SU(3)$ charges and, for example, color singlets can evolve into non-color singlets.
Equivalently, the Trotterized time evolution operator is not in the trivial representation of $SU(3)$.
To understand this point in more detail,
consider the transformation of 
$\left(T^a\right)^i_j  \ \left(T^a\right)^k_l$ for any given $a$. 
Because of the symmetry of this product of operators, each transforming as an ${\bf 8}$, the product must decompose into ${\bf 1}\oplus {\bf 8}\oplus {\bf 27} $,
where the elements of each of the irreps can be found from
\begin{equation}
\left(T^a\right)^i_j  \ \left(T^a\right)^k_l = 
\left(\hat {\cal O}_{27}^a\right)^{ik}_{jl} 
-\frac{2}{5}\left[ \delta^i_j \left(\hat {\cal O}^a_8\right)^k_l + \delta^k_l \left(\hat {\cal O}^a_8\right)^i_j \right]
+\frac{3}{5}
\left[\delta^i_l \left(\hat {\cal O}^a_8\right)^k_j + \delta^k_j \left(\hat {\cal O}^a_8\right)^i_l \right]
+ \frac{1}{8} \left( \delta^i_l \delta^k_j\ -\ \frac{1}{3} \delta^i_j \delta^k_l \right) \hat {\cal O}^a_1
\  ,
\end{equation}
where
\begin{align}
\left(\hat {\cal O}^a_{27}\right)^{ik}_{jl} 
 = & \
\frac{1}{2}
\left[ \left(T^a\right)^i_j \left(T^a\right)^k_l + \left(T^a\right)^i_l \left(T^a\right)^k_j \right]
-
\frac{1}{10}\left[ 
\delta^i_j \left(\hat {\cal O}^a_8\right)^k_l 
+ \delta^i_l \left(\hat {\cal O}^a_8\right)^k_j 
+ \delta^k_j \left(\hat {\cal O}^a_8\right)^i_l 
+ \delta^k_l \left(\hat {\cal O}^a_8\right)^i_j \right] \nonumber\\
& -\frac{1}{24} \left( \delta^i_j \delta^k_l + \delta^i_l \delta^k_j \right) \hat {\cal O}^a_1 \ ,
\nonumber\\
\left(\hat {\cal O}^a_8\right)^i_j 
 = & \
\left(T^a\right)^i_\beta \left(T^a\right)_j^\beta\ -\ \frac{1}{3} \delta^i_j \hat {\cal O}^a_1
\ ,\ \ 
\hat {\cal O}^a_1 \ =\  \left(T^a\right)^\alpha_\beta \left(T^a\right)_\alpha^\beta \ =\ \frac{1}{2}
\  .
\end{align}
When summed over $a=1,\ldots,8$, the contributions from the ${\bf 8}$ and ${\bf 27}$ vanish, leaving the familiar contribution from the ${\bf 1}$.
When only partials sums are available, as is the situation with individual contributions to the Trotterized evolution, 
each of the contributions is the exponential of 
${\bf 1}\oplus {\bf 8}\oplus 
{\bf 27} $, with only the singlet contributions leaving the lattice a color singlet.
The leading term in the expansion of the product of the four pairs of Trotterized evolution operators sum to leave only the singlet contribution. 
In contrast, higher-order terms do not cancel and 
non-singlet contributions are present.

This is a generic problem 
that will be encountered when satisfying Gauss's law
leads to non-local charge-charge interactions. 
This is not a problem for $U(1)$, and 
surprisingly, is not a problem for 
$SU(2)$ because $(Q^{(1)}_n \, Q^{(1)}_m, Q^{(2)}_n \, Q^{(2)}_m, Q^{(3)}_n \, Q^{(3)}_m )$ are in the Cartan sub-algebra of $SU(4)$ and therefore mutually commuting.  However, it is a problem for $N_c>2$.
One way around the breaking of global $SU(N_c)$ 
is through the co-design 
of unitaries that directly (natively) implement
${\exp}( i \alpha Q^{(a)}_n \, Q^{(a)}_m)$; see Sec.~\ref{sec:codesign}.
Without such a native unitary,
the breaking of $SU(N_c)$ appears as any other Trotter error, and can be systematically reduced in the same way. A potential caveat to this is if the time evolution operator took the system into a different phase, but our studies of $L=1$ show no evidence of this.

It is interesting to note that the terms generated by the Trotter commutators form a closed algebra.
In principle, a finite number of terms could be
included to define an effective Hamiltonian whose Trotterization exactly maps onto the desired evolution operator (without the extra terms).
It is straightforward to work out the terms generated order-by-order in the Baker-Campbell-Hausdorff formula.
Aside from re-normalizing the existing charges, there are $9$ new operator structures produced. 
For example, the leading-order commutators generate the three operators, ${\cal O}_i$, in Eq.~(\ref{eq:BCHOp}),
\begin{align}
{\cal O}_i = 
\begin{cases} 
      (\sigma^+ I \sigma^- \sigma^- Z \sigma^+ - \sigma^+ Z \sigma^- \sigma^- I \sigma^+) - {\rm h.c.}  \ ,\\
      (I \sigma^- \sigma^+ Z \sigma^+ \sigma^- - Z \sigma^- \sigma^+ I \sigma^+ \sigma^-) - {\rm h.c.}  \ , \\
       (\sigma^+ \sigma^- Z \sigma^- \sigma^+ I - \sigma^+ \sigma^- I \sigma^- \sigma^+ Z) - {\rm h.c.} \ .
\end{cases}
\label{eq:BCHOp}
\end{align}
In general, additional operators are constrained only by (anti)hermiticity, 
symmetry with respect to $n \leftrightarrow m$ and preservation of $(r,g,b)$, and should generically be included in the same spirit as terms in the  Symanzik-action~\cite{Symanzik:1983dc,Symanzik:1983gh} for lattice QCD.

With Trotterization of the gauge field introducing violations of gauge symmetry, and the presence of bit- and phase-flip errors within the device register, it is worth briefly considering a potential mitigation strategy. A single
bit-flip error will change isospin by $|\Delta I_3|=1/2$ and color charge by one unit of red or green or blue.
After each Trotter step on a real quantum device, such errors will be encountered and a mitigation or correction scheme is required.
Without the explicit gauge-field degrees of freedom and local charge conservation checks enabled by Gauss's law, such errors can only be detected globally, and hence, cannot be actively corrected during the evolution.\footnote{When local gauge fields are present,
previous works have found that including a quadratic ``penalty-term" in the Hamiltonian is effective in mitigating violation of Gauss's law~\cite{Hauke:2013jga,Zohar:2015hwa,Dalmonte:2016alw,Halimeh:2019svu}. See also Refs.~\cite{PhysRevLett.112.120406,Kasper:2020owz}.}
Motivated by this, consider introducing a twirling phase factor into the evolution, $\exp(-i \theta^a {\cal Q}^{(a)})$, where ${\cal Q}^{(a)}$ is the total charge on the lattice.
If applied after each Trotter step, with a randomly selected set of eight angles, $\theta^a$, 
the phases of color-nonsinglet states become random for each member of an ensemble, mitigating errors in some observables. 
Similar twirling phase factors could be included for the other charges that are conserved or approximately conserved.

\subsubsection{Quantum Resource Requirements for Time Evolution}
\noindent
It is straightforward to extend the circuits presented in the previous section to arbitrary $N_c$ and $N_f$. The quantum
resources required for time evolution can be quantified
for small, modest and asymptotically large systems. As discussed previously, a quantum register with $N_q=2 L N_c N_f$ qubits\footnote{The inclusion of an ancilla for the kinetic term increases the qubit requirement to $N_q = 2L N_c N_f + 1$.} is required to encode one-dimensional $SU(N_c)$ gauge theory
with $N_f$ flavors on $L$ spatial lattice sites using the JW transformation. For $SU(3)$ gauge theory, this leads to, for example, $N_q = 6L$ with only $u$-quarks and $N_q = 18L$ with $u,d,s$-quarks.
The five distinct contributions to the resource requirements, 
corresponding to application of the unitary operators providing 
a single Trotter step associated with the quark mass, $U_m$, the baryon chemical potential, $U_{\mu_B}$, the isospin chemical potential, $U_{\mu_I}$, the kinetic term, $U_{kin}$, and the chromo-electric field, $U_{el}$, are 
given in terms of the number of 
single-qubit rotations, denoted by ``$R_Z$'', the number of Hadamard gates, denoted by ``Hadamard'', and the number of CNOT gates, denoted by ``CNOT''.
It is found that\footnote{
For $N_c = 2$ only three of the $ZZ$ terms can be combined into $Q_{n,f}^{(a)} \, Q_{m,f'}^{(a)}$ and the number of CNOTs for one Trotter step of $U_{el}$ is 
\begin{equation}
U_{el}  \ : \ (2 L-1) N_f [9 (2 L-1) N_f-7] \ \ \ | \ \text{CNOT} \ .
\label{eq:Nc2CNOT}
\end{equation}
Additionally, for $N_c N_f < 4$, the Trotterization of $U_{{\rm kin}}$ is more efficient without an ancilla and the number of CNOTs required is
\begin{equation}
U_{ kin}  \ : \ 2 (2 L-1) N_c (N_c + 1) \ \ \ | \ \text{CNOT} \ .
\label{eq:UkinNoanc}
\end{equation}
The construction of the circuit that implements the time evolution of the hopping term for $N_c=3$ 
and $N_f=1$
is shown in Fig.~\ref{fig:Ukin1flavTrot}.
}
\begin{align}
   U_m  \ :& \ \ 2 N_c N_f L \ \ \ | \ R_Z \ ,\nonumber \\[5pt]
   U_{\mu_B}  \ :& \ \ 2 N_c N_f L \ \ \ | \ R_Z \ ,\nonumber \\[5pt]
   U_{\mu_I}  \ :& \ \ 2 N_c N_f L \ \ \ | \ R_Z \ ,\nonumber \\[5pt]
   U_{kin}  \ :& \ \ 2 N_c N_f(2L-1) \ \ \ | \ R_Z \ ,\nonumber \\
                & \ \ 2 N_c N_f (2L-1) \ \ \ | \ \text{Hadamard} \ ,\nonumber \\
                & \ \ 2 N_c N_f (8L-3) -4 \ \ \ | \ \text{CNOT} \ ,\nonumber \\[5pt]
    U_{el}  \ :& \ \ \frac{1}{2}(2L-1)N_c N_f\left [3-4N_c+N_f(2L-1)(5N_c-4)\right ] \ \ \ | \ R_Z \ ,\nonumber \\
                & \ \ \frac{1}{2}(2L-1)(N_c-1) N_c N_f\left [N_f(2L-1)-1\right ] \ \ \ | \ \text{Hadamard} \ ,\nonumber \\
                & \ \ \frac{1}{6} (2 L -1) (N_c-1) N_c N_f [(2 L-1) (2 N_c+17) N_f-2 N_c-11] \ \ \ | \ \text{CNOT} \ .
\label{eq:RHCN}
\end{align}

It is interesting to note the scaling of each of the contributions.  The mass, chemical potential and kinetic terms scale as ${\cal O}(L^1)$, while the non-local gauge-field contribution is ${\cal O}(L^2)$.
As anticipated from the outset, using Gauss's law to constrain the energy in the gauge field via the quark occupation has given rise to circuit depths that scale quadratically with the lattice extent, naively violating one of the criteria for quantum simulations at scale~\cite{Feynman:1981tf,DiVincenzo2000ThePI}.
This volume-scaling is absent for formulations that explicitly include the 
gauge-field locally, 
but with the trade-off of requiring a volume-scaling increase in the number of 
qubits or qudits or bosonic modes.\footnote{
The local basis on each link is spanned by the possible color irreps 
and the states of the left and right Hilbert spaces (see footnote~\ref{foot:irrep}). 
The possible irreps are built from the charges of the preceding fermion sites, 
and therefore the dimension of the link basis grows polynomially in $L$. 
This can be encoded in $\mathcal{O}(\log L)$ qubits per link and 
$\mathcal{O}(L \log L)$ qubits in total. 
The hopping and chromo-electric terms in the Hamiltonian are local, 
and therefore one Trotter step will require $\mathcal{O}(L)$ gate operations up to logarithmic corrections.} 
We expect that the architecture of quantum devices used for simulation 
and the resource requirements for the local construction will determine 
the selection of local versus non-local implementations.

For QCD with $N_f=2$, the total requirements are
\begin{align}
   R_Z  \ :& \ \ (2L-1)\left( 132 L -63 \right)+18 \ ,\nonumber \\
   \text{Hadamard} \ :& \ \ (2L-1)\left( 24L - 6 \right)  \ ,\nonumber \\
    \text{CNOT} \ :& \ \ (2L-1)\left( 184L - 78 \right) + 8 \ ,
\end{align}
and further, the CNOT requirements for a single Trotter step 
of $SU(2)$ and $SU(3)$ for $N_f = 1,2,3$ are shown in Table~\ref{tab:cnotA}.
\begin{table}[!ht]
\renewcommand{\arraystretch}{1.2}
\begin{tabularx}{0.48\textwidth}{||c | Y | Y | Y ||}
\hline
\multicolumn{4}{||c||}{Number of CNOT gates for one Trotter step of $SU(2)$} \\
 \hline
 $L$ & $N_f=1$ & $N_f=2$ & $N_f=3$ \\
 \hline\hline
 1 & 14 & 58 & 116 \\ 
 \hline
 2 & 96 & 382 & 818\\
 \hline
 5 & 774 & 3,082 & 6,812 \\
 \hline
 10 & 3,344 & 13,342 & 29,762 \\
 \hline
 100 & 357,404 & 1,429,222 & 3,213,062 \\
 \hline
\end{tabularx}
\renewcommand{\arraystretch}{1}
\ \
\renewcommand{\arraystretch}{1.2}
\begin{tabularx}{0.48\textwidth}{||c | Y | Y | Y ||}
\hline
\multicolumn{4}{||c||}{Number of CNOT gates for one Trotter step of $SU(3)$ } \\
 \hline
 $L$ & $N_f=1$ & $N_f=2$ & $N_f=3$ \\
 \hline\hline
 1 & 30 & 114 & 242 \\ 
 \hline
 2 & 228 & 878 & 1,940\\
 \hline
 5 & 1,926 & 7,586 & 16,970 \\
 \hline
 10 & 8,436 & 33,486 & 75,140 \\
 \hline
 100 & 912,216 & 3,646,086 & 8,201,600 \\
 \hline
\end{tabularx}
\renewcommand{\arraystretch}{1}
\caption{The CNOT requirements to perform one Trotter step of time evolution for a selection of simulation parameters.}
 \label{tab:cnotA}
 \end{table}
These resource requirements suggest that systems with up to $L=5$ could be simulated, with appropriate error mitigation protocols, using this non-local framework in the near future. Simulations beyond $L=5$ appear to present a challenge in the near term.

The resource requirements in Table~\ref{tab:cnotA} do not include those for a gauge-link beyond the end of the lattice.  As discussed previously, such additions to the time evolution could be used to move color-nonsinglet contributions to high frequency, allowing the possibility that they are filtered from observables.
Such terms contribute further to the quadratic volume scaling of resources.
Including chemical potentials in the time evolution does not increase the number of required entangling gates per Trotter step.  Their impact upon resource requirements may arise in preparing the initial state of the system.

\subsubsection{Elements for Future Co-Design Efforts}
\label{sec:codesign}
\noindent
Recent work has shown the capability of creating many-body entangling gates natively~\cite{Andrade:2021pil,Katz:2022czu} which have similar fidelity to two qubit gates.
This has multiple benefits. First, it allows for (effectively) deeper circuits to be run within coherence times. 
Second, it can eliminate some of the Trotter errors due to non-commuting terms. 
The possibility of using native gates for these calculations is particularly interesting from the standpoint of eliminating or mitigating the Trotter errors that violate the global $SU(3)$ symmetry, as discussed in Sec.~\ref{sec:colorBreak}.
Specifically, 
it would be advantageous to have a ``black box" unitary operation of the form,
\begin{align}
   e^{-i \alpha Q_n^{(a)} \, Q_m^{(a)}} =& \ \exp \bigg \{-i \frac{\alpha}{2} \bigg [\sigma^+_{n} \sigma^-_{n+1}\sigma^-_{m}\sigma^+_{m+1} + \sigma^-_{n}\sigma^+_{n+1}\sigma^+_{m}\sigma^-_{m+1} +  \sigma^+_{n+1}\sigma^-_{n+2}\sigma^-_{m+1}\sigma^+_{m+2} 
   + \sigma^-_{n+1}\sigma^+_{n+2}\sigma^+_{m+1}\sigma^-_{m+2} \nonumber \\
   &+ \sigma^+_{n}\sigma^z_{n+1}\sigma^-_{n+2}\sigma^-_{m}\sigma^z_{m+1}\sigma^+_{m+2} + \sigma^-_{n}\sigma^z_{n+1}\sigma^+_{n+2}\sigma^+_{m}\sigma^z_{m+1}\sigma^-_{m+2} + \frac{1}{6}(\sigma^z_n \sigma^z_m + \sigma^z_{n+1} \sigma^z_{m+1} + \sigma^z_{n+2} \sigma^z_{m+2})\nonumber \\
   &- \frac{1}{12}(\sigma^z_n \sigma^z_{m+1} + \sigma^z_n \sigma^z_{m+2} + \sigma^z_{n+1} \sigma^z_m + \sigma^z_{n+1} \sigma^z_{m+2} + \sigma^z_{n+2} \sigma^z_m + \sigma^z_{n+2 }\sigma^z_{m+1})
   \bigg ] \bigg \} \ ,
\end{align}
for arbitrary $\alpha$ and pairs of sites, $n$ and $m$ (sum on $a$ is implied).
A more detailed discussion of co-designing 
interactions for quantum simulations of these theories is clearly warranted.

\subsection{Results from Quantum Simulators}
\begin{figure}[!ht]
    \centering
    \includegraphics[width=\columnwidth]{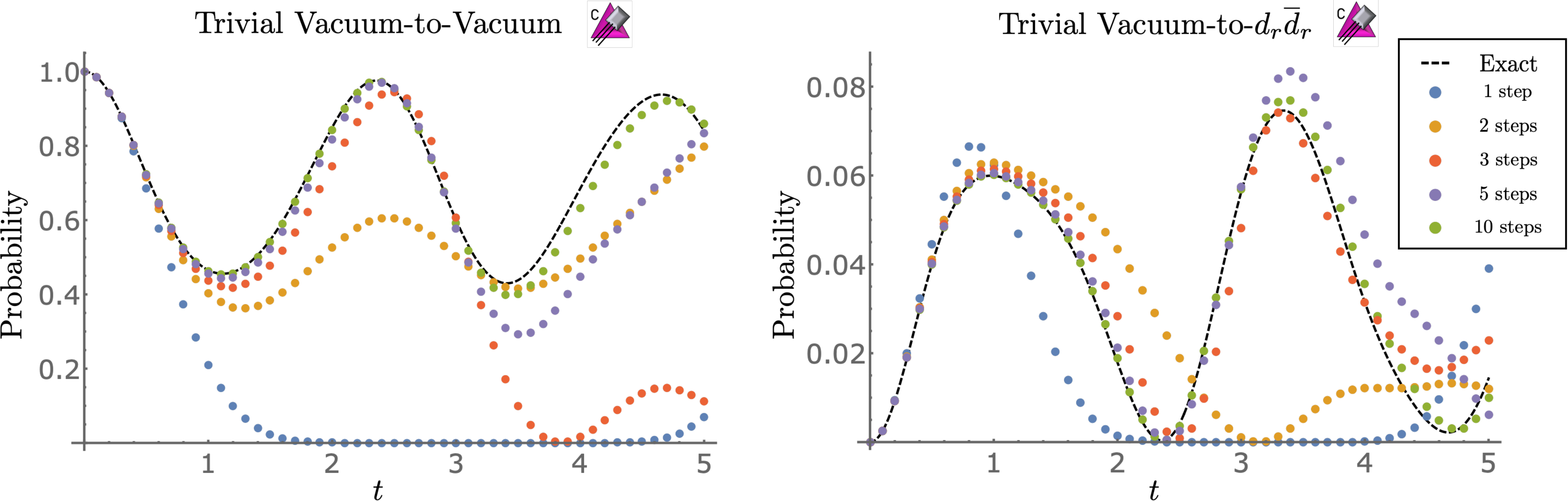}
    \caption{
    The trivial vacuum-to-vacuum 
    and trivial vacuum-to-$d_r \overline{d}_r$ probabilities in QCD with $N_f=2$
    for $m=g=L=1$. 
    Shown are the results obtained from exact exponentiation of the Hamiltonian (dashed black curve) and from the Trotterized implementation with $1$, $2$, $3$, $5$ and $10$ Trotter steps using the (classical) quantum simulators in {\tt cirq} and {\tt qiskit} (denoted by the purple icons~\cite{Klco:2019xro}).
    }
    \label{fig:VacTo}
\end{figure}

\noindent
The circuits laid out in Sec.~\ref{sec:Circuits} are too deep to be executed on currently available quantum devices,  
but can be readily implemented with quantum simulators such as {\tt cirq} and {\tt qiskit}.
This allows for an estimate of the number of Trotter steps required to achieve a desired precision in the determination of any given observable as a function of time.
Figure~\ref{fig:VacTo} shows results for the 
trivial vacuum-to-vacuum and trivial vacuum-to-$d_r \overline{d}_r$ probabilities as a function of time for $L=1$. See App.~\ref{app:Nf1SU3circs} for the full circuit which
implements a single Trotter step, and App.~\ref{app:MDTD} for the decomposition of the energy starting in the trivial vacuum.

The number of Trotter steps, 
$N_{\rm Trott}$, required to evolve out to a given $t$ within a specified (systematic) error, 
$\epsilon_{\rm Trott}$, was also investigated. 
$\epsilon_{\rm Trott}$ is defined as the 
maximum fractional error between the 
Trotterized and exact time evolution in two quantities, the vacuum-to-vacuum persistence probability 
and the vacuum-to-$d_r\overline{d}_r$ transition probability. For demonstrative purposes, an analysis at leading order in the Trotter expansion is sufficient.
Naive expectations based upon global properties of the Hamiltonian defining the evolution operators indicate that an upper bound for $\epsilon_{\rm Trott}$ scales as 
\begin{equation}
\Big\lvert \Big\lvert e^{-i H t} - \left[ U_1\left (\frac{t}{N_{\rm Trott}}\right ) \right]^{N_{\rm Trott}} \Big\rvert \Big\rvert \ \le \ \frac{1}{2} \sum_i \sum_{j>i} \Big\lvert \Big\lvert \left[ H_i , H_j \right] \Big\rvert \Big\rvert \frac{t^2}{N_{\rm Trott}} \ ,
\label{eq:LOTrottbound}
\end{equation}
where the Hamiltonian has been divided into sets of mutually commuting terms, $H = \sum_i H_i$. This upper bound indicates that the required number of Trotter steps to maintain a fixed error scales as  $N_{\rm Trott}\sim t^2$~\cite{Childs_2021}.

\begin{figure}[!ht]
    \centering
    \includegraphics[width=\columnwidth]{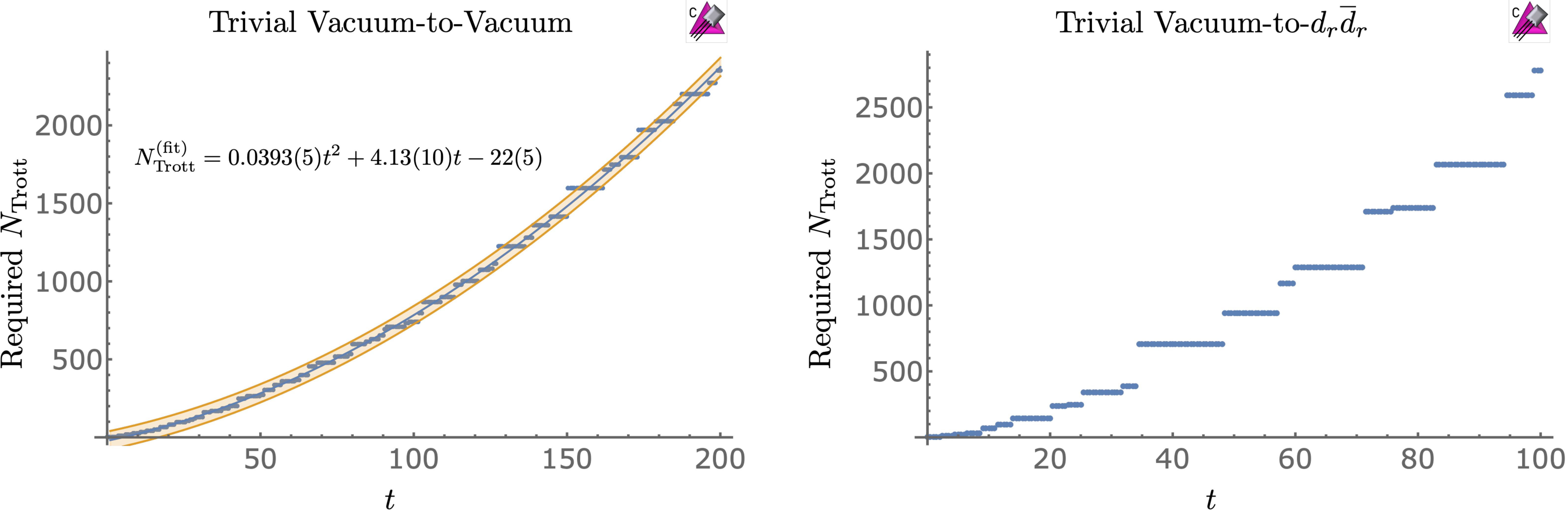}
    \caption{
    The number of Trotter steps, 
    $N_{\rm Trott}$,
   required to 
   achieve a systematic fractional error of 
   $\epsilon_{{\rm Trott}}\le 0.1$
    at time $t$ in the trivial vacuum-to-vacuum  probability (left panel) and the trivial vacuum-to-$d_r\overline{d}_r$ probability
    (right panel) for QCD with $N_f=2$ and
$m=g=L=1$. The blue points are results obtained by direct calculation.
   }
    \label{fig:TrotErrorB}
\end{figure}

To explore the resource requirements for simulation based upon explicit calculations between exclusive states, as opposed to upper bounds for inclusive processes, given in Eq.~(\ref{eq:LOTrottbound}), 
a series of calculations was performed requiring $\epsilon_{\rm Trott}\le0.1$ for a range of times, $t$.
Figure~\ref{fig:TrotErrorB} shows the required 
$N_{\rm Trott}$ as a function of $t$ for $m=g=L=1$.
The plateaus 
observed in Fig.~\ref{fig:TrotErrorB} arise from
resolving upper bounds from oscillating functions, 
and introduce a limitation in fitting to extract scaling behavior.  This is less of a limitation 
for the larger vacuum-to-vacuum probabilities which are fit well by a quadratic polynomial, starting from $t=1$, with coefficients,
\begin{equation}
    N_{{\rm Trott}} = 0.0393(5) t^2 + 4.13(10) t - 22(5) \ .
    \label{eq:TrotErrorFit}
\end{equation}
The uncertainty represents a 95\% confidence interval in the fit parameters and corresponds to the shaded orange region in
Fig.~\ref{fig:TrotErrorB}. The weak quadratic scaling with $t$ implies that, even out to $t \sim 100$, the number of Trotter
steps scales approximately linearly, and a constant error in the observables can be achieved with a fixed Trotter step size. 
We have been unable to distinguish between fits with and without logarithmic terms.

These results can be contrasted with those obtained for the Schwinger model in Weyl gauge. The authors of Ref.~\cite{Shaw:2020udc} estimate a resource
requirement, as quantified by the number of $T$-gates, that scales as $\sim (L t)^{3/2}\log L t$, increasing to
$\sim L^{5/2} t^{3/2}\log L t \log L$ if the maximal value of the gauge fields is accommodated within the Hilbert space.
The results obtained in this section suggest that resource requirements in axial gauge, as quantified by the number of CNOTs,
effectively scale as $\sim L^2 t$ up to intermediate times and as $\sim L^2 t^2$ asymptotically. In a scattering process with localized
wave-packets, it is appropriate to take $L\sim t$ 
(for the speed of light taken to be $c=1$),
as the relevant non-trivial time evolution is bounded by the light cone. 
This suggests that the required resources scale asymptotically as $\sim t^4$, independent of the chosen gauge to define the simulation. 
This could have been
anticipated at the outset by assuming that the minimum change in complexity for a process has physical meaning~\cite{https://doi.org/10.48550/arxiv.quant-ph/0701004,doi:10.1126/science.1121541,https://doi.org/10.48550/arxiv.quant-ph/0502070,Jefferson:2017sdb}.

\section{Simulating \texorpdfstring{\boldmath$1+1$}{1+1}D QCD with \texorpdfstring{\boldmath$N_f=1$}{Nf=1} and \texorpdfstring{\boldmath$L=1$}{L=1}}
\label{sec:Nc3Nf1}
\noindent
With the advances in quantum devices, algorithms and mitigation strategies, quantum simulations of $1+1$D QCD can now begin, and this section presents results for $N_f=1$ and $L=1$. Both state preparation and time evolution will be discussed.

\subsection{State Preparation with VQE}
\begin{figure}[!ht]
    \centering
    \includegraphics[width=6cm]{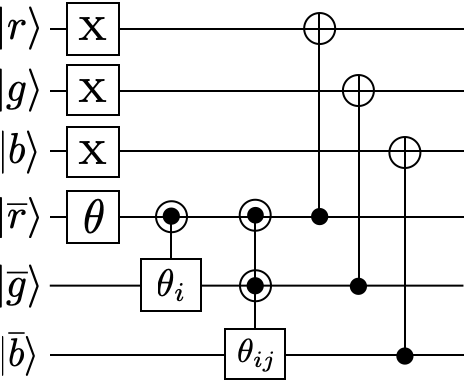}
    \caption{
    Building upon the trivial vacuum, 
    this circuit initializes the 
    most general real wavefunction 
    (with 7 independent rotation angles for 3 qubits)
    in the $\overline{q}$-sector, which is subsequently
    mirrored into the $q$-sector by 3 CNOTs. Gates labelled by ``$\theta$" are shorthand for $R_Y(\theta)$ and half-filled circles denote a control on $0$ 
    and a different control on $1$.
    }
    \label{circ:GeneralVQE}
\end{figure}

\noindent
Restricting the states of the lattice to be color singlets reduces the complexity of state preparation significantly.  
Transformations in the quark sector are mirrored in the antiquark sector.
A circuit that
prepares the most general state with $r=g=b=0$ is shown in Fig.~\ref{circ:GeneralVQE}. 
The (multiply-)controlled $\theta$ gates are short-hand for (multiply-)controlled $R_Y(\theta)$ gates with half-filled circles denoting a control on $0$ 
and a different control on $1$.
The subscripts on $\theta_{ij}$ signify that there are different angles for each controlled rotation. For example,
$\theta_{i}$ has two components, $\theta_{0}$ and $\theta_{1}$, corresponding to a rotation controlled on $0$ and $1$, respectively. 
The CNOTs at the end of the circuit
enforce that there are equal numbers of quarks and antiquarks with the same color,
i.e., that $r=g=b=0$. 
This circuit can be further simplified by constraining the angles to only parameterize color singlet states. The color singlet subspace is spanned by\footnote{
The apparent asymmetry between $q_r,q_g,q_b$ is due to the charge operators generating hops over different numbers of quarks or antiquarks. 
For example, $Q^{(1)}$ hops $q_r$ to $q_g$ without passing over any intermediate quarks, but $Q^{(4)}$ hops $q_r$ to $q_b$ passing over $q_g$. 
Also note that when $m=0$ the $\mathbb{Z}_2$ spin-flip symmetry reduces the space of states to be two-dimensional.}
\begin{align}
    \ket{{\Omega_0}} \ &, \ \ \frac{1}{\sqrt{3}}\left (\ket{q_r \overline{q}_r} - \ket{q_g \overline{q}_g} + \ket{q_b \overline{q}_b}\right ) \ , \nonumber \\
    \ket{q_r \overline{q}_r \, q_g \overline{q}_g \, q_b \overline{q}_b} \ &, \ \  \frac{1}{\sqrt{3}} \left (\ket{q_r \overline{q}_r \, q_g\overline{q}_g} - \ket{q_r \overline{q}_r \, q_b\overline{q}_b} + \ket{q_g \overline{q}_g \, q_b \overline{q}_b}\right ) \ , 
    \label{eq:vacBasis}
\end{align}
where $\ket{{\Omega_0}} = \ket{{000111}}$ is the trivial vacuum.
This leads to the following relations between angles,
\begin{align}
    \theta_{10} &= \theta_{01}\ ,  & \theta_{00} &= -2 \sin^{-1}\left[ \tan(\theta_{0}/2) \, \cos(\theta_{01}/2) \right] \ ,\nonumber \\
    \theta_{01} &= -2 \sin^{-1}\left[ \cos(\theta_{11}/2)\, \tan(\theta_{1}/2) \right]\ , & \theta_0 &= -2 \sin^{-1} \left[ \tan(\theta/2) \, \cos(\theta_{1}/2) \right] \ .
    \label{eq:angleconst}
\end{align}

\begin{figure}[!ht]
    \centering
    \includegraphics[width=\columnwidth]{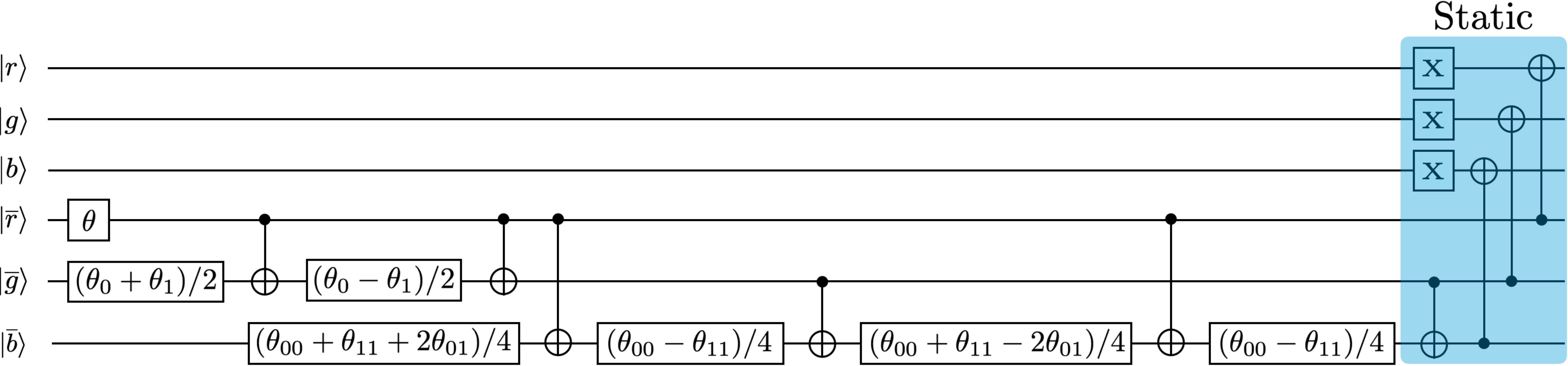}
    \caption{
    A circuit that initializes the most general 
    $B=0$ color singlet state for $N_f=1$ and $L=1$. Gates labelled by ``$\theta$" are shorthand for $R_Y(\theta)$
    and the $X$s at the end are to initialize the trivial vacuum. The color singlet constraint, $\theta_{10} = \theta_{01}$, has been used and the other angles are related by Eq.~(\ref{eq:angleconst}).}
    \label{circ:VQEImp}
\end{figure}

The circuit in Fig.~\ref{circ:GeneralVQE} can utilize the strategy outlined in Ref.~\cite{Atas:2021ext}  to
separate into a ``variational" part and a ``static" part. 
If the VQE circuit can be written as 
$U_{var}(\theta) U_s$, 
where $U_s$ is independent
of the variational parameters, 
then $U_s$ can be absorbed by a redefinition of
the Hamiltonian. 
Specifically, matrix elements of the Hamiltonian can be written as 
\begin{equation}
    \bra{{\Omega_0}} U_{var}^{\dagger}(\theta) \tilde{H} U_{var}(\theta) \ket{{\Omega_0}}
    \ ,
\end{equation}
where $\tilde{H}= U_s^{\dagger} H U_s$.
Table~\ref{tab:PUHU} shows the 
transformations of various Pauli strings under conjugation by a CNOT controlled on the smaller index qubit. 
Note that the $\mathbb{Z}_2$ nature of this transformation is manifest.
\begin{table}[!ht]
\renewcommand{\arraystretch}{1.2}
\begin{tabular}{||c | c | c | c | c | c | c | c ||} 
 \hline
 $XX \to IX$ & $XY \to IY$ & $YX \to ZY$ & $XZ \to XZ$ & $ZZ \to ZI$ & $YZ \to YI$ & $ZY \to YX$ & $YY \to (-)ZX$ \\
 \hline
 $IX \to XX$ & $IY \to XY$ & $XI \to XI$  & $IZ \to IZ$ & $ZI \to ZZ$  & $YI \to YZ$ &  $ZX \to (-)YY$  & $II \to II$\\
 \hline
\end{tabular}
\renewcommand{\arraystretch}{1}
\caption{The transformation of Pauli strings under conjugation by a CNOT controlled on the smaller index qubit.}
\label{tab:PUHU}
\end{table}
In essence, entanglement is
traded for a larger number of  correlated measurements.  
Applying the techniques in Ref.~\cite{Klco:2019xro}, the VQE circuit of Fig.~\ref{circ:GeneralVQE} can be put into the form of Fig.~\ref{circ:VQEImp},
which requires $5$ CNOTs along with all-to-all connectivity between the three $\overline{q}$s.
%

\subsection{Time Evolution Using IBM's 7-Qubit Quantum Computers}
\begin{figure}[!ht]
    \centering
    \includegraphics[height=0.25\textheight]{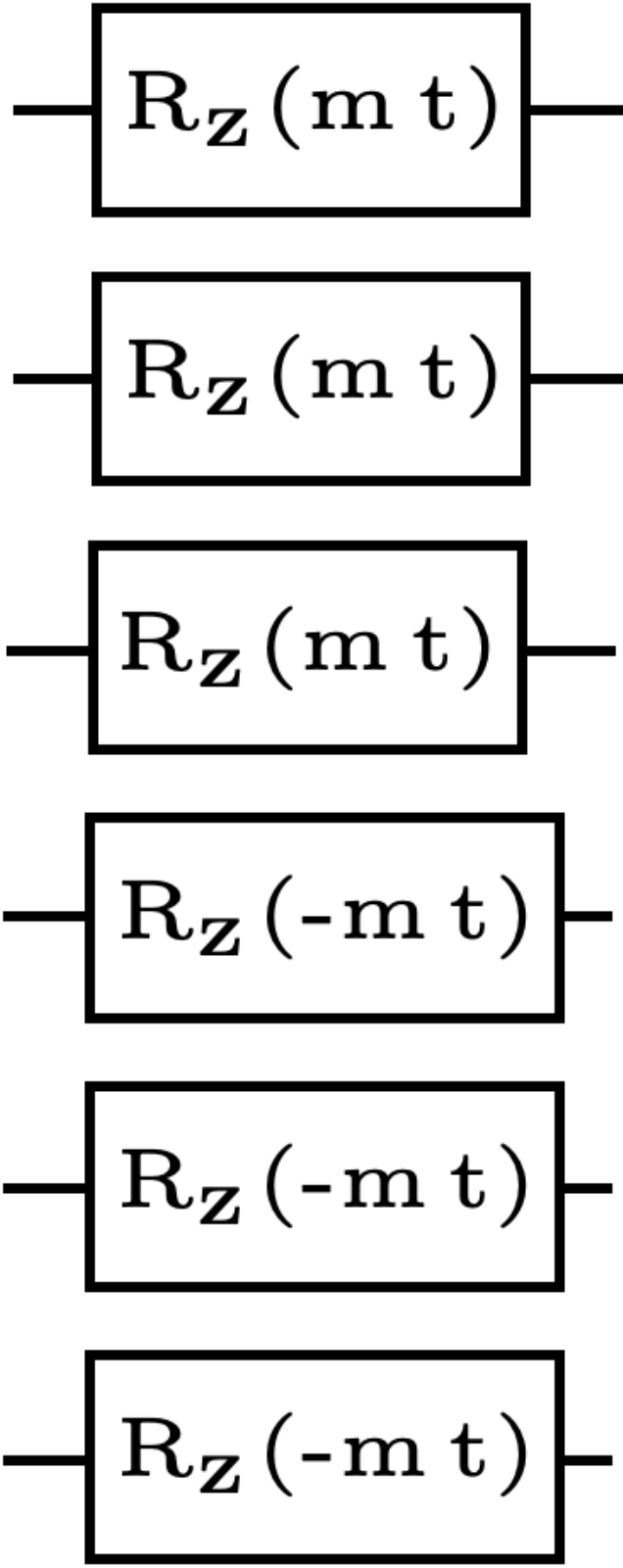}
    \caption{A circuit that implements 
    $U_m(t) = \exp(- i H_m t)$ for $N_f=1$ and $L=1$.}
    \label{circ:Um1flav}
\end{figure}

\begin{figure}[!ht]
    \centering
    \includegraphics[width=12cm]{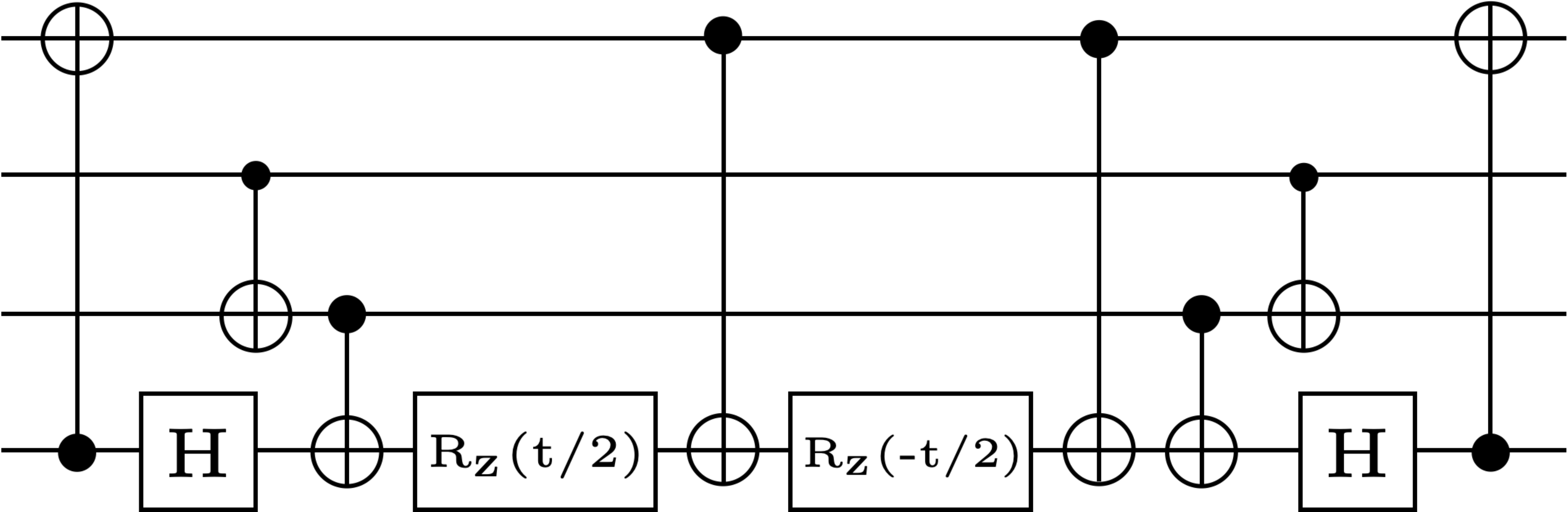}
    \caption{A circuit that implements $\exp[-i \frac{t}{2} (\sigma^+ Z Z \sigma^- + {\rm h.c.})]$.}
    \label{fig:Ukin1flavTrot}
\end{figure}

\noindent
A single leading-order Trotter step of $N_f=1$ QCD with $L=1$ requires 28 CNOTs.\footnote{By evolving with $U_{el}$ before $U_{kin}$ in the Trotterized time evolution, two of the CNOTs
become adjacent in the circuit and can be canceled.}
A circuit that implements one Trotter step of the mass term is shown in Fig.~\ref{circ:Um1flav}.
As discussed around Eq.~(\ref{eq:UkinNoanc}), it is more efficient to not use an ancilla qubit in the Trotterization of the kinetic part of the Hamiltonian. 
A circuit that implements one Trotter step of a single hopping term is shown in Fig.~\ref{fig:Ukin1flavTrot}~\cite{Stetina:2020abi}.
Similarly, for this system,
the only contribution to $H_{el}$ is $Q^{(a)}_n \, Q^{(a)}_n$, which contains three $ZZ$ terms that are Trotterized using the standard two CNOT implementation.
The complete set of circuits required for Trotterized time evolution are given in App.~\ref{app:Nf1SU3circs}.

\begin{figure}[!ht]
    \centering
    \includegraphics[width=12cm]{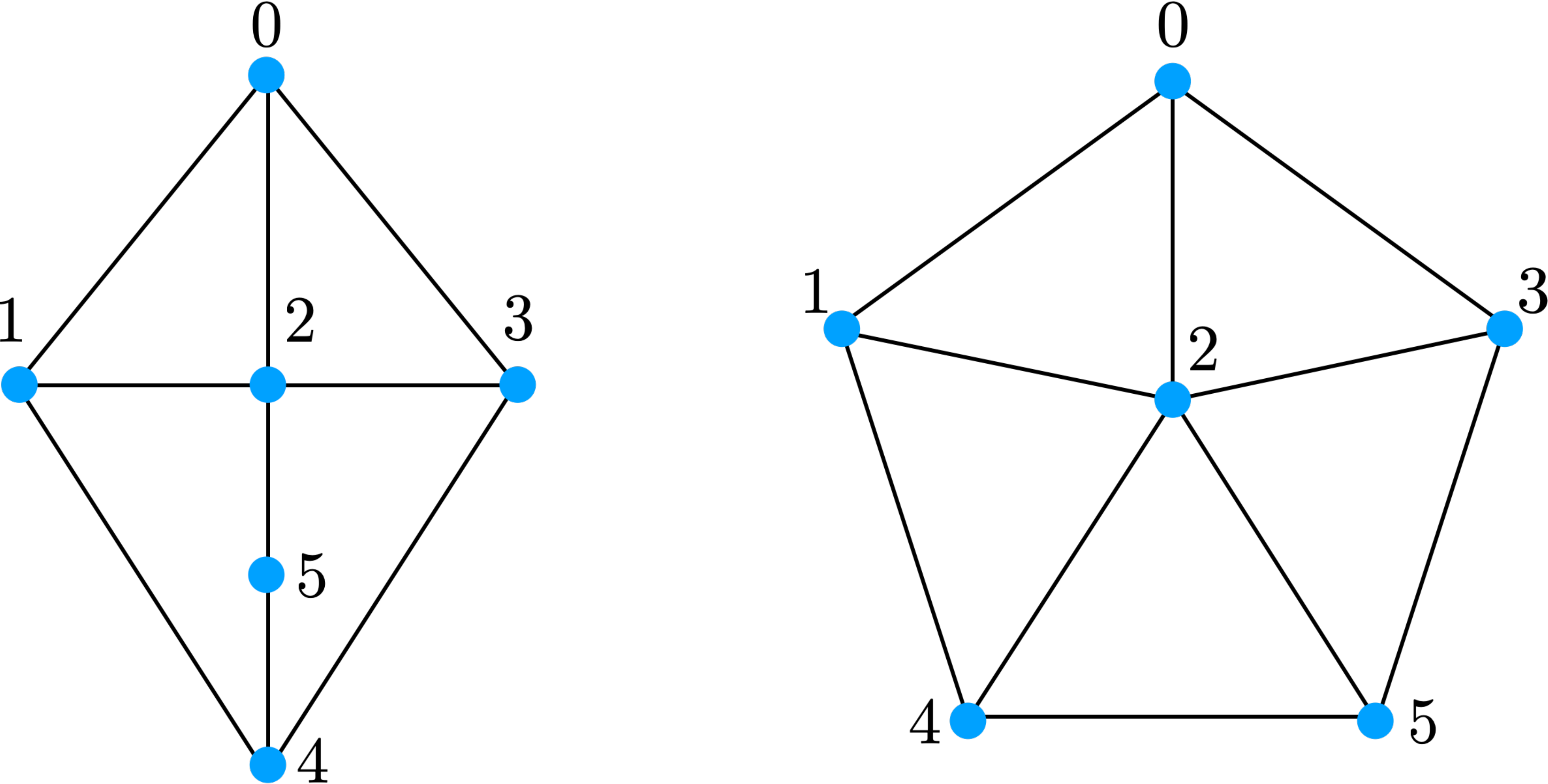}
    \caption{
    Two potential quantum device topologies for the  implementation of Trotterized time evolution.
    }
    \label{fig:TrotTopo}
\end{figure}

To map the system onto a quantum device, it is necessary to understand the required connectivity for efficient simulation. 
Together, the hopping and chromo-electric terms require connectivity between nearest neighbors as well as between $q_r$ and $q_b$ and
$q$s and $\overline{q}$s of the same color. 
The required device topology is planar and two embedding options are
shown in Fig.~\ref{fig:TrotTopo}.
The ``kite'' topology follows from the above circuits, 
while the ``wagon wheel'' topology makes use of the identities $CX(q_a,q_b) \cdot CX(q_b,q_c) = CX(q_a,q_c) \cdot CX(q_b,q_c) = CX(q_b,q_a) \cdot CX(q_a,q_c)$ where $CX(q_a,q_b)$ denotes a CNOT controlled on qubit $q_a$.
Both topologies can be employed on devices with all-to-all connectivity, such as trapped-ion systems, but
neither topology exists natively on available superconducting-qubit devices.

We performed leading-order Trotter evolution to study the trivial vacuum persistence and transition probability using IBM's quantum computers {\tt ibmq\_jakarta} and {\tt ibm\_perth}, each a {\tt r5.11H} quantum processor with 7 qubits and ``H"-connectivity.
The circuits developed for this system require a higher degree of connectivity than available with these devices, and so SWAP-gates were necessary for implementation.   
The IBM {\tt transpiler} was used to first compile the circuit for the H-connectivity and then again to compile the Pauli twirling (discussed next).
An efficient use of SWAP-gates allows for a single Trotter step to be executed with 34 CNOTs.

A number of error-mitigation techniques were employed to minimize associated systematic uncertainties in our calculations: randomized compiling of the CNOTs (Pauli twirling)~\cite{PhysRevA.94.052325} combined with decoherence renormalization~\cite{Urbanek:2021oej,Rahman:2022rlg}, measurement error mitigation, post-selecting on physical states and dynamical decoupling~\cite{PhysRevA.58.2733,DUAN1999139,ZANARDI199977,PhysRevLett.82.2417}.\footnote{A recent detailed study of the stability of some of IBM's quantum devices using a system of physical interest can be found in Ref.~\cite{Yeter-Aydeniz:2022vuy}.}
The circuits were randomly complied with each CNOT Pauli-twirled as a mechanism to transform coherent errors in the CNOT gates into statistical noise in the ensemble.
This has been shown to be effective in improving the quality of results in other simulations, for example, Refs.~\cite{Kim2021ScalableEM,Rahman:2022rlg}.
Pauli twirling involves multiplying the right side of each CNOT by a randomly chosen
element of the two-qubit Pauli group, $G_2$,
and the left side by $G'_2$ 
such that $G'_2 \, CX \, G_2 = CX$ (up to a phase). 
For an ideal CNOT gate, this would have no effect on the circuit. 
A table of required CNOT identities is given,
for example,
in an appendix in Ref.~\cite{Rahman:2022rlg}.
Randomized Pauli-twirling is combined with performing measurements of a ``non-physics", mitigation circuit, which is the time evolution circuit evaluated at $t=0$, and is the identity in the absence of noise.
Assuming that the randomized-compiling of the Pauli-twirled CNOTs transforms coherent noise into depolarizing noise, 
the fractional deviation of the noiseless and computed results 
from the asymptotic limit of complete decoherence
are expected to be approximately equal for both the physics and mitigation ensembles. Assuming linearity, it follows that
\begin{equation}
\left ( P_{\text{pred}}^{(\text{phys})}-\frac{1}{8} \right ) = \left ( P_{\text{meas}}^{(\text{phys})}-\frac{1}{8} \right ) \times \left ( \frac{1-\frac{1}{8}}{ P_{\text{meas}}^{(\text{mit})}-\frac{1}{8} } \right )\ ,
\label{eq:mit}
\end{equation}
where $P_{\text{meas}}^{(\text{phys})}$ and $P_{\text{meas}}^{(\text{mit})}$ are post-processed probabilities and
$P_{\text{pred}}^{(\text{phys})}$ is an estimate of the probability once the effects of depolarizing noise have been removed. 
The ``$\frac{1}{8}$" represents the fully decohered probability after post-selecting on physical states (described next) and the ``$1$" is the probability of measuring the initial state from the mitigation circuit in the absence of noise. 

The computational basis of 6 qubits contains $2^6$ states but time evolution only connects those with the same $r$, $g$ and $b$. Starting from the trivial vacuum, this
implies that only the $8$ states with $r=g=b=0$ are accessible through time evolution.
The results off the quantum computer were post-processed to only select events that populated 1 of the 8 physically
allowed states, discarding outcomes that were unphysical. Typically, this resulted in a retention rate of $\sim 30\%$. The
workflow interspersed physics and mitigation circuits to provide a correlated calibration of the quantum devices. This enabled the detection (and removal) of
out-of-specs device performance during post-processing. We explored using the same twirling sequences for both physics and
mitigation circuits and found that it had no significant impact. 
The impact of dynamical decoupling of idle qubits using {\tt qiskit}'s built in functionality was also investigated and found to have little effect. 
The results of each run were corrected for measurement error using IBM's available function, {\tt TensoredMeasFitter}, and associated downstream operations.

\begin{figure}[!ht]
    \centering
    \includegraphics[width=\columnwidth]{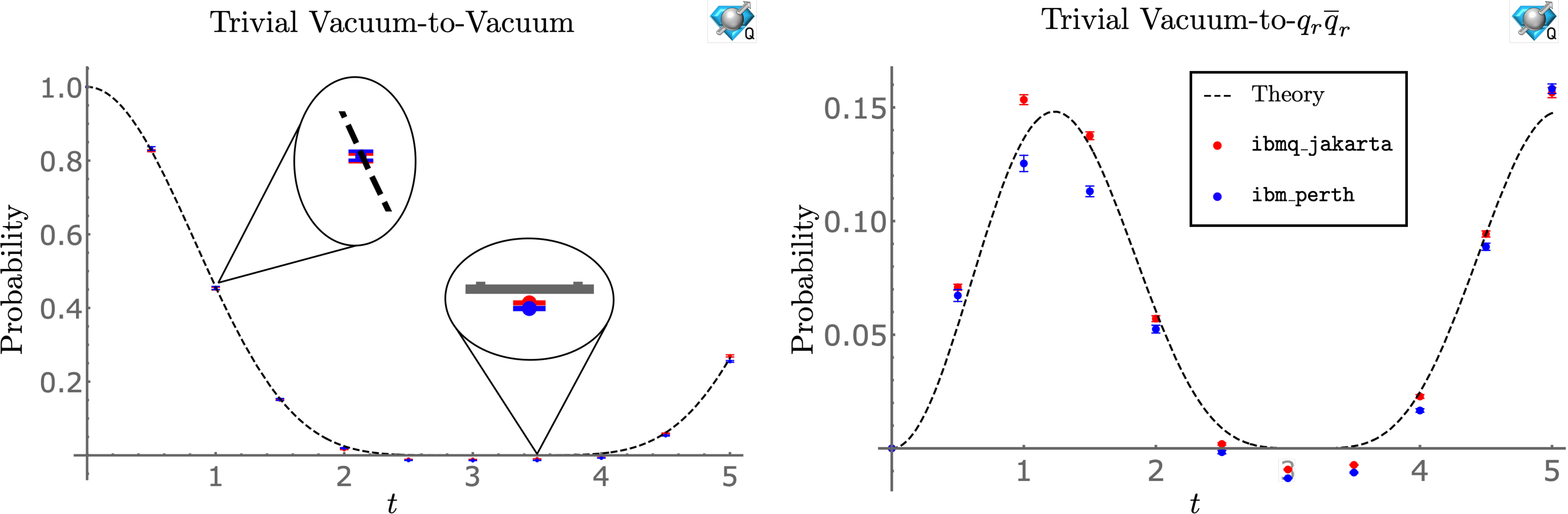}
    \caption{
    The trivial vacuum-to-vacuum (left panel) and trivial vacuum-to-$q_r \overline{q}_r$ (right panel) probabilities for $N_f=1$ QCD and $m=g=L=1$. 
    The dashed-black curve shows the expected result from one step of leading-order Trotter evolution.
    The results, given in Tables~\ref{tab:IBMvacresultsLO} and~\ref{tab:IBMrrbresultsLO}, were obtained by using $10^3$ shots for 447 Pauli-twirled circuits using IBM's quantum computers {\tt ibmq\_jakarta} (red) and {\tt ibm\_perth} (blue).
    }
    \label{fig:IBMresults}
\end{figure}

The results obtained for the trivial vacuum-to-vacuum and trivial vacuum-to-$q_r \overline{q}_r$ probabilities from one step of leading-order Trotter time evolution are shown in Fig.~\ref{fig:IBMresults}.
For each time, 447 Pauli-twirled physics circuits 
and 447 differently twirled circuits with zeroed angles (mitigation) were analyzed using $10^3$ shots on both {\tt ibmq\_jakarta} and {\tt ibm\_perth} (to estimate device systematics).
After post-selecting on physical states, correlated Bootstrap Resampling was used to form the final result.\footnote{As the mitigation and physics circuits were executed as adjacent jobs on the devices, the same Bootstrap sample was used to select results from both ensembles to account for temporal correlations.}
Tables~\ref{tab:IBMvacresultsLO} and~\ref{tab:IBMrrbresultsLO} display the results of the calculations performed using {\tt ibmq\_jakarta} and {\tt ibm\_perth} quantum computers.
The same mitigation data was used for both the trivial vacuum-to-vacuum and trivial vacuum-to-$q_r\overline{q}_r$ calculations, and is provided in columns 2 and 4 of Table~\ref{tab:IBMvacresultsLO}. 
See App.~\ref{app:LOTrott} for an extended discussion of leading-order Trotter. 
Note that the negative probabilities seen in Fig.~\ref{fig:IBMresults} indicate that additional non-linear terms are needed in Eq.~(\ref{eq:mit}).
\begin{table}[!ht]
\renewcommand{\arraystretch}{1.2}
\begin{tabularx}{\textwidth}{||c | Y | Y | Y | Y | Y | Y | Y ||}
\hline
\multicolumn{8}{||c||}{Vacuum-to-Vacuum Probabilities for $N_f=1$ QCD from IBM's {\tt ibmq\_jakarta} and {\tt ibm\_perth}} \\
 \hline
 $t$ & 
 \makecell{Mitigation \\ {\tt jakarta}} & 
 \makecell{Physics \\ {\tt jakarta}} & 
 \makecell{Mitigation \\ {\tt perth}} & 
 \makecell{Physics \\ {\tt perth}} & 
 \makecell{Results \\ {\tt jakarta}} & 
 \makecell{Results \\ {\tt perth}} & 
 Theory
 \\
 \hline\hline
 0 & - & - & - & - & - & - & 1  \\
 \hline
 0.5 & 
 0.9176(10) & 0.7607(24)  &
 0.8744(23) & 0.7310(42)  & 
 0.8268(27) & 0.8326(52) & 0.8274 \\
 \hline
 1.0 & 
 0.9059(12) & 0.4171(32) & 
 0.9118(16) & 0.4211(39) & 
 0.4523(36) & 0.4543(43) & 0.4568 \\
 \hline
 1.5 & 
 0.9180(12) & 0.1483(16) & 
 0.9077(17) & 0.1489(23) &  
 0.1507(17)  & 0.1518(25) & 0.1534 \\
 \hline
 2.0 & 
 0.8953(15) & 0.0292(08) & 
 0.8953(21) & 0.0324(10) &  
 0.0162(09)  &  0.0198(11) & 0.0249 \\
 \hline
 2.5 & 
 0.9169(12) & 0.0020(01) &
 0.8938(21) & 0.0032(02) & 
 -0.0109(03)   &  -0.0136(04) & 0.0010 \\
 \hline
 3.0 &
 0.9282(13) & 0.00010(2) &
 0.9100(13) & 0.00017(3) &  
 -0.0111(02)  &  -0.0140(02) & $1.3\times 10^{-7}$ \\
 \hline
 3.5 &
 0.9357(10)& 0.00017(3) & 
 0.9109(14) & 0.00037(4)& 
 -0.0097(02) &  -0.0138(02) & $3.2\times 10^{-5}$ \\
 \hline
 4.0 & 
 0.9267(13) & 0.0081(03) & 
 0.9023(14) & 0.0076(03) & 
 -0.0026(04) &  -0.0072(04) & 0.0052 \\
 \hline
 4.5 & 
 0.9213(12) & 0.0653(10) &
 0.8995(16) & 0.0619(11) & 
 0.0594(11)  & 0.0537(13) & 0.0614 \\
 \hline
 5.0 & 
 0.9105(12) & 0.2550(26) & 
 0.9031(14) & 0.2405(21) &  
 0.2698(29)  & 0.2550(23) & 0.2644 \\
 \hline
\end{tabularx}
\renewcommand{\arraystretch}{1}
\caption{
The trivial vacuum-to-vacuum probabilities for $m=g=L=1$ using {\tt ibmq\_jakarta} and {\tt ibm\_perth}, the underlying distributions of which are displayed in Fig.~\ref{fig:IBMhistos}.
The 2nd through 5th columns are the results after selecting only physical states and columns $6$ and $7$ are the results after using the mitigation circuit to account for depolarizing noise.
}
\label{tab:IBMvacresultsLO}
\end{table}
\begin{table}[!ht]
\renewcommand{\arraystretch}{1.2}
\begin{tabularx}{0.8\textwidth}{||c | Y | Y | Y | Y | Y ||}
\hline
\multicolumn{6}{||c||}{Vacuum-to-$q_r\overline{q}_r$ Probabilities for $N_f=1$ QCD from IBM's {\tt ibmq\_jakarta} and {\tt ibm\_perth}} \\
 \hline
 $t$ & 
 \makecell{Physics \\ {\tt jakarta}} & 
 \makecell{Physics \\ {\tt perth}} & 
 \makecell{Results \\ {\tt jakarta}} & 
 \makecell{Results \\ {\tt perth}} & 
 Theory
 \\
 \hline\hline
 0 & - & - & - & - & 0  \\
 \hline
 0.5 & 
 0.0760(12) & 0.0756(22)  & 
 0.0709(13) & 0.0673(26) & 0.0539 \\
 \hline
 1.0 & 
 0.1504(19) & 0.1253(32) & 
 0.1534(22) & 0.1254(36) & 0.1363 \\
 \hline
 1.5 & 
 0.1364(15) & 0.1144(21) &  
 0.1376(17)  & 0.1131(23) & 0.1332 \\
 \hline
 2.0 & 
 0.0652(11) & 0.0611(15) &  
 0.0571(13)  &  0.0525(17) & 0.0603 \\
 \hline
 2.5 & 
 0.0136(04) & 0.0137(06) & 
 0.0019(05)   &  -0.0017(07) & 0.0089 \\
 \hline
 3.0 &
 0.0017(01) & 0.0011(01) &  
 -0.0093(02)  &  -0.0132(02) & $2.5\times 10^{-5}$ \\
 \hline
 3.5 &
 0.0024(01) & 0.0032(02)& 
 -0.0073(02) &  -0.0107(03) & 0.0010 \\
 \hline
 4.0 & 
 0.0314(07) & 0.0288(07) & 
 0.0228(08) &  0.0167(08) & 0.0248 \\
 \hline
 4.5 & 
 0.0971(12) & 0.0929(14) & 
 0.0943(13)  & 0.0887(16) & 0.0943 \\
 \hline
 5.0 & 
 0.1534(20) & 0.1546(19) &  
 0.1566(22)  & 0.1583(21) & 0.1475 \\
 \hline
\end{tabularx}
\renewcommand{\arraystretch}{1}
\caption{
The trivial vacuum-to-$q_r \overline{q}_r$ probabilities for $m=g=L=1$ using {\tt ibmq\_jakarta} and {\tt ibm\_perth}.
The 2nd and 3rd columns are the results after selecting only physical states and columns 4 and 5 are the results after using the mitigation circuit to account for depolarizing noise.
}
\label{tab:IBMrrbresultsLO}
\end{table}

\begin{figure}[!ht]
    \centering
    \includegraphics[width=0.9 \columnwidth]{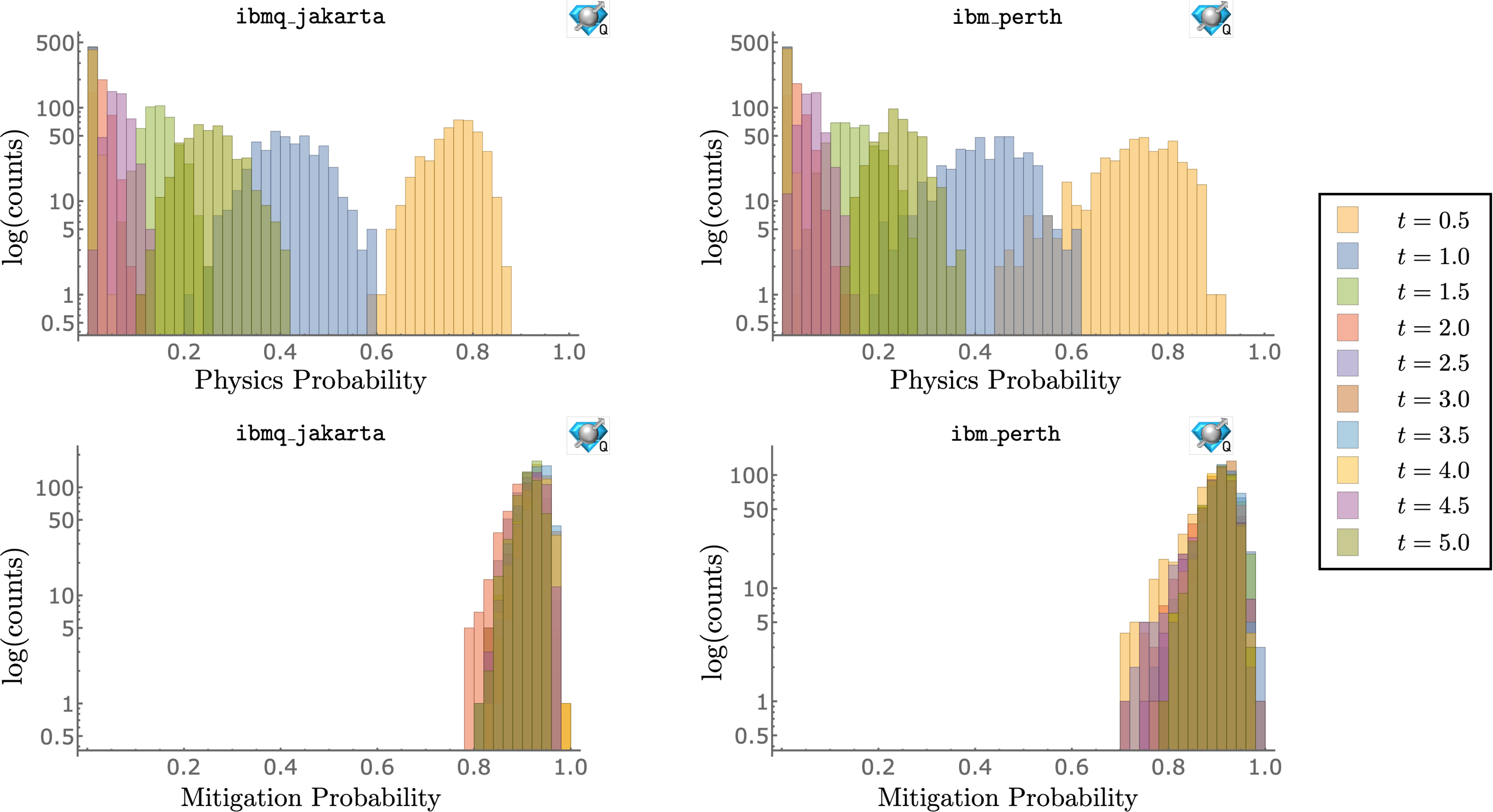}
    \caption{
    Histograms of the post-processed vacuum-to-vacuum results obtained using {\tt ibmq\_jakarta} and {\tt ibm\_perth}.
    The horizontal axes show the value of the vacuum-to-vacuum probability, and the vertical axes show bin counts on a log-scale.
    The top panels display the results obtained from the physics circuits for the range of evolution times and the bottom panels display the results obtained for the corresponding mitigation circuits.
    }
    \label{fig:IBMhistos}
\end{figure}

It is interesting to consider the distributions of events obtained from the Pauli-twirled circuits, as shown in Fig.~\ref{fig:IBMhistos}.
The distributions are not Gaussian and, in a number of instances, exhibit heavy tails particularly near the boundaries.\footnote{For a study of heavy-tailed distributions in Euclidean-space lattice QCD calculations, see Refs.~\cite{Wagman:2016bam,Wagman:2017gqi}.}
The spread of the distributions, associated with non-ideal CNOT gates, is seen to reach a maximum of $\sim 0.4$, but with a full-width at half-max that is $\sim 0.2$. These distributions are already broad with a 34 CNOT circuit, and we probed the limit
of these devices by time-evolving with two first-order Trotter steps,\footnote{Under a particular ordering of terms, two steps of first- and second-order Trotter time evolution are equivalent.}
which requires 91 CNOTs after accounting for SWAPs. 
Using the aforementioned techniques, this was found to be beyond the capabilities of {\tt ibmq\_jakarta}, {\tt ibmq\_lagos} and {\tt ibm\_perth}.

\section{Arbitrary \texorpdfstring{\boldmath$N_c$}{Nc} and \texorpdfstring{\boldmath$N_f$}{Nf}}
\label{sec:NcNf}
\noindent
In this section, the structure of the Hamiltonian for $N_f$ flavors of quarks in the fundamental representation of $SU(N_c)$ is developed. 
The mapping to spins has the same structure as for 
$N_f=2$ QCD, but now, there are $N_c\times N_f$ $q$s and $N_c\times N_f$ $\overline{q}$s per spatial lattice site.
While the mass and kinetic terms generalize straightforwardly, the energy in the chromo-electric field is more tricky.
After enforcing Gauss's law, it is 
\begin{equation}
    H_{el} = \frac{g^2}{2} \sum_{n=0}^{2L-2} \left ( \sum_{m \leq n} Q^{(a)}_m \right ) ^2
    \ ,\ \ 
    Q^{(a)}_m = \phi^{\dagger}_m T^a \phi_m
    \ ,
\end{equation}
where $T^a$ are now the generators of $SU(N_c)$.
The Hamiltonian,
including chemical potentials for baryon number (chemical potentials for other flavor combinations can be included as needed), is found to be
\begin{subequations}
    \label{eq:HNfNc}
    \begin{align}
        H = & \ H_{kin}\ +\ H_m\ +\ H_{el} \ +\ H_{\mu_B}  \ , \\[4pt]
        H_{kin} =& \ \frac{1}{2}\sum_{n=0}^{2L-2}\sum_{f=0}^{N_f-1}\sum_{c=0}^{N_c-1} \left[ \sigma_{i(n,f,c)}^+ \left ( \bigotimes_{j=1}^{N_cN_f-1}(-\sigma_{i(n,f,c)+j}^z ) \right ) \sigma_{i(n,f,c) + N_c N_f}^- +\rm{h.c.} \right] \ ,
        \label{eq:HkinN}\\[4pt]
        H_m =& \ \frac{1}{2} \sum_{n=0}^{2L-1} \sum_{f=0}^{N_f-1}\sum_{c=0}^{N_c-1} m_f \left[ (-1)^n \sigma^z_{i(n,f,c)} + 1 \right] \ ,
        \label{eq:HmN}\\[4pt]
        H_{el} =& \ \frac{g^2}{2} \sum_{n=0}^{2L-2}(2L-1-n)\left( \sum_{f=0}^{N_f-1} Q_{n,f}^{(a)} \, Q_{n,f}^{(a)} \ \ + \ \  
            2 \sum_{f=0}^{N_f-2} \sum_{f'=f+1}^{N_f-1}Q_{n,f}^{(a)} \, Q_{n,f'}^{(a)}
             \right)   \nonumber \\[4pt]
            & + g^2 \sum_{n=0}^{2L-3} \sum_{m=n+1}^{2L-2}(2L-1-m) \sum_{f=0}^{N_f-1} \sum_{f'=0}^{N_f-1} Q_{n,f}^{(a)} \, Q_{m,f'}^{(a)}  \ ,
        \label{eq:HelN}\\[4pt]
        H_{\mu_B} =& \ -\frac{\mu_B}{2 N_c} \sum_{n=0}^{2L-1} \sum_{f=0}^{N_f-1} \sum_{c=0}^{N_c-1} \sigma^z_{i(n,f,c)}   \ ,
        \label{eq:HmuBN}
    \end{align}
\end{subequations}
where, $i(n,f,c) = (N_c N_f n + N_f f + c)$, 
and the products of the charges are
\begin{align}
    4 Q_{n,f}^{(a)} \, Q_{n,f}^{(a)} =& \ \frac{N_c^2-1}{2} - \left (1+\frac{1}{N_c} \right )\sum_{c=0}^{N_c-2}\sum_{c' = c+1}^{N_c-1} \sigma^z_{i(n,f,c)} \sigma^z_{i(n,f,c')} \ ,  \nonumber \\[4pt]
    8 Q_{n,f}^{(a)} \, Q_{m,f'}^{(a)} =& \ 4 \sum_{c=0}^{N_c-2}\sum_{c'=c+1}^{N_c-1} \left[ \sigma^+_{i(n,f,c)} \ \left(\otimes Z \right)_{(n,f,c,c')} \ \sigma^-_{i(n,f,c')} \sigma^-_{i(m,f',c)} \ \left(\otimes Z \right)_{(m,f',c,c')} \ \sigma^+_{i(m,f',c')} + {\rm h.c.}
    \right] \  \nonumber \\
    &+ \sum_{c=0}^{N_c-1} \sum_{c'=0}^{N_c-1}\left (\delta_{cc'} - \frac{1}{N_c} \right)\sigma^z_{i(n,f,c)}\sigma^z_{i(m,f',c')}\ , \nonumber \\
    \left(\otimes Z \right)_{(n,f,c,c')}  \equiv & \ \bigotimes_{k=1}^{c'-c-1} \sigma^z_{i(n,f,c)+k}
    \ .
    \label{eq:QnfQmfpN}
\end{align}
The resource requirements for implementing Trotterized time evolution 
using generalizations of the circuits in Sec.~\ref{sec:Circuits} are given in Eq.~(\ref{eq:RHCN}). 

It is interesting to consider the large-$N_c$ limit of the Hamiltonian, 
where quark loops are parametrically suppressed and 
the system can be described semi-classically~\cite{tHooft:1973alw,tHooft:1974pnl,Witten:1979kh,RevModPhys.54.407}.
Unitarity requires rescaling the strong coupling, 
$g^2 \to g^2/N_c$ and leading terms in the Hamiltonian scale as $\mathcal{O}(N_c)$.
The leading order contribution to the product of charges is
\begin{align}
    4 Q_{n,f}^{(a)} \, Q_{n,f}^{(a)} =&\ \sum_{c=0}^{N_c-2}\sum_{c' = c+1}^{N_c-1} \left (1 - \sigma^z_{i(n,f,c)} \sigma^z_{i(n,f,c')}\right ) \ ,  \nonumber \\[4pt]
    8 Q_{n,f}^{(a)} \, Q_{m,f'}^{(a)} =&\ 4 \sum_{c=0}^{N_c-2}\sum_{c'=c+1}^{N_c-1} 
    \left[ \sigma^+_{i(n,f,c)} \ 
    \left(\otimes Z \right)_{(n,f,c,c')} \ 
    \sigma^-_{i(n,f,c')} \sigma^-_{i(m,f',c)} \ 
    \left(\otimes Z \right)_{(m,f',c,c')} \ 
    \sigma^+_{i(m,f',c')} + {\rm h.c.}
    \right] \ .
\end{align}
Assuming that the number of $q\overline{q}$ pairs that contribute to the meson wavefunctions do not scale with $N_c$, 
as expected in the large-$N_c$ limit,
$H_{el} \propto N_c$ 
and mesons are non-interacting, a well known consequence of the large-$N_c$ limit~\cite{tHooft:1973alw,tHooft:1974pnl}.
Baryons on the other hand are expected to have strong interactions at leading order in $N_c$~\cite{Witten:1979kh}. This is a semi-classical limit and we expect that there exists a basis
where states factorize into localized tensor products, and the time evolution operator is non-entangling.
The latter result has been observed in the large-$N_c$ limit of hadronic scattering~\cite{Beane:2018oxh,Beane:2021zvo,Low:2021ufv,Aoude:2020mlg,Cervera-Lierta:2017tdt}.

\section{Summary and Discussion}
\label{sec:SandC}
\noindent
Important for future quantum simulations of processes that can be meaningfully compared to experiment, the real-time dynamics of strongly-interacting systems are predicted to be efficiently computable with quantum computers of sufficient capability.
Building upon foundational work in quantum chemistry and in low-dimensional $U(1)$ and $SU(2)$ gauge theories, this work has developed the tools necessary for the quantum simulation of
$1+1$D QCD (in axial gauge) using open boundary conditions, with arbitrary numbers of quark flavors and colors and including chemical potentials for baryon number and isospin.
Focusing largely on QCD with $N_f=2$, which shares many of the complexities of QCD in $3+1$D, we have performed a detailed analysis of the required quantum resources for simulation of real-time dynamics, including efficient quantum circuits and associated gate counts, and the scaling of the number of Trotter steps for a fixed-precision time evolution.
The structure and dynamics of small systems, with $L=1,2$ for $N_c=3$ and $N_f=1,2$ have been detailed using classical computation, quantum simulators, D-Wave's {\tt Advantage} and IBM's 7-qubit devices {\tt ibmq\_jakarta} and {\tt ibm\_perth}. Using recently developed error mitigation strategies, relatively small uncertainties were obtained for a single Trotter step with $34$ CNOT gates after transpilation onto the QPU connectivity.

Through a detailed study of the low-lying spectrum, both the relevant symmetries and the color-singlets in the mesonic and baryonic sectors, including a bound two-baryon nucleus, have been identified.  
Open boundary conditions also permit low-lying color edge-states that penetrate into the lattice volume by a distance set by the confinement scale.
By examining quark entanglement in the hadrons, a transition from the mesons being primarily composed of quark-antiquarks to baryon-antibaryons was found.
We have presented the relative contributions of each of the terms in the Hamiltonian to the energy of the vacuum, mesons and baryons.  

This work has provided an estimate for the number of CNOT-gates required to implement one Trotter step in $N_f=2$, $1+1$D axial-gauge QCD. For $L = 10$ spatial sites, $\sim 3 \times 10^4$ CNOTs
are required, while $\sim 4 \times 10^6$ CNOTs are required for $L = 100$.
Realistically, quantum simulations with $L=10$ are a beginning toward providing results with a complete quantification of uncertainties, including lattice-spacing and finite-volume
artifacts, and $L=100$ will likely yield high-precision results. It was found that, in the axial-gauge formulation, resources for time evolution effectively scale as $L^2 t$ for intermediate times and $L^2 t^2$ for
asymptotic times. With $L\sim t$, this asymptotic scaling is the same as in the Schwinger model, suggesting no differences in scaling between Weyl and axial gauges.

\begin{acknowledgements}
We would like to thank Fabio Anza, Anthony Ciavarella, Stephan Caspar, David B.~Kaplan,
Natalie Klco, Sasha Krassovsky and Randy Lewis for very helpful discussions and insightful comments.
We would also like to thank
Christian Bauer,
Ewout van den Berg,
Alaina Green,
Abhinav Kandala,
Antonio Mezzacapo,
Mohan Sarovar and 
Prasanth Shyamsundar
for very helpful discussions during the IQuS-INT workshop on
{\it Quantum Error Mitigation for Particle and Nuclear Physics}, May 9-13, 2022 (\url{https://iqus.uw.edu/events/iqus-workshop-22-1b}).

This work was supported, in part, by 
the U.S. Department of Energy grant DE-FG02-97ER-41014 (Farrell), 
the U.S. Department of Energy,
Office of Science, Office of Nuclear Physics, InQubator for Quantum Simulation (IQuS) 
(\url{https://iqus.uw.edu})
under Award Number DOE
(NP) Award DE-SC0020970 (Chernyshev, Farrell, Powell, Savage, Zemlevskiy), 
and the 
Quantum Science Center (QSC)
(\url{https://qscience.org}), 
a National Quantum Information Science Research Center of the U.S.  Department of Energy (DOE) (Illa).
This work is also supported, in part, through the Department of Physics 
(\url{https://phys.washington.edu}) 
and the College of Arts and Sciences 
(\url{https://www.artsci.washington.edu})
at the University of Washington.

We acknowledge the use of IBM Quantum services~\cite{IBMQ} for this work. The views expressed are those of the authors, and do not reflect the official policy or position of IBM or the IBM Quantum team.  
In this paper we used 
{\tt ibm\_lagos},
{\tt ibm\_perth} and
{\tt ibmq\_jakarta}, which are three  of the IBM's r5.11H Quantum Processors.
All calculations performed on D-Wave's QAs were through cloud access~\cite{DwaveLeap}.
We have made extensive use of Wolfram {\tt Mathematica}~\cite{Mathematica},
{\tt python}~\cite{python3,Hunter:2007}, {\tt julia}~\cite{Julia-2017},
{\tt jupyter} notebooks~\cite{PER-GRA:2007} 
in the {\tt Conda} environment~\cite{anaconda},
and the quantum programming environments: Google's {\tt cirq}~\cite{cirq_developers_2022_6599601} and IBM's {\tt qiskit}~\cite{gadi_aleksandrowicz_2019_2562111}.

This work was enabled, in part, by
the use of advanced computational, storage and networking infrastructure provided by the Hyak supercomputer system at the University of Washington 
(\url{https://itconnect.uw.edu/research/hpc}).
\end{acknowledgements}

\clearpage
\appendix

\section{Mapping to Qubits}
\label{app:hamConst}
\noindent 
This appendix outlines how the qubit Hamiltonian in Eq.~(\ref{eq:H2flav}) is obtained from the lattice Hamiltonian in Eq.~(\ref{eq:GFHam}). 
For this system, 
the constraint of Gauss's law is sufficient to uniquely determine the chromo-electric field carried by the links between lattice sites in terms of a background chromo-electric field and the distribution of color charges.  The difference between adjacent chromo-electric fields at a site with charge 
$Q^{(a)}$ 
is
\begin{equation}
    {\bf E}^{(a)}_{n+1} - {\bf E}^{(a)}_n = Q^{(a)}_n \ ,
    \label{eq:GaussLaw}
\end{equation}
for $a=1$ to $8$, resulting in a
chromo-electric field 
\begin{equation}
    {\bf E}^{(a)}_{n} = {\bf F}^{(a)}
    \: + \: \sum_{i\leq n} Q^{(a)}_i \  .
    \label{eq:GaussLawSol}
\end{equation}
In general, there can be a non-zero background chromo-electric field, ${\bf F}^{(a)}$,
which in this paper has been set to zero.
Inserting the chromo-electric field in terms of the charges into Eq.~(\ref{eq:KSHam}) yields Eq.~(\ref{eq:GFHam}). 

The color and flavor degrees of freedom of each $q$ and $\overline{q}$ are then distributed over 
$6$ ($=N_c N_f$) sites as illustrated in Fig.~(\ref{fig:2flavLayout}). 
There are now creation and annihilation operators for each quark, and the Hamiltonian is
\begin{align}
   H =&\ \sum_{n=0}^{2L-1} \sum_{f=0}^1 \sum_{c=0}^2  \left [ m_f (-1)^n \psi^{\dagger}_{6n+3f+c} \psi_{6n+3f+c} \: - \: \frac{\mu_B}{3} \psi^{\dagger}_{6n+3f+c} \psi_{6n+3f+c} \: - \: \frac{\mu_I}{2}(-1)^f \psi^{\dagger}_{6n+3f+c} \psi_{6n+3f+c} \right ] \nonumber \\[4pt]
   &+  \frac{1}{2} \sum_{n=0}^{2L-2}\sum_{f=0}^1 \sum_{c=0}^2  \left (\psi^{\dagger}_{6n+3f+c} \psi_{6(n+1)+3f+c} + \: {\rm h.c.} \right ) \: + \: \frac{g^2}{2} \sum_{n=0}^{2L-2} \left ( \sum_{m\leq n} \sum_{f=0}^1 Q^{(a)}_{m,f} \right ) ^2 \ ,
    \label{eq:FockHam}
\end{align}
where the color charge is evaluated over three $(r,g,b)$ occupation sites with the same flavor,
\begin{equation}
   Q_{m,f}^{(a)} = \sum_{c=0}^{2}\sum_{c'=0}^2 \psi^{\dagger}_{6m+3f+c} \ T^a_{cc'}\  \psi_{6m+3f+c'} \ ,
\end{equation}
and the $T^a$ are the eight generators of $SU(3)$. 
The fermionic  operators in Fock space are mapped onto spin operators via the JW transformation,
\begin{equation}
    \psi_n =  \bigotimes_{l<n}( -\sigma^z_l ) \sigma^-_n \ , \ \ \psi_n^{\dagger} =  \bigotimes_{l<n}( -\sigma^z_l ) \sigma^+_n \ .
    \label{eq:JW}
\end{equation}
In terms of spins, the eight $SU(3)$ charge operators become\footnote{Calculations of quadratics of the gauge charges are simplified by the Fierz identity,
\begin{equation}
    \left ( T^{(a)} \right )^{\alpha}_{\beta} \, \left (T^{(a)}\right )^{\gamma}_{\delta} = \frac{1}{2} (\delta^{\alpha}_{\delta} \delta^{\gamma}_{\beta} - \frac{1}{N_c} \delta^{\alpha}_{\beta}\delta^{\gamma}_{\delta}) \ .
    \label{eq:Fierz}
\end{equation}}
\begin{align}
    Q_{m,f}^{(1)} = & \ \frac{1}{2}\sigma^+_{6m+3f} \sigma^-_{6m+3f+1} + \rm{h.c.} \ , \nonumber \\
    Q_{m,f}^{(2)} = & \ -\frac{i}{2}\sigma^+_{6m+3f} \sigma^-_{6m+3f+1} + \rm{h.c.} \ , \nonumber \\
    Q_{m,f}^{(3)} = & \ \frac{1}{4}(\sigma^z_{6m+3f} - \sigma^z_{6m+3f+1}) \ , \nonumber \\
    Q_{m,f}^{(4)} = & \ -\frac{1}{2}\sigma^+_{6m+3f} \sigma^z_{6m+3f+1} \sigma^-_{6m+3f+2} + \rm{h.c.} \ , \nonumber \\
    Q_{m,f}^{(5)} = & \ \frac{i}{2}\sigma^+_{6m+3f} \sigma^z_{6m+3f+1} \sigma^-_{6m+3f+2} + \rm{h.c.} \ , \nonumber \\
    Q_{m,f}^{(6)} = & \ \frac{1}{2}\sigma^+_{6m+3f+1} \sigma^-_{6m+3f+2} + \rm{h.c.} \ , \nonumber \\
    Q_{m,f}^{(7)} = & \ -\frac{i}{2}\sigma^+_{6m+3f+1} \sigma^-_{6m+3f+2} + \rm{h.c.} \ , \nonumber \\
    Q_{m,f}^{(8)} = & \ \frac{1}{4 \sqrt{3}}(\sigma^z_{6m+3f} + \sigma^z_{6m+3f+1} - 2\sigma^z_{6m+3f+2})  \ .
    \label{eq:SU3chargesFull}
\end{align}
Substituting Eqs.~(\ref{eq:JW}) and~(\ref{eq:SU3chargesFull}) into Eq.~(\ref{eq:FockHam}) gives the Hamiltonian in Eq.~(\ref{eq:H2flav}). For reference, the expanded Hamiltonian for $L=1$ is
\begin{subequations}
    \label{eq:H2flavL1}
    \begin{align}
    H = & \ H_{kin}\ +\ H_m\ +\ H_{el} \ +\ 
    H_{\mu_B}\ +\ H_{\mu_I}\ ,\\[4pt]
    H_{kin} = & \ -\frac{1}{2} (\sigma^+_6 \sigma^z_5 \sigma^z_4 \sigma^z_3 \sigma^z_2 \sigma^z_1 \sigma^-_0 + \sigma^-_6 \sigma^z_5 \sigma^z_4 \sigma^z_3 \sigma^z_2 \sigma^z_1 \sigma^+_0 + \sigma^+_7 \sigma^z_6 \sigma^z_5 \sigma^z_4 \sigma^z_3 \sigma^z_2 \sigma^-_1 + \sigma^-_7 \sigma^z_6 \sigma^z_5 \sigma^z_4 \sigma^z_3 \sigma^z_2 \sigma^+_1  \nonumber \\
    &+\, \sigma^+_8 \sigma^z_7 \sigma^z_6 \sigma^z_5 \sigma^z_4 \sigma^z_3 \sigma^-_2 + \sigma^-_8 \sigma^z_7 \sigma^z_6 \sigma^z_5 \sigma^z_4 \sigma^z_3 \sigma^+_2 + \sigma^+_9 \sigma^z_8 \sigma^z_7 \sigma^z_6 \sigma^z_5 \sigma^z_4 \sigma^-_3 + \sigma^-_9 \sigma^z_8 \sigma^z_7 \sigma^z_6 \sigma^z_5 \sigma^z_4 \sigma^+_3 \nonumber \\
    &+\, \sigma^+_{10} \sigma^z_9 \sigma^z_8 \sigma^z_7 \sigma^z_6 \sigma^z_5 \sigma^-_4 + \sigma^-_{10} \sigma^z_9 \sigma^z_8 \sigma^z_7 \sigma^z_6 \sigma^z_5 \sigma^+_4 + \sigma^+_{11} \sigma^z_{10} \sigma^z_9 \sigma^z_8 \sigma^z_7 \sigma^z_6 \sigma^-_5 + \sigma^-_{11} \sigma^z_{10} \sigma^z_9 \sigma^z_8 \sigma^z_7 \sigma^z_6 \sigma^+_5 ) \ ,
        \label{eq:Hkin2flavL1}\\[4pt]
    H_m = & \ \frac{1}{2} \left [ m_u\left (\sigma^z_0 + \sigma^z_1 + \sigma^z_2 -\sigma^z_6 - \sigma^z_7 - \sigma^z_8 + 6\right )+ m_d\left (\sigma^z_3 + \sigma^z_4 + \sigma^z_5 -\sigma^z_9 - \sigma^z_{10} - \sigma^z_{11} + 6\right ) \right ]\ ,
        \label{eq:Hm2flavL1}\\[4pt]
    H_{el} = & \ \frac{g^2}{2} \bigg [ \frac{1}{3}(3 - \sigma^z_1 \sigma^z_0 - \sigma^z_2 \sigma^z_0 - \sigma^z_2 \sigma^z_1) + \sigma^+_4\sigma^-_3\sigma^-_1\sigma^+_0  + \sigma^-_4\sigma^+_3\sigma^+_1\sigma^-_0  + \sigma^+_5\sigma^z_4\sigma^-_3\sigma^-_2\sigma^z_1\sigma^+_0 + \sigma^-_5\sigma^z_4\sigma^+_3\sigma^+_2\sigma^z_1\sigma^-_0 \nonumber \\
    & +\, \sigma^+_5\sigma^-_4\sigma^-_2\sigma^+_1 + \sigma^-_5\sigma^+_4\sigma^+_2\sigma^-_1  \nonumber \\
    & +\,\frac{1}{12}\left (2 \sigma^z_3 \sigma^z_0 + 2\sigma^z_4 \sigma^z_1 + 2\sigma^z_5 \sigma^z_2 - \sigma^z_5 \sigma^z_0 - \sigma^z_5 \sigma^z_1 - \sigma^z_4 \sigma^z_2 - \sigma^z_4 \sigma^z_0 - \sigma^z_3 \sigma^z_1  - \sigma^z_3 \sigma^z_2 \right ) \bigg ] \ ,
    \label{eq:Hel2flavL1}\\[4pt] 
    H_{\mu_B} = & \ -\frac{\mu_B}{6} \left ( \sigma^z_0 + \sigma^z_1 + \sigma^z_2 + \sigma^z_3 + \sigma^z_4 + \sigma^z_5
    - \sigma^z_6 + \sigma^z_7 + \sigma^z_8 + \sigma^z_9 + \sigma^z_{10} + \sigma^z_{11} \right )\ ,
        \label{eq:HmuB2flavL1}\\[4pt]
    H_{\mu_I} = & \ -\frac{\mu_I}{4} \left ( \sigma^z_0 + \sigma^z_1 + \sigma^z_2 - \sigma^z_3 - \sigma^z_4 - \sigma^z_5
    + \sigma^z_6 + \sigma^z_7 + \sigma^z_8 - \sigma^z_9 - \sigma^z_{10} - \sigma^z_{11} \right ) \ .
        \label{eq:HmuI2flavL1}
    \end{align}
\end{subequations}
%

\section{Symmetries of the Free-Quark Hamiltonian}
\label{app:freeSym}
\noindent 
Here the symmetries of the free-quark Hamiltonian are identified to better understand the degeneracies observed in the spectrum of $1+1$D QCD with $N_f=2$ and $L=1$ as displayed in Figs.~\ref{fig:specDegenh} and~\ref{fig:specDegeng}.
Specifically, the Hamiltonian with $g=h=\mu_B=\mu_I=0$, leaving only the hopping and mass terms ($m = m_u = m_d$), is
\begin{equation}
   H =   \sum_{f=0}^1 \sum_{c=0}^2  \left [ m \sum_{n=0}^{2L-1} (-1)^n \psi^{\dagger}_{6n+3f+c} \psi_{6n+3f+c} \: + \: \frac{1}{2} \sum_{n=0}^{2L-2} \left (\psi^{\dagger}_{6n+3f+c} \psi_{6(n+1)+3f+c} + \: {\rm h.c.} \right ) \right ] \ .
\end{equation}
The mapping of degrees of freedom is taken to be as shown in Fig.~\ref{fig:2flavLayout}, but it will be convenient to work with Fock-space quark operators instead of spin operators. 
In what follows the focus will be on $L=1$, and larger systems follow similarly.

The creation operators can be assembled into a 12-component vector, 
$\Psi^{\dagger}_i = (\psi_0^\dagger, \, \psi_1^\dagger, \ldots ,\psi_{10}^{\dagger}, \, \psi_{11}^{\dagger})$, 
in terms of which the Hamiltonian  becomes
\begin{equation}
    H = \Psi^{\dagger}_i M_{ij} \Psi_j \ ,
\end{equation}
where $M$ is a $12 \times 12$ block matrix of the form,
\begin{equation}
   M  = 
   \left[
   \begin{array}{c|c}
       m  & 1/2 \\
       \hline
    1/2  & -m 
\end{array}
\right]
\ ,
\end{equation}
with each block a $6 \times 6$  diagonal matrix. 
Diagonalizing $M$, gives rise to 
\begin{equation}
   \tilde M = 
   \left[
   \begin{array}{c|c}
       \lambda  & 0 \\
       \hline
    0  & -\lambda 
\end{array}
\right]
\ ,\ \ 
\lambda = \frac{1}{2}\sqrt{1+4m^2} \ ,
\end{equation}
with associated eigenvectors,
\begin{equation}
    \tilde{\psi}_i = \frac{1}{\sqrt{2}}\left (\sqrt{1+\frac{\lambda}{m}}\, \psi_i \ + \ \sqrt{1-\frac{\lambda}{m}}\, \psi_{6+i} \right ) \ , \ \tilde{\psi}_{6+i} = \frac{1}{\sqrt{2}}\left (-\sqrt{1-\frac{\lambda}{m}}\,\psi_i \ + \ \sqrt{1+\frac{\lambda}{m}}\, \psi_{6+i} \right )
    \ ,
\end{equation}
where $\tilde{\psi}_i$ ($\tilde{\psi}_{6+i}$) corresponds to the positive (negative) eigenvalue
and the index $i$ takes values $0$ to $5$.
These eigenvectors create superpositions of quarks and antiquarks with the same color and flavor, which are the OBC analogs of momentum plane-waves. 
In this basis, the Hamiltonian becomes
\begin{equation}
    H = \sum_{i=0}^{5} \lambda\left ( \tilde{\psi}^{\dagger}_i \tilde{\psi}_i - \tilde{\psi}^{\dagger}_{6+i} \tilde{\psi}_{6+i} \right )
    \ ,
    \label{eq:Hamifree0}
\end{equation}
which has a vacuum state,
\begin{equation}
    \lvert \Omega_0 \rangle = \prod_{i=0}^{i=5}\tilde{\psi}^{\dagger}_{6+i} \ket{\omega_0} \ ,
\end{equation}
where $\ket{\omega_0}$ is the unoccupied state,
and
$\lvert \Omega_0 \rangle$ corresponds to 
$\lvert 000000111111 \rangle$ (in binary)
in this transformed basis.
Excited states are formed by acting with either $\tilde{\psi}^{\dagger}_i$ or $\tilde{\psi}_{6+i}$ on 
$\lvert \Omega_0 \rangle$ which raises the energy of the system by $\lambda$. 
A further transformation is required for the $SU(12)$ symmetry to be manifest.
In terms of the 12-component vector, $\tilde{\Psi}^{\dagger} = (\tilde{\psi}^{\dagger}_0, \, \ldots, \, \tilde{\psi}^{\dagger}_5, \, \tilde{\psi}_6, \, \ldots, \, \tilde{\psi}_{11})$, the Hamiltonian in Eq.~(\ref{eq:Hamifree0}) becomes,
\begin{equation}
    H = 
    \sum_{i=0}^{5} \lambda\left ( \tilde{\psi}^{\dagger}_i \tilde{\psi}_i - \tilde{\psi}^{\dagger}_{6+i} \tilde{\psi}_{6+i} \right )
    \ =\ 
    \lambda\left( 
    \tilde{\Psi}^{\dagger} \tilde{\Psi} - 6 
    \right)
    \ ,
\end{equation}
where the canonical anticommutation relations have been used to obtain the final equality.
This is invariant under a $SU(12)$ symmetry, where $\tilde{\Psi}$ transforms in the fundamental representation. 
The free-quark spectrum ($g=h=0$) is therefore described by states with degeneracies corresponding to the ${\bf 1}$ and ${\bf 12}$ of $SU(12)$ as well as
the antisymmetric combinations of fundamental irreps, ${\bf 66}, {\bf 220}, \ldots$ as illustrated in Figs.~\ref{fig:specDegenh} and~\ref{fig:specDegeng}.
The vacuum state corresponds to the singlet of $SU(12)$. The lowest-lying {\bf 12} corresponds to single quark or antiquark excitations, which are color ${\bf 3}_c$s for quarks and $\overline{\bf 3}_c$s for antiquarks and will each appear as isodoublets, i.e., ${\bf 12}\rightarrow {\bf 3}_c\otimes {\bf 2}_f \oplus \overline{\bf 3}_c\otimes {\bf 2}_f$.
The {\bf 66} arises from double excitations of quarks and antiquarks.  The possible color-isospin configurations are, based upon totally-antisymmetric wavefunctions for $qq$, $\overline{q}\overline{q}$ and $\overline{q}q$,  
${\bf 66} =
{\bf 1}_c\otimes {\bf 1}_f
\oplus
{\bf 1}_c\otimes {\bf 3}_f
\oplus
{\bf 8}_c\otimes {\bf 1}_f
\oplus
{\bf 8}_c\otimes {\bf 3}_f
\oplus
{\bf 6}_c\otimes {\bf 1}_f
\oplus
\overline{\bf 6}_c\otimes {\bf 1}_f
\oplus
{\bf 3}_c\otimes {\bf 3}_f
\oplus
\overline{\bf 3}_c\otimes {\bf 3}_f
$.
The OBCs split the naive symmetry between quarks and antiquarks and, for $g\ne 0$, the lowest-lying color edge-states are from the antiquark sector with degeneracies $6$ from a single excitation and $6,9$ from double excitations. 
Larger lattices possess an analogous global 
$SU(12)$ symmetry, coupled between spatial sites by the hopping term, and the spectrum is again one of non-interacting quasi-particles.

\section{Details of the D-Wave Implementations}
\label{app:dwave}
\noindent
In this appendix, additional details are provided on the procedure used in Sec.~\ref{sec:dwave_spectrum} to extract the lowest three eigenstates and corresponding energies using D-Wave's {\tt Advantage}, (a more complete description can be found in Ref.~\cite{Illa:2022jqb}). The objective function $F$ to be minimized can be written in terms of binary variables and put into QUBO form. Defining $F=\langle \Psi \rvert \tilde{H} \lvert \Psi \rangle -\eta \langle \Psi| \Psi \rangle$~\cite{doi:10.1021/acs.jctc.9b00402}, and expanding the wavefunction with a finite dimensional orthonormal basis $\psi_{\alpha}$, $\lvert \Psi \rangle =\sum^{n_s}_{\alpha} a_\alpha |\psi_{\alpha}\rangle$, it is found
\begin{equation}
    F=\langle \Psi \rvert \tilde{H} \lvert \Psi \rangle -\eta \langle \Psi| \Psi \rangle = \sum_{\alpha\beta}^{n_s} a_\alpha a_\beta[\langle \psi_\alpha \rvert \tilde{H} \lvert \psi_\beta \rangle -\eta \langle \psi_\alpha| \psi_\beta \rangle] =\sum_{\alpha\beta}^{n_s} a_\alpha a_\beta (\tilde{H}_{\alpha\beta} -\eta \delta_{\alpha\beta})=\sum_{\alpha\beta}^{n_s} a_\alpha a_\beta h_{\alpha\beta}\ ,
\end{equation}
where $h_{\alpha\beta}$ are the matrix elements of the Hamiltonian that can be computed classically. The coefficients $a_\alpha$ are then expanded in a fixed-point representation using $K$ bits~\cite{doi:10.1021/acs.jctc.9b00402,Chang:2019,ARahman:2021ktn}, 
\begin{equation}
    a^{(z+1)}_\alpha=a^{(z)}_\alpha+\sum_{i=1}^{K}2^{i-K-z}(-1)^{\delta_{iK}}q^{\alpha}_i \ ,
\end{equation}
where $z$ is the zoom parameter. The starting point is $a_\alpha^{(z=0)}=0$, and for each consecutive value of $z$, the range of values that $a_\alpha^{(z+1)}$ is allowed to explore is reduced by a factor of $2$, centered around the previous solution $a_\alpha^{(z)}$. Now $F$ takes the following form,
\begin{equation}
    F=\sum_{\alpha,\beta}^{n_s}\sum_{i,j}^K Q_{\alpha,i;\beta,j} q^{\alpha}_i q^{\beta}_j\ , \ Q_{\alpha,i;\beta,j}=2^{i+j-2K-2z} (-1)^{\delta_{iK}+\delta_{jK}} h_{\alpha\beta} + 2 \delta_{\alpha\beta} \delta_{ij} 2^{i-K-z} (-1)^{\delta_{iK}} \sum_\gamma^{n_s} a^{(z)}_\gamma h_{\gamma\beta} \ .
\end{equation}
The iterative procedure used to improve the precision of the results is based on the value $a^{(z)}_\alpha$ obtained after $14$ zoom steps (starting from $a_\alpha^{(z_0=0)}=0$), and then launching a new annealing workflow with $z_1 \neq 0$ (e.g., $z_1=4$), with $a^{(z=z_0+14)}_\alpha$ as the starting point. After another 14 zoom steps, the final value $a^{(z=z_1+14)}_\alpha$ can be used as the new starting point for $a^{(z=z_2)}_\alpha$, with $z_2 > z_1$. This process can be repeated until no further improvement is seen in the convergence of the energy and wavefunction.

In Table~\ref{tab:QAresults2}, the difference between the exact energy of the vacuum and masses of the $\sigma$- and $\pi$-mesons and the ones computed with the QA, for each iteration of this procedure after 14 zoom steps, are given, together with the overlap of the wavefunctions $1-|\langle \Psi^{\rm exact}| \Psi^{\tt Adv.}\rangle|^2$. See also Fig.~\ref{fig:QAresults}.
\begin{table}[!ht]
\renewcommand{\arraystretch}{1.2}
\begin{tabular}{|| c | r | r@{\ \ \ \ \ } | r | r@{\ \ \ \ \ } | r | r@{\ \ \ \ \ } ||} 
\hline
  & \multicolumn{2}{c|}{$\ket{\Omega}$} & \multicolumn{2}{c|}{$\ket{\sigma}$} & \multicolumn{2}{c|}{$\ket{\pi}$} \\ \hline
  Step & \multicolumn{1}{c|}{$\delta E_\Omega$} & \multicolumn{1}{c|}{$1-|\langle \Psi_\Omega^{\rm exact}| \Psi_\Omega^{\tt Adv.}\rangle|^2$} & \multicolumn{1}{c|}{$\delta M_\sigma$} & \multicolumn{1}{c|}{$1-|\langle \Psi_\sigma^{\rm exact}| \Psi_\sigma^{\tt Adv.}\rangle|^2$} & \multicolumn{1}{c|}{$\delta M_\pi$} & \multicolumn{1}{c|}{$1-|\langle \Psi_\pi^{\rm exact}| \Psi_\pi^{\tt Adv.}\rangle|^2$} \\
 \hline \hline 
 0 & $4^{\,+2}_{\,-2}\times 10^{-1}$ & $10^{\,+3}_{\,-5}\times 10^{-2}$ & $4^{\,+2}_{\,-2}\times 10^{-1}$ & $11^{\,+7}_{\,-5}\times 10^{-2}$& $3^{\,+1}_{\,-1}\times 10^{-1}$ & $11^{\,+71}_{\,-4}\times 10^{-2}$\\[0.2ex]
 1 & $9^{\,+4}_{\,-3}\times 10^{-3}$ & $2^{\,+6}_{\,-5}\times 10^{-3}$  & $3^{\,+1}_{\,-1}\times 10^{-2}$ & $7^{\,+2}_{\,-2}\times 10^{-3}$ & $9^{\,+4}_{\,-3}\times 10^{-3}$ & $3^{\,+3}_{\,-1}\times 10^{-3}$\\[0.2ex]
 2 & $6^{\,+2}_{\,-2}\times 10^{-4}$ & $12^{\,+3}_{\,-5}\times 10^{-5}$ & $4^{\,+1}_{\,-1}\times 10^{-3}$ & $12^{\,+3}_{\,-4}\times 10^{-4}$& $7^{\,+2}_{\,-3}\times 10^{-4}$ & $3^{\,+2}_{\,-2}\times 10^{-4}$\\[0.2ex]
 3 & $4^{\,+1}_{\,-2}\times 10^{-5}$ & $9^{\,+3}_{\,-4}\times 10^{-6}$  & $2^{\,+1}_{\,-1}\times 10^{-4}$ & $6^{\,+1}_{\,-2}\times 10^{-5}$ & $4^{\,+2}_{\,-2}\times 10^{-5}$ & $12^{\,+6}_{\,-3}\times 10^{-6}$\\[0.2ex]
 4 & $16^{\,+6}_{\,-6}\times 10^{-7}$& $3^{\,+2}_{\,-1}\times 10^{-7}$  & $10^{\,+6}_{\,-3}\times 10^{-6}$& $9^{\,+1}_{\,-1}\times 10^{-6}$ & $7^{\,+9}_{\,-5}\times 10^{-7}$ & $8^{\,+2}_{\,-2}\times 10^{-6}$\\
 \hline
\end{tabular}
\renewcommand{\arraystretch}{1}
\caption{Convergence of the energy, masses and wavefunctions of the three lowest-lying states in the $B=0$ sector of $1+1$D QCD with $N_f=2$ and $m=g=L=1$, between exact results from diagonalization of the Hamiltonian and those 
obtained from D-Wave's {\tt Advantage}.}
\label{tab:QAresults2}
\end{table}

\begin{figure}[!ht]
    \centering
    \includegraphics[width=\columnwidth]{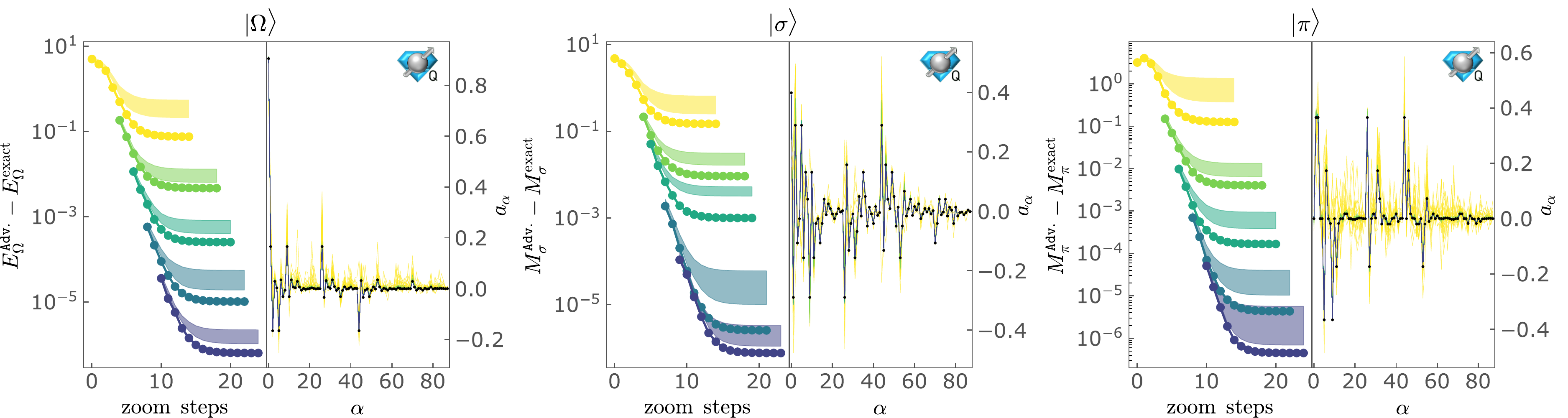}
    \caption{
    Iterative convergence of the energy, masses and wavefunctions for the three lowest-lying states in the $B=0$ sector of $1+1$D QCD with $N_f=2$ and $m=g=L=1$: vacuum (left), $\sigma$-meson (center) and $\pi$-meson (right). Compared to Fig.~\ref{fig:QAresults}, the exact wavefunctions are used when projecting the Hamiltonian to study the excited states.}
    \label{fig:QAresults_exact}
\end{figure}

Focusing on the lowest line of the last panel of Fig.~\ref{fig:QAresults}, which shows the convergence as a function of zoom steps for the pion mass, it can be seen that it displays some oscillatory behavior compared to the rest, which are smooth. This is expected, since the wavefunctions used to project out the lower eigenstates from the Hamiltonian are known with a finite precision (obtained from previous runs). For example, the vacuum state is extracted at the $10^{-6}$ precision level. Then, when looking at the excited states with increased precision (like for the pion, around $10^{-7}$), the variational principle might not hold, and the computed energy level might be below the ``true'' one (and not above). To support this argument, the same calculation has been pursued, but using the exact wavefunctions when projecting the Hamiltonian to study the excited states (instead of the ones computed using {\tt Advantage}), and no oscillatory behavior is observed, as displayed in Fig.~\ref{fig:QAresults_exact}.

\section{Quantum Circuits Required for Time Evolution by the Gauge-Field Interaction}
\label{app:circ}
\begin{figure}[!ht]
    \centering
    \includegraphics[width=8cm]{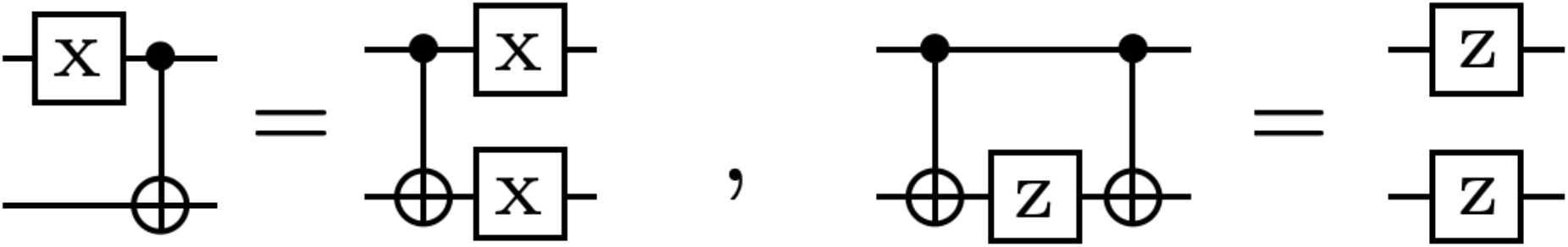}
    \caption{The $X$ and $Z$ circuit identities.}
    \label{circ:XZError}
\end{figure}

\begin{figure}
    \centering
    \includegraphics[width=17cm]{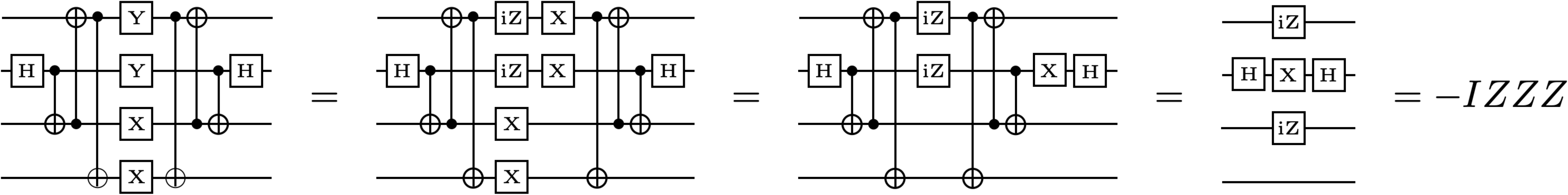}
    \caption{The diagonalization of $XXYY$ via a GHZ state-preparation circuit.}
    \label{fig:XXYYDiag}
\end{figure}

\noindent
This appendix provides more detail about the construction of the quantum circuits which implement the Trotterized time evolution of the chromo-electric terms of the Hamiltonian. 
It closely follows the presentation in the appendix of Ref.~\cite{Stetina:2020abi}.
The four-qubit interaction in $H_{el}$ has the form
\begin{equation}
\sigma^+ \sigma^- \sigma^- \sigma^+ + {\rm h.c.} = \frac{1}{8}(XXXX + XXYY + XYXY - XYYX + YXYX - YXXY +YYXX + YYYY) \ .
\label{eq:pmmpApp}
\end{equation}
Since the 8 Pauli strings are mutually commuting, they can be simultaneously diagonalized by a unitary transformation. The strategy for
identifying the quantum circuit(s) to implement this term will be to first change to a basis where every term is diagonal, then apply the diagonal unitaries and finally
return back to the computational basis.
The GHZ state-preparation circuits,
shown in Fig.~\ref{circ:GHZ}, 
diagonalize all 8 of the Pauli strings, for example,
\begin{align}
G^{\dagger} \ &( XXXX + YYXX + YXYX - YXXY - XYYX + XYXY + XXYY + YYYY) \ G \ =  \nonumber \\[4pt]
 & IIZI - ZIZZ - ZZZZ + ZIZI + IZZI - IIZZ - IZZZ + ZZZI  \ .
\label{eq:GHZDiag}
\end{align}
This can be verified by using the identities that are shown in Fig.~\ref{circ:XZError} to simplify the circuits formed by conjugating each Pauli string by $G$. 
As an example, the diagonalization of $XXYY$ is displayed in Fig.~\ref{fig:XXYYDiag}. 
The first equality uses $Y = i Z X$ and the second equality uses the $X$
circuit identity to move all $X$s past the CNOTs. The third equality moves the $Z$s past
the controls of the CNOTs and uses the $Z$ circuit identity. The other Pauli strings are diagonalized in a similar manner.

It is also straightforward to show that, for example, 
\begin{equation}
    G^{\dagger}(IZZI + IZIZ + ZIIZ)G = IZII + IIIZ + ZIII \ .
    \label{eq:ZZGHZG}
\end{equation}
In general, a $ZZ$ in the computational basis becomes a single $Z$ in the GHZ basis if the state-preparation circuit has a CNOT that connects the
original two $Z$s. The two GHZ state-preparation circuits, $G$ and $\tilde{G}$, were chosen so that all $9$ of the $ZZ$ terms in Eq.~(\ref{eq:QnfQmfp}) are mapped to single qubit rotations.
Once in the GHZ basis, the diagonal unitaries are performed, e.g., $\exp(-i IZZZ)$. 
They are arranged to minimize the number of CNOTs required, and the optimal circuit layouts are shown in Fig.~\ref{circ:UpmmpZZ}.

\section{Complete Circuits for \texorpdfstring{\boldmath$N_f=1,2$}{Nf=1,2} QCD with \texorpdfstring{\boldmath$L=1$}{L=1}}
\label{app:Nf1SU3circs}
\begin{figure}[!ht]
    \centering
    \includegraphics[width=\columnwidth]{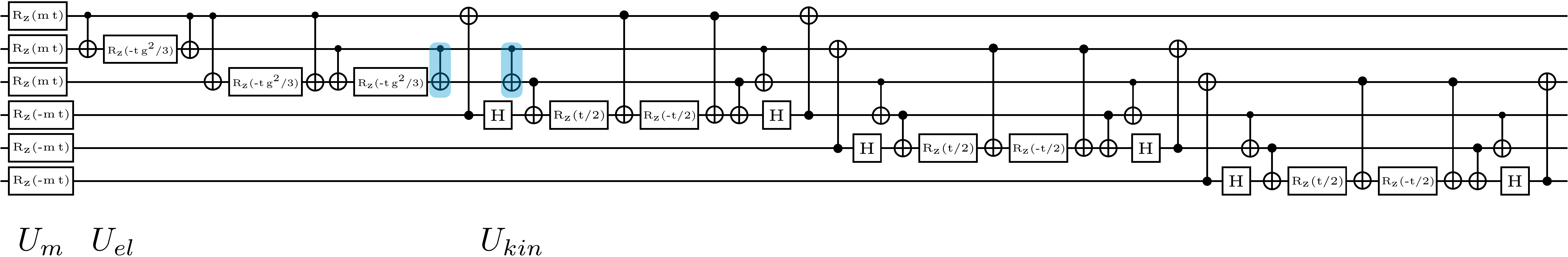}
    \caption{The complete circuit that implements a single Trotter step for $N_f=1$ QCD with $L=1$.
    }
    \label{fig:Nf1Trot}
\end{figure}

\begin{figure}[!ht]
    \centering
    \includegraphics[height=0.95\textheight]{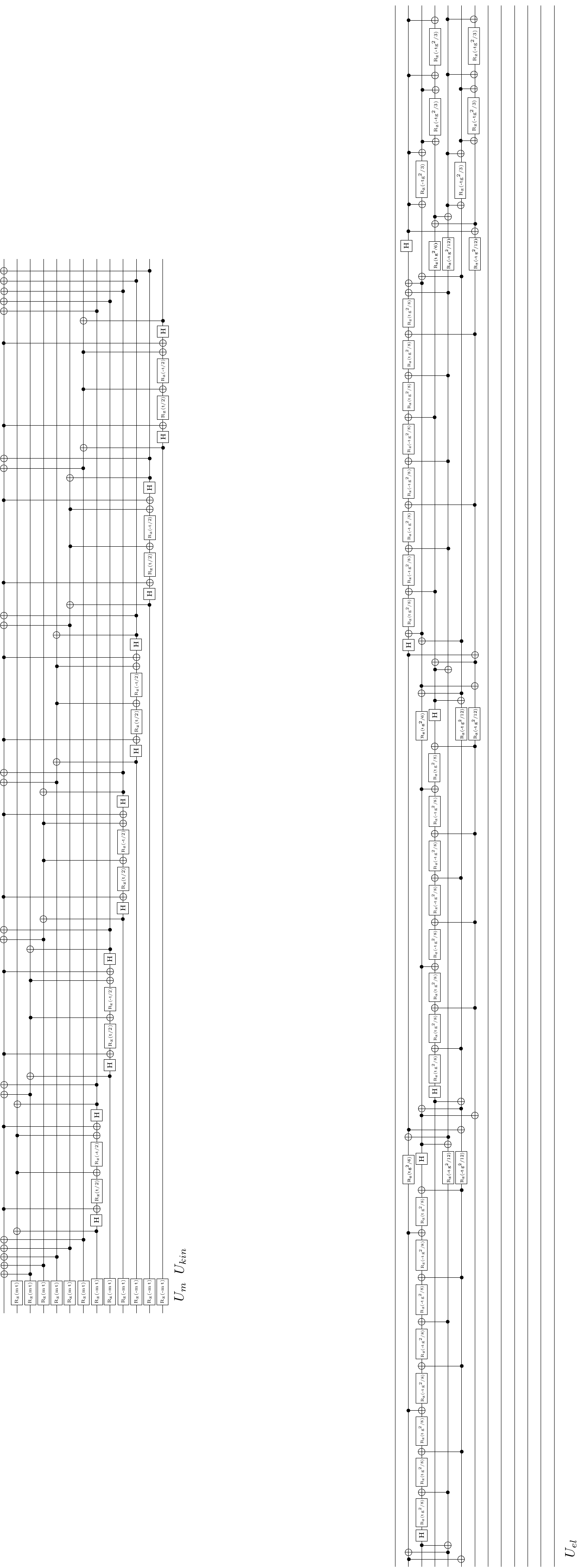}
    \caption{The complete circuit that implements a single Trotter step for $N_f=2$ QCD with $L=1$.
    }
    \label{fig:Nf2Trot}
\end{figure}

\noindent 
This appendix provides the complete set of circuits required to
implement one Trotter step for 
$N_f=1$ and $N_f=2$ QCD with $L=1$. 
The composite circuit for $N_f=1$ is shown in 
Fig.~\ref{fig:Nf1Trot} where, by ordering $U_{el}$ before $U_{kin}$, the CNOTs highlighted in blue cancel. The composite circuit for $N_f=2$ is shown in 
Fig.~\ref{fig:Nf2Trot}, 
where the ordering in the Trotterization
is $U_m$
followed by $U_{kin}$ 
and then by $U_{el}$.

%

\section{Energy Decomposition Associated with Time Evolution from the Trivial Vacuum}
\label{app:MDTD}
\begin{figure}[!ht]
    \centering
    \includegraphics[width=0.9\columnwidth]{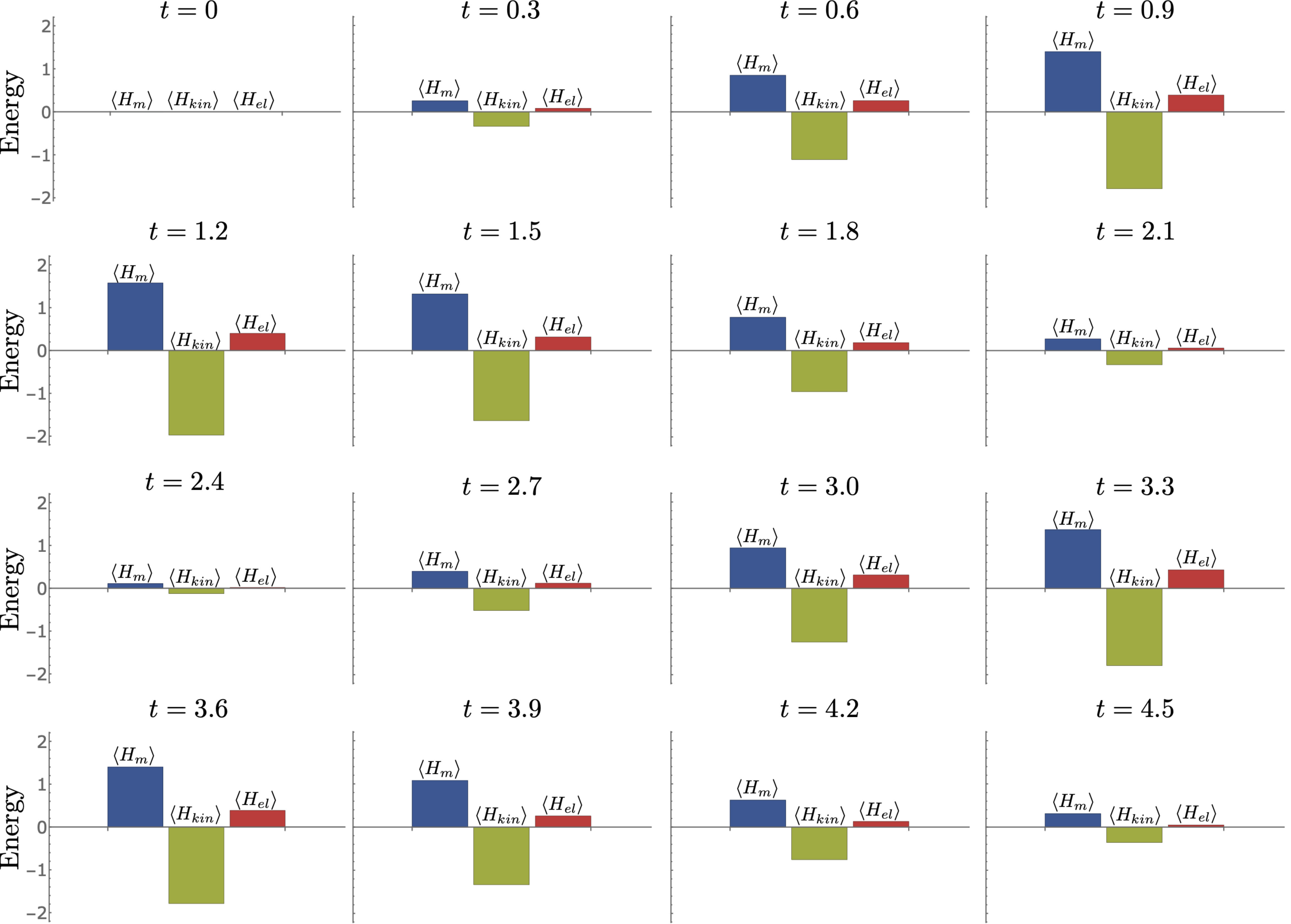}
    \caption{The time evolution of the decomposition of the energy starting from the trivial vacuum starting at $t=0$ for $N_f=2$ QCD with $m=g=L=1$.
    }
    \label{fig:Hanim}
\end{figure}

\noindent 
This appendix shows,
in Fig.~\ref{fig:Hanim}, the time evolution of the decomposition of the expectation value of the Hamiltonian starting with the trivial vacuum at $t=0$ for $N_f=2$ QCD with $m=g=L=1$. 
Notice that the sum of all three terms equals zero for all times as required by energy conservation and that the period of oscillations is the same as the period of the persistence amplitude shown in Fig.~\ref{fig:VacTo}. 
%

\section{Details on One First-Order Trotter Step of \texorpdfstring{\boldmath$N_f=1$}{Nf=1} QCD with \texorpdfstring{\boldmath$L=1$}{L=1}}
\label{app:LOTrott}
\begin{figure}[!ht]
    \centering
    \includegraphics[width=0.6\columnwidth]{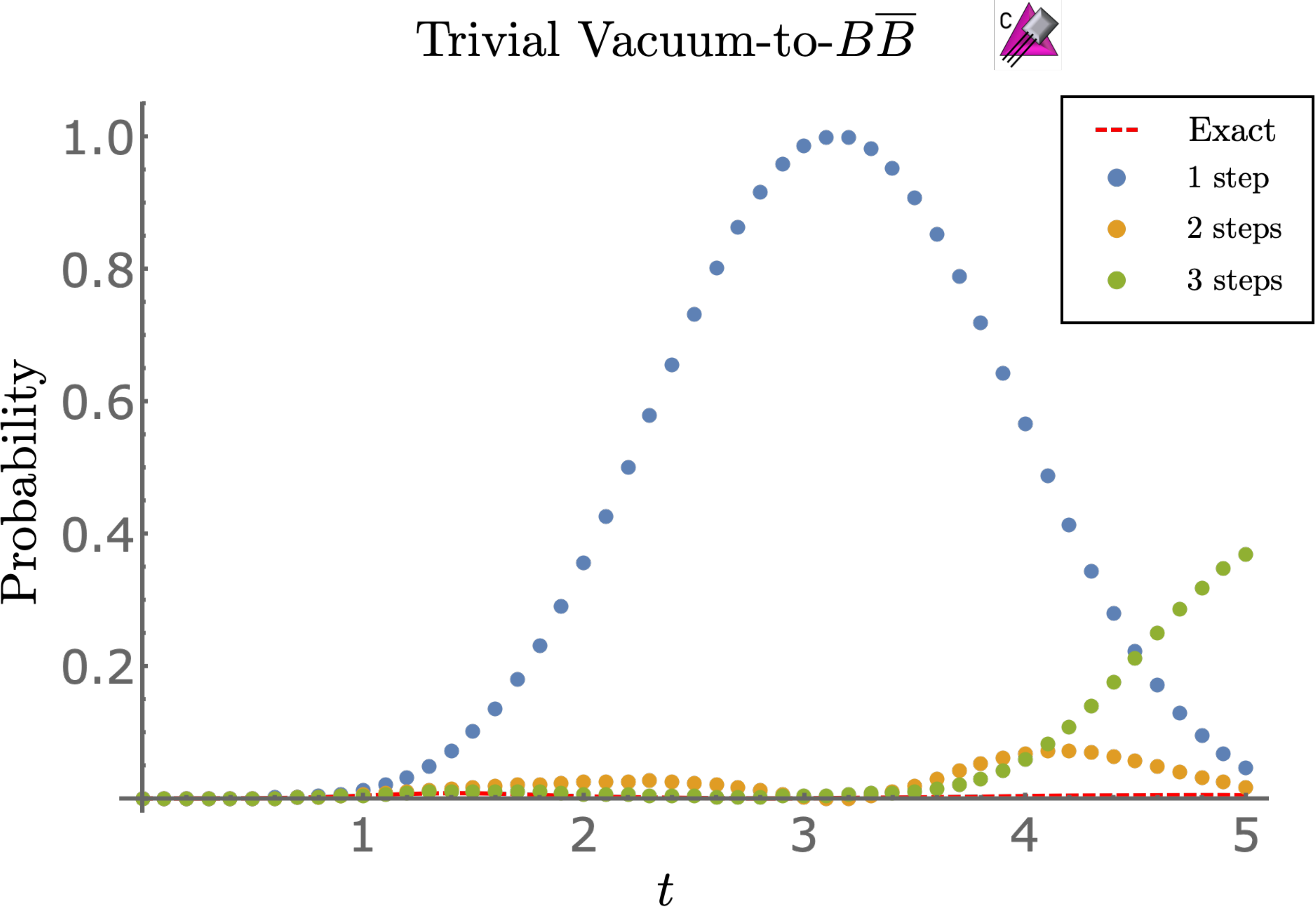}
    \caption{The trivial vacuum-to-$B\overline{B}$ probability for $1+1$D QCD with $m=g=L=1$. Shown are the results obtained from exact exponentiation of the Hamiltonian (dashed red curve) and from the Trotterized implementation with $1$, $2$ and $3$ Trotter steps.}
    \label{fig:VacToBBbar}
\end{figure}

\noindent 
This appendix discusses the theoretical expectations for one step of first-order Trotter time evolution for $N_f=1$ QCD with $L=1$.
The time evolution operator 
is decomposed
into $U_1(t) = U_{kin}(t) U_{el}(t) U_m(t)$ where the subscript ``$1$'' is to denote first-order Trotter. Both the trivial vacuum-to-vacuum and trivial vacuum-to-$q_r\overline{q}_r$ probabilities involve measurements in the computational basis where $U_m(t)$ and $U_{el}(t)$ are diagonal and have no effect. 
Thus, the time-evolution operator is effectively $U_1(t) = U_{kin}(t)$, which is exact (no Trotter errors) over a single spatial site. The trivial vacuum-to-vacuum, trivial vacuum-to-$q_r \overline{q}_r$ and trivial vacuum-to-$B \overline{B}$ probabilities are found to be,
\begin{equation}
\lvert \langle \Omega_0 \rvert e^{-i H_{kin} t} \lvert \Omega_0 \rangle\rvert ^2 = \cos^6(t/2) \ , \ \lvert\langle q_r \overline{q}_r \rvert e^{-i H_{kin} t} \lvert \Omega_0 \rangle\rvert ^2 = \cos^4(t/2)\sin^2(t/2) \ , \ \lvert\langle B \overline{B} \rvert e^{-i H_{kin} t} \lvert \Omega_0 \rangle\rvert ^2 = \sin^6(t/2) \ .
\end{equation}
For large periods of the evolution, the wavefunction is dominated by $B\overline{B}$ as shown in Fig.~\ref{fig:VacToBBbar}. Exact time evolution, on the other hand, has a  small probability of $B\overline{B}$
which suggests that detecting  
$B \overline{B}$ could lead to an additional way to mitigate Trotter errors.
It is interesting that the kinetic term alone favors transitioning the trivial vacuum into color singlets on each site. This same behavior holds
for $N_f=2$ where the dominant transition is to $\Delta \Delta \overline{\Delta} \overline{\Delta}$.

\section{Supplementary Data}
\label{app:SuppData}
\noindent 
This appendix contains the tabulated data used to produce the figures in the text. The splitting between the $\pi$- and $\sigma$-meson is given in Table~\ref{tab:MpiMsigSplitData}.
\begin{table}[!ht]
\renewcommand{\arraystretch}{1.2}
\begin{tabularx}{0.8\textwidth}{||c | Y | Y ||c | Y | Y ||}
\hline
\multicolumn{6}{||c||}{$M_{\pi} - M_{\sigma}$ } \\
 \hline
 $g$ & 
 \makecell{$L=1$} & 
 \makecell{$L=2$} &
 $g$ & 
 \makecell{$L=1$} & 
 \makecell{$L=2$} 
 \\
 \hline\hline
0 & 0.0 & 0.0 & 
2.1 & 0.1085 & 0.09954\\ \hline
 0.1 & $2.373\times 10^{-6}$ & $1.364\times 10^{-6}$ & 
 2.2 & 0.1227 & 0.1133\\ \hline
 0.2 & 0.0000374 & 0.00002335 & 
 2.3 & 0.1374 & 0.1287\\ \hline
 0.3 & 0.0001847 & 0.0001285 & 
 2.4 & 0.1482 & 0.1431\\ \hline
 0.4 & 0.0005637 & 0.0004306 & 
 2.5 & 0.1468 & 0.1475\\ \hline
 0.5 & 0.001317 & 0.001066 & 
 2.6 & 0.1321 & 0.1352\\ \hline
 0.6 & 0.002589 & 0.002163 & 
 2.7 & 0.1131 & 0.1156\\ \hline
 0.7 & 0.004508 & 0.00383 & 
 2.8 & 0.09662 & 0.09813\\ \hline
 0.8 & 0.007173 & 0.006158 & 
 2.9 & 0.08372 & 0.08456\\ \hline
 0.9 & 0.01064 & 0.009207 & 
 3.0 & 0.07375 & 0.07422\\ \hline
 1.0 & 0.01493 & 0.01301 & 
 3.1 & 0.06592 & 0.06617\\ \hline
 1.1 & 0.02002 & 0.01755 & 
 3.2 & 0.05961 & 0.05975\\ \hline
 1.2 & 0.02586 & 0.02281 & 
 3.3 & 0.05442 & 0.05448\\ \hline
 1.3 & 0.0324 & 0.02874 & 
 3.4 & 0.05005 & 0.05008\\ \hline
 1.4 & 0.03956 & 0.03529 & 
 3.5 & 0.04631 & 0.04632\\ \hline
 1.5 & 0.04731 & 0.04243 & 
 3.6 & 0.04307 & 0.04306\\ \hline
 1.6 & 0.05562 & 0.05011 & 
 3.7 & 0.04022 & 0.04021\\ \hline
 1.7 & 0.0645 & 0.05837 & 
 3.8 & 0.0377 & 0.03769\\ \hline
 1.8 & 0.07405 & 0.06725 & 
 3.9 & 0.03545 & 0.03543\\ \hline
 1.9 & 0.0844 & 0.0769 & 
 4.0 & 0.03342 & 0.0334\\ \hline
 2.0 & 0.0958 & 0.08755 & & & \\ 
 \hline
\end{tabularx}
\renewcommand{\arraystretch}{1}
\caption{The mass splitting between the $\sigma$- and $\pi$-mesons for $m=1$ and $L=1,2$.
}
\label{tab:MpiMsigSplitData}
\end{table}
The decomposition of the vacuum energy, hadronic masses and deuteron binding energy is given in Table~\ref{tab:SpecBreakdownData}.
\begin{table}[!ht]
\renewcommand{\arraystretch}{1.2}
\begin{tabularx}{0.8\textwidth}{||c | Y | Y | Y | Y | Y ||}
\hline
\multicolumn{6}{||c||}{Decomposition of the vacuum energy, hadronic masses and deuteron binding energy} \\
 \hline
 & 
 \makecell{$E_{\Omega}$} & 
 \makecell{$M_{\sigma}$} & 
 \makecell{$M_{\pi}$} &
 \makecell{$M_{\Delta}$} & 
 \makecell{$B_{\Delta \Delta}$}
 \\
 \hline\hline
 $\langle H_m \rangle$ & 1.0566 & 2.056 & 2.032 & 2.855 & -0.001596 \\
 \hline
 $\langle H_{kin} \rangle$ & -2.975 & 0.1271 & 0.1425 & 0.4182 & 0.002399\\
 \hline
 $\langle H_{el} \rangle$ & 0.3374 & 0.5401 & 0.5609 & -0.03099 & -0.0003429\\
 \hline
\end{tabularx}
\renewcommand{\arraystretch}{1}
\caption{The decomposition of vacuum energy ($E_{\Omega}$), the masses of the lightest hadrons ($M_{\sigma}$, $M_{\pi}$ and $M_{\Delta}$) and the deuteron binding energy ($B_{\Delta \Delta}$) into contributions from the mass, kinetic and chromo-electric field terms in the Hamiltonian for $L=2$ and $m=g=1$.
}
\label{tab:SpecBreakdownData}
\end{table}
The binding energy of the deuteron is given in Table~\ref{tab:deutBE}.
\begin{table}[!ht]
\renewcommand{\arraystretch}{1.2}
\begin{tabularx}{0.4\textwidth}{||c | Y || c | Y ||}
 \hline
 $g$ & 
 \makecell{$B_{\Delta \Delta}$} &
 $g$ & 
 \makecell{$B_{\Delta \Delta}$}
 \\
 \hline\hline
0 & 0.0  & 1.6 & 0.0005388 \\ \hline
 0.1 & 0.00005099 & 1.7 & 0.000541 \\ \hline
 0.2 & 0.0006768 & 1.8 & 0.0005332\\ \hline
 0.3 & 0.002351 & 1.9 & 0.0005172\\ \hline
 0.4 & 0.003947 & 2.0 & 0.0004948 \\ \hline
 0.5 & 0.003905 & 2.1 & 0.0004677\\ \hline
 0.6 & 0.002716 & 2.2 & 0.0004378\\ \hline
 0.7 & 0.001592 & 2.3 & 0.0004063\\ \hline
 0.8 & 0.0009178 & 2.4 & 0.0003745\\ \hline
 0.9 & 0.0005902 & 2.5 & 0.0003432\\ \hline
 1.0 & 0.0004599 & 2.6 & 0.0003129\\ \hline
 1.1 & 0.000429 & 2.7 & 0.0002842\\ \hline
 1.2 & 0.000443 & 2.8 & 0.0002574\\ \hline
 1.3 & 0.0004727 & 2.9 & 0.0002324\\ \hline
 1.4 & 0.0005029 & 3.0 & 0.0002095\\ \hline
 1.5 & 0.000526 & & \\
 \hline
\end{tabularx}
\renewcommand{\arraystretch}{1}
\caption{The binding energy of the deuteron, $B_{\Delta \Delta}$, for $m=1$ and $L=2$.
}
\label{tab:deutBE}
\end{table}

The linear entropy between quarks and antiquarks in the vacuum, the $\sigma$- and $\pi$-meson and the $\Delta$ are given in Table~\ref{tab:linentData}.
\begin{table}[!ht]
\renewcommand{\arraystretch}{0.93}
\begin{tabularx}{0.6\textwidth}{||c | Y  | Y  | Y  | Y ||}
\hline
\multicolumn{5}{||c||}{The linear entropy between quarks and antiquarks} \\
 \hline
 $g$ & 
 \makecell{$\ket{\Omega}$}
 & 
 \makecell{$\ket{\sigma}$}
 & 
 \makecell{$\ket{\pi_{I_3=1}}$}
 & 
 \makecell{$\ket{\Delta_{I_3=3/2}}$}
 \\
 \hline\hline
0.1 & 0.4668 & 0.9115 & 0.7786 & 0.2698 \\ \hline
 0.2 & 0.4617 & 0.9124 & 0.7786 & 0.2663 \\ \hline
 0.3 & 0.4532 & 0.9137 & 0.7786 & 0.2605 \\ \hline
 0.4 & 0.4416 & 0.9154 & 0.7786 & 0.2527 \\ \hline
 0.5 & 0.4271 & 0.9174 & 0.7787 & 0.243 \\ \hline
 0.6 & 0.41 & 0.9194 & 0.7787 & 0.2318 \\ \hline
 0.7 & 0.3908 & 0.9214 & 0.7789 & 0.2194 \\ \hline
 0.8 & 0.3699 & 0.9232 & 0.7791 & 0.206 \\ \hline
 0.9 & 0.3478 & 0.9248 & 0.7794 & 0.1921 \\ \hline
 1.0 & 0.3248 & 0.926 & 0.7798 & 0.1779 \\ \hline
 1.1 & 0.3015 & 0.9269 & 0.7803 & 0.1638 \\ \hline
 1.2 & 0.2784 & 0.9274 & 0.7811 & 0.15 \\ \hline
 1.3 & 0.2557 & 0.9277 & 0.7821 & 0.1367 \\ \hline
 1.4 & 0.2339 & 0.9277 & 0.7834 & 0.1241 \\ \hline
 1.5 & 0.213 & 0.9277 & 0.7851 & 0.1123 \\ \hline
 1.6 & 0.1935 & 0.9278 & 0.7875 & 0.1014 \\ \hline
 1.7 & 0.1752 & 0.928 & 0.7906 & 0.0913 \\ \hline
 1.8 & 0.1584 & 0.9287 & 0.7949 & 0.0821 \\ \hline
 1.9 & 0.1429 & 0.93 & 0.801 & 0.07375 \\ \hline
 2.0 & 0.1288 & 0.9322 & 0.8097 & 0.06622 \\ \hline
 2.1 & 0.116 & 0.9355 & 0.8225 & 0.05944 \\ \hline
 2.2 & 0.1045 & 0.9398 & 0.8416 & 0.05337 \\ \hline
 2.3 & 0.09411 & 0.9434 & 0.8675 & 0.04794 \\ \hline
 2.4 & 0.08479 & 0.9402 & 0.8889 & 0.04309 \\ \hline
 2.5 & 0.07643 & 0.9212 & 0.8731 & 0.03876 \\ \hline
 2.6 & 0.06896 & 0.8872 & 0.8209 & 0.03491 \\ \hline
 2.7 & 0.06227 & 0.8526 & 0.7729 & 0.03148 \\ \hline
 2.8 & 0.05629 & 0.8263 & 0.7414 & 0.02842 \\ \hline
 2.9 & 0.05095 & 0.8081 & 0.7216 & 0.02569 \\ \hline
 3.0 & 0.04617 & 0.7956 & 0.7086 & 0.02326 \\ \hline
 3.1 & 0.0419 & 0.7867 & 0.6994 & 0.02108 \\ \hline
 3.2 & 0.03807 & 0.7803 & 0.6928 & 0.01914 \\ \hline
 3.3 & 0.03464 & 0.7754 & 0.6877 & 0.01741 \\ \hline
 3.4 & 0.03157 & 0.7716 & 0.6837 & 0.01586 \\ \hline
 3.5 & 0.02881 & 0.7686 & 0.6805 & 0.01446 \\ \hline
 3.6 & 0.02633 & 0.7662 & 0.6779 & 0.01321 \\ \hline
 3.7 & 0.02409 & 0.7642 & 0.6757 & 0.01209 \\ \hline
 3.8 & 0.02208 & 0.7625 & 0.6739 & 0.01107 \\ \hline
 3.9 & 0.02027 & 0.7611 & 0.6723 & 0.01016 \\ \hline
 4.0 & 0.01863 & 0.7599 & 0.671 & 0.009334 \\ \hline
 4.1 & 0.01714 & 0.7588 & 0.6699 & 0.008589 \\ \hline
 4.2 & 0.0158 & 0.7579 & 0.6689 & 0.007913 \\ \hline
 4.3 & 0.01458 & 0.7572 & 0.668 & 0.007301 \\ \hline
 4.4 & 0.01347 & 0.7565 & 0.6673 & 0.006745 \\ \hline
 4.5 & 0.01246 & 0.7559 & 0.6666 & 0.006239 \\ \hline
 4.6 & 0.01154 & 0.7553 & 0.666 & 0.005779 \\ \hline
 4.7 & 0.01071 & 0.7549 & 0.6655 & 0.005359 \\ \hline
 4.8 & 0.00994 & 0.7545 & 0.665 & 0.004975 \\ \hline
 4.9 & 0.00924 & 0.7541 & 0.6646 & 0.004625 \\ \hline
 5.0 & 0.0086 & 0.7538 & 0.6642 & 0.004304 \\ \hline
 5.1 & 0.008012 & 0.7535 & 0.6639 & 0.004009 \\ \hline
 5.2 & 0.007473 & 0.7532 & 0.6636 & 0.003739 \\ \hline
 5.3 & 0.006978 & 0.753 & 0.6633 & 0.003491 \\ \hline
 5.4 & 0.006522 & 0.7527 & 0.6631 & 0.003263 \\ \hline
 5.5 & 0.006101 & 0.7525 & 0.6629 & 0.003052 \\ \hline
 5.6 & 0.005714 & 0.7524 & 0.6627 & 0.002858 \\ \hline
 5.7 & 0.005356 & 0.7522 & 0.6625 & 0.002679 \\ \hline
 5.8 & 0.005026 & 0.752 & 0.6623 & 0.002514 \\ \hline
 5.9 & 0.00472 & 0.7519 & 0.6621 & 0.002361 \\ \hline
 6.0 & 0.004436 & 0.7518 & 0.662 & 0.002219 \\
 \hline
\end{tabularx}
\renewcommand{\arraystretch}{1}
\caption{The linear entropy between quarks and antiquarks in the vacuum, $\ket{\Delta_{I_3=3/2}}$, $\ket{\sigma}$ and $\ket{\pi_{I_3=1}}$ for $m=L=1$.}
\label{tab:linentData}
\end{table}
The quark occupation (total number of quarks plus antiquarks) in the $\sigma$- and $\pi$-mesons is given in Table~\ref{tab:qoccData}.
\begin{table}[!ht]
\renewcommand{\arraystretch}{1.2}
\begin{tabularx}{0.6\textwidth}{||c | Y  | Y  || c | Y  | Y ||}
\hline
\multicolumn{6}{||c||}{The quark occupation} \\
 \hline
 $g$ 
 & 
 \makecell{$\ket{\sigma}$}
 & 
 \makecell{$\ket{\pi_{I_3 = 1}}$}
 &
 $g$ 
 & 
 \makecell{$\ket{\sigma}$}
 & 
 \makecell{$\ket{\pi_{I_3 = 1}}$}
 \\
 \hline\hline
\ \ 0 \ \ & 2.422 & 2.422 & 
\ 3.1\  & 5.884 & 5.915\\ \hline
 0.1 & 2.422 & 2.422 & 
 3.2 & 5.913 & 5.936\\ \hline
 0.2 & 2.422 & 2.422 & 
 3.3 & 5.934 & 5.95\\ \hline
 0.3 & 2.423 & 2.422 & 
 3.4 & 5.948 & 5.961\\ \hline
 0.4 & 2.424 & 2.423 & 
 3.5 & 5.959 & 5.968\\ \hline
 0.5 & 2.426 & 2.423 & 
 3.6 & 5.967 & 5.974\\ \hline
 0.6 & 2.429 & 2.423 & 
 3.7 & 5.973 & 5.979\\ \hline
 0.7 & 2.434 & 2.424 & 
 3.8 & 5.978 & 5.982\\ \hline
 0.8 & 2.44 & 2.425 & 
 3.9 & 5.981 & 5.985\\ \hline
 0.9 & 2.449 & 2.427 & 
 4.0 & 5.984 & 5.988\\ \hline
 1.0 & 2.46 & 2.43 & 
 4.1 & 5.987 & 5.989\\ \hline
 1.1 & 2.473 & 2.434 & 
 4.2 & 5.989 & 5.991\\ \hline
 1.2 & 2.489 & 2.439 & 
 4.3 & 5.99 & 5.992\\ \hline
 1.3 & 2.507 & 2.445 & 
 4.4 & 5.992 & 5.993\\ \hline
 1.4 & 2.529 & 2.454 & 
 4.5 & 5.993 & 5.994\\ \hline
 1.5 & 2.555 & 2.466 & 
 4.6 & 5.994 & 5.995\\ \hline
 1.6 & 2.586 & 2.483 & 
 4.7 & 5.995 & 5.996\\ \hline
 1.7 & 2.626 & 2.505 & 
 4.8 & 5.995 & 5.996\\ \hline
 1.8 & 2.676 & 2.537 & 
 4.9 & 5.996 & 5.997\\ \hline
 1.9 & 2.744 & 2.584 & 
 5.0 & 5.996 & 5.997\\ \hline
 2.0 & 2.839 & 2.655 & 
 5.1 & 5.997 & 5.997\\ \hline
 2.1 & 2.979 & 2.769 & 
 5.2 & 5.997 & 5.998\\ \hline
 2.2 & 3.193 & 2.966 & 
 5.3 & 5.997 & 5.998\\ \hline
 2.3 & 3.524 & 3.318 & 
 5.4 & 5.998 & 5.998\\ \hline
 2.4 & 4.004 & 3.911 & 
 5.5 & 5.998 & 5.998\\ \hline
 2.5 & 4.579 & 4.66 & 
 5.6 & 5.998 & 5.999\\ \hline
 2.6 & 5.091 & 5.249 & 
 5.7 & 5.998 & 5.999\\ \hline
 2.7 & 5.439 & 5.577 & 
 5.8 & 5.999 & 5.999\\ \hline
 2.8 & 5.646 & 5.743 & 
 5.9 & 5.999 & 5.999\\ \hline
 2.9 & 5.766 & 5.832 & 
 6.0 & 5.999 & 5.999\\ \hline
 3.0 & 5.838 & 5.883 & & & \\ 
 \hline
\end{tabularx}
\renewcommand{\arraystretch}{1}
\caption{The expectation value of quark occupation in the $\ket{\sigma}$ and $\ket{\pi_{I_3 = 1}}$ for $m=L=1$.}
\label{tab:qoccData}
\end{table}

The trivial vacuum-to-vacuum probabilities, as obtained by {\tt circ} and {\tt qiskit}, are given in Table~\ref{tab:vactovacData}. 
\begin{table}[!ht]
\renewcommand{\arraystretch}{1.06}
\begin{tabularx}{0.6\textwidth}{||c | Y  | Y  | Y  | Y | Y ||}
\hline
\multicolumn{6}{||c||}{The trivial vacuum-to-vacuum probabilities} \\
 \hline
 $t$ & 
 \makecell{\text{1 Step}}
 & 
 \makecell{\text{2 Steps}}
 & 
 \makecell{\text{3 Steps}}
 & 
 \makecell{\text{5 Steps}}
 & 
 \makecell{\text{10 Steps}}
 \\
 \hline\hline
\ \ 0 \ \  & 1 & 1 & 1 & 1 & 1 \\ \hline
 0.1 & 0.9851 & 0.9852 & 0.9852 & 0.9852 & 0.9852
   \\ \hline
 0.2 & 0.9417 & 0.9427 & 0.9429 & 0.943 & 0.943 \\ \hline
 0.3 & 0.8733 & 0.878 & 0.8789 & 0.8793 & 0.8795 \\ \hline
 0.4 & 0.7854 & 0.799 & 0.8014 & 0.8027 & 0.8032 \\ \hline
 0.5 & 0.6846 & 0.7141 & 0.7193 & 0.7219 & 0.723 \\ \hline
 0.6 & 0.5779 & 0.6309 & 0.6401 & 0.6447 & 0.6466
   \\ \hline
 0.7 & 0.4721 & 0.5554 & 0.5697 & 0.5769 & 0.5799
   \\ \hline
 0.8 & 0.3728 & 0.4914 & 0.5118 & 0.522 & 0.5263 \\ \hline
 0.9 & 0.2841 & 0.4405 & 0.4678 & 0.4815 & 0.4872
   \\ \hline
 1.0 & 0.2087 & 0.4032 & 0.438 & 0.4556 & 0.4629 \\ \hline
 1.1 & 0.1474 & 0.3787 & 0.422 & 0.4438 & 0.4529 \\ \hline
 1.2 & 0.0999 & 0.3658 & 0.4188 & 0.4454 & 0.4565
   \\ \hline
 1.3 & 0.0648 & 0.3634 & 0.4275 & 0.4596 & 0.4729
   \\ \hline
 1.4 & 0.0401 & 0.3699 & 0.4473 & 0.4855 & 0.5012
   \\ \hline
 1.5 & 0.0235 & 0.3843 & 0.4775 & 0.5225 & 0.5408
   \\ \hline
 1.6 & 0.0131 & 0.4053 & 0.5174 & 0.5696 & 0.5904
   \\ \hline
 1.7 & 0.0068 & 0.4316 & 0.566 & 0.6252 & 0.6483 \\ \hline
 1.8 & 0.0033 & 0.4616 & 0.6218 & 0.6873 & 0.712 \\ \hline
 1.9 & 0.0015 & 0.4935 & 0.6827 & 0.7529 & 0.7782
   \\ \hline
 2.0 & 0.0006 & 0.5253 & 0.7458 & 0.8179 & 0.8423 \\ \hline
 2.1 & 0.0002 & 0.5546 & 0.8075 & 0.8775 & 0.8992
   \\ \hline
 2.2 & 0.0001 & 0.579 & 0.8634 & 0.9262 & 0.9433 \\ \hline
 2.3 & 0.0000 & 0.5965 & 0.9084 & 0.9583 & 0.9691 \\ \hline
 2.4 & 0.0000 & 0.6055 & 0.9374 & 0.969 & 0.9727 \\ \hline
 2.5 & 0.0000 & 0.6055 & 0.9453 & 0.9549 & 0.9518 \\ \hline
 2.6 & 0.0000 & 0.5965 & 0.9281 & 0.915 & 0.907 \\ \hline
 2.7 & 0.0000 & 0.5798 & 0.8839 & 0.8513 & 0.8418 \\ \hline
 2.8 & 0.0000 & 0.5571 & 0.8129 & 0.7688 & 0.7624 \\ \hline
 2.9 & 0.0000 & 0.5305 & 0.7187 & 0.6748 & 0.6766 \\ \hline
 3.0 & 0.0000 & 0.5026 & 0.6077 & 0.5778 & 0.5925 \\ \hline
 3.1 & 0.0000 & 0.4756 & 0.4886 & 0.4865 & 0.5176 \\ \hline
 3.2 & 0.0000 & 0.4518 & 0.3708 & 0.408 & 0.4581 \\ \hline
 3.3 & 0.0000 & 0.4329 & 0.2629 & 0.3477 & 0.4179 \\ \hline
 3.4 & 0.0000 & 0.4204 & 0.1713 & 0.3087 & 0.3991 \\ \hline
 3.5 & 0.0000 & 0.4152 & 0.0997 & 0.2921 & 0.4017 \\ \hline
 3.6 & 0.0000 & 0.4179 & 0.0488 & 0.2967 & 0.424 \\ \hline
 3.7 & 0.0000 & 0.4283 & 0.0175 & 0.3196 & 0.4626 \\ \hline
 3.8 & 0.0000 & 0.4457 & 0.0035 & 0.3561 & 0.5133 \\ \hline
 3.9 & 0.0000 & 0.4689 & 0.0039 & 0.4009 & 0.5714 \\ \hline
 4.0 & 0.0000 & 0.4964 & 0.0158 & 0.4485 & 0.6323 \\ \hline
 4.1 & 0.0001 & 0.5265 & 0.0363 & 0.4947 & 0.6925
   \\ \hline
 4.2 & 0.0003 & 0.5578 & 0.062 & 0.5373 & 0.7496 \\ \hline
 4.3 & 0.0007 & 0.5891 & 0.0893 & 0.576 & 0.8015 \\ \hline
 4.4 & 0.0017 & 0.6198 & 0.1144 & 0.6125 & 0.8467
   \\ \hline
 4.5 & 0.0038 & 0.6498 & 0.1339 & 0.6489 & 0.8833
   \\ \hline
 4.6 & 0.0077 & 0.6794 & 0.1454 & 0.687 & 0.909 \\ \hline
 4.7 & 0.0145 & 0.709 & 0.148 & 0.7269 & 0.9213 \\ \hline
 4.8 & 0.0258 & 0.7389 & 0.1422 & 0.7671 & 0.9177
   \\ \hline
 4.9 & 0.0436 & 0.7689 & 0.1295 & 0.8042 & 0.8969
   \\ \hline
 5.0 & 0.0699 & 0.7986 & 0.1125 & 0.8339 & 0.8586 \\
 \hline
\end{tabularx}
\renewcommand{\arraystretch}{1}
\caption{The trivial vacuum-to-vacuum probabilities for $N_f=2$ and $m=g=L=1$. Results are shown for 1, 2, 3, 5 and 10 Trotter steps.}
\label{tab:vactovacData}
\end{table}
The trivial vacuum-to-$d_r \overline{d}_r$ probabilities, as obtained by {\tt circ} and {\tt qiskit}, are given in Table~\ref{tab:vactodrdrbData}. 
\begin{table}[!ht]
\renewcommand{\arraystretch}{1.06}
\begin{tabularx}{0.6\textwidth}{||c | Y  | Y  | Y  | Y | Y ||}
\hline
\multicolumn{6}{||c||}{The trivial vacuum-to-$d_r \overline{d}_r$ probabilities} \\
 \hline
 $t$ & 
 \makecell{\text{1 Step}}
 & 
 \makecell{\text{2 Steps}}
 & 
 \makecell{\text{3 Steps}}
 & 
 \makecell{\text{5 Steps}}
 & 
 \makecell{\text{10 Steps}}
 \\
 \hline\hline
\ \ 0\ \  & 0 & 0 & 0 & 0 & 0 \\ \hline
 0.1 & 0.0025 & 0.0025 & 0.0025 & 0.0025 & 0.0025
   \\ \hline
 0.2 & 0.0095 & 0.0093 & 0.0093 & 0.0093 & 0.0093
   \\ \hline
 0.3 & 0.0199 & 0.0192 & 0.0191 & 0.019 & 0.019 \\ \hline
 0.4 & 0.0323 & 0.0304 & 0.0301 & 0.0299 & 0.0298
   \\ \hline
 0.5 & 0.0446 & 0.0411 & 0.0405 & 0.0401 & 0.04 \\ \hline
 0.6 & 0.0553 & 0.05 & 0.049 & 0.0486 & 0.0483 \\ \hline
 0.7 & 0.0629 & 0.0565 & 0.0553 & 0.0546 & 0.0543
   \\ \hline
 0.8 & 0.0666 & 0.0605 & 0.0591 & 0.0583 & 0.058 \\ \hline
 0.9 & 0.0663 & 0.0625 & 0.061 & 0.0601 & 0.0597 \\ \hline
 1.0 & 0.0623 & 0.0629 & 0.0614 & 0.0605 & 0.0601 \\ \hline
 1.1 & 0.0554 & 0.0623 & 0.0608 & 0.0598 & 0.0594
   \\ \hline
 1.2 & 0.0468 & 0.0612 & 0.0597 & 0.0586 & 0.058 \\ \hline
 1.3 & 0.0374 & 0.0599 & 0.0581 & 0.0567 & 0.0561
   \\ \hline
 1.4 & 0.0284 & 0.0585 & 0.0561 & 0.0543 & 0.0535
   \\ \hline
 1.5 & 0.0204 & 0.0569 & 0.0537 & 0.0512 & 0.0501
   \\ \hline
 1.6 & 0.0139 & 0.0552 & 0.0505 & 0.0472 & 0.0458
   \\ \hline
 1.7 & 0.0088 & 0.0531 & 0.0464 & 0.0422 & 0.0403
   \\ \hline
 1.8 & 0.0053 & 0.0505 & 0.0414 & 0.0361 & 0.0339
   \\ \hline
 1.9 & 0.0029 & 0.0473 & 0.0353 & 0.029 & 0.0265 \\ \hline
 2.0 & 0.0015 & 0.0434 & 0.0283 & 0.0213 & 0.0188 \\ \hline
 2.1 & 0.0007 & 0.039 & 0.0208 & 0.0137 & 0.0113 \\ \hline
 2.2 & 0.0003 & 0.0341 & 0.0134 & 0.007 & 0.0051 \\ \hline
 2.3 & 0.0001 & 0.0291 & 0.0069 & 0.0022 & 0.0012
   \\ \hline
 2.4 & 0.0000 & 0.0239 & 0.0024 & 0.0005 & 0.0006 \\ \hline
 2.5 & 0.0000 & 0.0189 & 0.0008 & 0.0025 & 0.0036 \\ \hline
 2.6 & 0.0000 & 0.0142 & 0.0031 & 0.0085 & 0.0105 \\ \hline
 2.7 & 0.0000 & 0.0099 & 0.0098 & 0.0183 & 0.0206 \\ \hline
 2.8 & 0.0000 & 0.0061 & 0.0204 & 0.0308 & 0.0326 \\ \hline
 2.9 & 0.0000 & 0.0031 & 0.0339 & 0.0444 & 0.0452 \\ \hline
 3.0 & 0.0000 & 0.0011 & 0.0482 & 0.0574 & 0.0568 \\ \hline
 3.1 & 0.0000 & 0.0001 & 0.061 & 0.0686 & 0.0663 \\ \hline
 3.2 & 0.0000 & 0.0002 & 0.0701 & 0.0768 & 0.073 \\ \hline
 3.3 & 0.0000 & 0.0012 & 0.0741 & 0.0818 & 0.0766 \\ \hline
 3.4 & 0.0000 & 0.003 & 0.0728 & 0.0834 & 0.0769 \\ \hline
 3.5 & 0.0000 & 0.0052 & 0.0673 & 0.082 & 0.0741 \\ \hline
 3.6 & 0.0000 & 0.0074 & 0.0594 & 0.0777 & 0.0687 \\ \hline
 3.7 & 0.0000 & 0.0094 & 0.0508 & 0.0712 & 0.0613 \\ \hline
 3.8 & 0.0000 & 0.0108 & 0.0429 & 0.0632 & 0.0527 \\ \hline
 3.9 & 0.0000 & 0.0117 & 0.0366 & 0.0546 & 0.0439 \\ \hline
 4.0 & 0.0001 & 0.0121 & 0.0315 & 0.0467 & 0.0356 \\ \hline
 4.1 & 0.0003 & 0.0122 & 0.0275 & 0.0401 & 0.0283
   \\ \hline
 4.2 & 0.0008 & 0.0121 & 0.024 & 0.0351 & 0.0221 \\ \hline
 4.3 & 0.0017 & 0.0121 & 0.0209 & 0.0316 & 0.0167
   \\ \hline
 4.4 & 0.0033 & 0.0123 & 0.0183 & 0.0289 & 0.012 \\ \hline
 4.5 & 0.0058 & 0.0126 & 0.0166 & 0.0262 & 0.008 \\ \hline
 4.6 & 0.0096 & 0.013 & 0.0162 & 0.0229 & 0.0049 \\ \hline
 4.7 & 0.0149 & 0.0132 & 0.0169 & 0.0188 & 0.003 \\ \hline
 4.8 & 0.0217 & 0.0131 & 0.0185 & 0.0142 & 0.003 \\ \hline
 4.9 & 0.0299 & 0.0127 & 0.0207 & 0.0097 & 0.0052
   \\ \hline
 5.0 & 0.039 & 0.0119 & 0.0228 & 0.0062 & 0.0099 \\
 \hline
\end{tabularx}
\renewcommand{\arraystretch}{1}
\caption{The trivial vacuum-to-$d_r \overline{d}_r$ probabilities for $N_f=2$ and $m=g=L=1$. Results are shown for 1, 2, 3, 5 and 10 Trotter steps.}
\label{tab:vactodrdrbData}
\end{table}
The required $N$\textsubscript{Trott} for a $\epsilon_{{\rm Trott}} < 0.1$ in the trivial vacuum-to-vacuum probability is given in Table~\ref{tab:vacNTrottData}.
\begin{table}[!ht]
\renewcommand{\arraystretch}{1.06}
\begin{tabularx}{\textwidth}{||c | Y  | Y  | Y  | Y | Y | Y  | Y  | Y  | Y | Y| Y  | Y  | Y  | Y | Y| Y  | Y  | Y  | Y  ||}
\hline
\multicolumn{16}{||c||}{Required number of Trotter steps, $N$\textsubscript{Trott}} \\
 \hline
 $t$ & $N$\textsubscript{Trott} &
 $t$ & $N$\textsubscript{Trott} &
 $t$ & $N$\textsubscript{Trott} &
 $t$ & $N$\textsubscript{Trott} &
 $t$ & $N$\textsubscript{Trott} &
 $t$ & $N$\textsubscript{Trott} &
 $t$ & $N$\textsubscript{Trott} &
 $t$ & $N$\textsubscript{Trott}
 \\
 \hline\hline
0 & 1 & 25 & 95 & 50 & 271 & 75 & 521 & 100 & 740 & 125 & 1080 & 150 & 1417 & 175 & 1967 \\ \hline
 0.5 & 1 & 25.5 & 95 & 50.5 & 271 & 75.5 & 521 & 100.5 & 740 & 125.5 & 1080 & 150.5 & 1598 & 175.5 & 1967 \\ \hline
 1 & 3 & 26 & 103 & 51 & 271 & 76 & 521 & 101 & 740 & 126 & 1113 & 151 & 1598 & 176 & 1967 \\ \hline
 1.5 & 4 & 26.5 & 103 & 51.5 & 302 & 76.5 & 521 & 101.5 & 740 & 126.5 & 1113 & 151.5 & 1598 & 176.5 & 1967 \\ \hline
 2 & 4 & 27 & 114 & 52 & 302 & 77 & 521 & 102 & 794 & 127 & 1113 & 152 & 1598 & 177 & 1967 \\ \hline
 2.5 & 4 & 27.5 & 114 & 52.5 & 302 & 77.5 & 521 & 102.5 & 794 & 127.5 & 1224 & 152.5 & 1598 & 177.5 & 1967 \\ \hline
 3 & 4 & 28 & 114 & 53 & 302 & 78 & 521 & 103 & 868 & 128 & 1224 & 153 & 1598 & 178 & 1967 \\ \hline
 3.5 & 10 & 28.5 & 121 & 53.5 & 302 & 78.5 & 521 & 103.5 & 868 & 128.5 & 1224 & 153.5 & 1598 & 178.5 & 1967 \\ \hline
 4 & 10 & 29 & 130 & 54 & 302 & 79 & 535 & 104 & 868 & 129 & 1224 & 154 & 1598 & 179 & 2023 \\ \hline
 4.5 & 10 & 29.5 & 130 & 54.5 & 335 & 79.5 & 535 & 104.5 & 868 & 129.5 & 1224 & 154.5 & 1598 & 179.5 & 2023 \\ \hline
 5 & 10 & 30 & 130 & 55 & 335 & 80 & 597 & 105 & 868 & 130 & 1224 & 155 & 1598 & 180 & 2023 \\ \hline
 5.5 & 10 & 30.5 & 130 & 55.5 & 335 & 80.5 & 597 & 105.5 & 868 & 130.5 & 1224 & 155.5 & 1598 & 180.5 & 2023 \\ \hline
 6 & 19 & 31 & 130 & 56 & 335 & 81 & 597 & 106 & 868 & 131 & 1224 & 156 & 1598 & 181 & 2023 \\ \hline
 6.5 & 19 & 31.5 & 159 & 56.5 & 335 & 81.5 & 597 & 106.5 & 868 & 131.5 & 1224 & 156.5 & 1598 & 181.5 & 2023 \\ \hline
 7 & 19 & 32 & 159 & 57 & 363 & 82 & 597 & 107 & 868 & 132 & 1224 & 157 & 1598 & 182 & 2023 \\ \hline
 7.5 & 19 & 32.5 & 159 & 57.5 & 363 & 82.5 & 597 & 107.5 & 868 & 132.5 & 1224 & 157.5 & 1598 & 182.5 & 2023 \\ \hline
 8 & 19 & 33 & 159 & 58 & 363 & 83 & 597 & 108 & 868 & 133 & 1224 & 158 & 1598 & 183 & 2023 \\ \hline
 8.5 & 27 & 33.5 & 159 & 58.5 & 363 & 83.5 & 597 & 108.5 & 868 & 133.5 & 1224 & 158.5 & 1598 & 183.5 & 2023 \\ \hline
 9 & 27 & 34 & 167 & 59 & 363 & 84 & 597 & 109 & 896 & 134 & 1224 & 159 & 1598 & 184 & 2023 \\ \hline
 9.5 & 27 & 34.5 & 167 & 59.5 & 363 & 84.5 & 613 & 109.5 & 896 & 134.5 & 1224 & 159.5 & 1598 & 184.5 & 2023 \\ \hline
 10 & 27 & 35 & 167 & 60 & 363 & 85 & 613 & 110 & 896 & 135 & 1224 & 160 & 1598 & 185 & 2137 \\ \hline
 10.5 & 29 & 35.5 & 167 & 60.5 & 363 & 85.5 & 613 & 110.5 & 896 & 135.5 & 1224 & 160.5 & 1598 & 185.5 & 2137 \\ \hline
 11 & 34 & 36 & 167 & 61 & 363 & 86 & 631 & 111 & 896 & 136 & 1224 & 161 & 1598 & 186 & 2137 \\ \hline
 11.5 & 34 & 36.5 & 167 & 61.5 & 370 & 86.5 & 631 & 111.5 & 896 & 136.5 & 1279 & 161.5 & 1598 & 186.5 & 2137 \\ \hline
 12 & 34 & 37 & 167 & 62 & 370 & 87 & 631 & 112 & 896 & 137 & 1279 & 162 & 1719 & 187 & 2137 \\ \hline
 12.5 & 34 & 37.5 & 182 & 62.5 & 370 & 87.5 & 631 & 112.5 & 896 & 137.5 & 1279 & 162.5 & 1719 & 187.5 & 2203 \\ \hline
 13 & 42 & 38 & 182 & 63 & 397 & 88 & 631 & 113 & 896 & 138 & 1279 & 163 & 1719 & 188 & 2203 \\ \hline
 13.5 & 42 & 38.5 & 186 & 63.5 & 397 & 88.5 & 631 & 113.5 & 975 & 138.5 & 1279 & 163.5 & 1719 & 188.5 & 2203 \\ \hline
 14 & 42 & 39 & 186 & 64 & 397 & 89 & 650 & 114 & 975 & 139 & 1356 & 164 & 1719 & 189 & 2203 \\ \hline
 14.5 & 42 & 39.5 & 186 & 64.5 & 397 & 89.5 & 650 & 114.5 & 975 & 139.5 & 1356 & 164.5 & 1747 & 189.5 & 2203 \\ \hline
 15 & 42 & 40 & 197 & 65 & 397 & 90 & 650 & 115 & 975 & 140 & 1356 & 165 & 1747 & 190 & 2203 \\ \hline
 15.5 & 51 & 40.5 & 197 & 65.5 & 397 & 90.5 & 692 & 115.5 & 975 & 140.5 & 1356 & 165.5 & 1747 & 190.5 & 2203 \\ \hline
 16 & 51 & 41 & 197 & 66 & 453 & 91 & 692 & 116 & 1000 & 141 & 1356 & 166 & 1747 & 191 & 2203 \\ \hline
 16.5 & 51 & 41.5 & 197 & 66.5 & 453 & 91.5 & 705 & 116.5 & 1000 & 141.5 & 1356 & 166.5 & 1747 & 191.5 & 2203 \\ \hline
 17 & 51 & 42 & 197 & 67 & 453 & 92 & 705 & 117 & 1000 & 142 & 1356 & 167 & 1747 & 192 & 2203 \\ \hline
 17.5 & 67 & 42.5 & 197 & 67.5 & 453 & 92.5 & 705 & 117.5 & 1000 & 142.5 & 1356 & 167.5 & 1798 & 192.5 & 2203 \\ \hline
 18 & 67 & 43 & 248 & 68 & 453 & 93 & 707 & 118 & 1000 & 143 & 1356 & 168 & 1798 & 193 & 2203 \\ \hline
 18.5 & 67 & 43.5 & 248 & 68.5 & 475 & 93.5 & 707 & 118.5 & 1000 & 143.5 & 1356 & 168.5 & 1798 & 193.5 & 2203 \\ \hline
 19 & 67 & 44 & 248 & 69 & 475 & 94 & 707 & 119 & 1000 & 144 & 1356 & 169 & 1798 & 194 & 2203 \\ \hline
 19.5 & 67 & 44.5 & 248 & 69.5 & 475 & 94.5 & 707 & 119.5 & 1000 & 144.5 & 1417 & 169.5 & 1798 & 194.5 & 2203 \\ \hline
 20 & 85 & 45 & 248 & 70 & 475 & 95 & 707 & 120 & 1000 & 145 & 1417 & 170 & 1798 & 195 & 2203 \\ \hline
 20.5 & 85 & 45.5 & 264 & 70.5 & 475 & 95.5 & 707 & 120.5 & 1000 & 145.5 & 1417 & 170.5 & 1798 & 195.5 & 2203 \\ \hline
 21 & 85 & 46 & 264 & 71 & 475 & 96 & 707 & 121 & 1000 & 146 & 1417 & 171 & 1798 & 196 & 2203 \\ \hline
 21.5 & 85 & 46.5 & 264 & 71.5 & 475 & 96.5 & 707 & 121.5 & 1075 & 146.5 & 1417 & 171.5 & 1798 & 196.5 & 2273 \\ \hline
 22 & 85 & 47 & 264 & 72 & 475 & 97 & 707 & 122 & 1075 & 147 & 1417 & 172 & 1798 & 197 & 2273 \\ \hline
 22.5 & 95 & 47.5 & 264 & 72.5 & 475 & 97.5 & 735 & 122.5 & 1075 & 147.5 & 1417 & 172.5 & 1798 & 197.5 & 2273 \\ \hline
 23 & 95 & 48 & 264 & 73 & 475 & 98 & 735 & 123 & 1075 & 148 & 1417 & 173 & 1798 & 198 & 2273 \\ \hline
 23.5 & 95 & 48.5 & 264 & 73.5 & 475 & 98.5 & 740 & 123.5 & 1075 & 148.5 & 1417 & 173.5 & 1967 & 198.5 & 2273 \\ \hline
 24 & 95 & 49 & 264 & 74 & 475 & 99 & 740 & 124 & 1075 & 149 & 1417 & 174 & 1967 & 199 & 2352 \\ \hline
 24.5 & 95 & 49.5 & 264 & 74.5 & 521 & 99.5 & 740 & 124.5 & 1075 & 149.5 & 1417 & 174.5 & 1967 & 199.5 & 2352 \\
 \hline
\end{tabularx}
\renewcommand{\arraystretch}{1}
\caption{The required $N$\textsubscript{Trott} for a $\epsilon_{{\rm Trott}} < 0.1$ in the trivial vacuum-to-vacuum probability.}
\label{tab:vacNTrottData}
\end{table}
The required $N$\textsubscript{Trott} for a $\epsilon_{{\rm Trott}} < 0.1$ in the trivial vacuum-to-$d_r \overline{d}_r$ probability is given in Table~\ref{tab:drdrbNTrottData}.
\begin{table}[!ht]
\renewcommand{\arraystretch}{1.06}
\begin{tabularx}{.6\textwidth}{||c | Y  | Y  | Y  | Y | Y | Y  | Y ||}
\hline
\multicolumn{8}{||c||}{Required number of Trotter steps, $N$\textsubscript{Trott}} \\
 \hline
 $t$ & $N$\textsubscript{Trott} &
 $t$ & $N$\textsubscript{Trott} &
 $t$ & $N$\textsubscript{Trott} &
 $t$ & $N$\textsubscript{Trott} 
 \\
 \hline\hline
 0 & 1 & 25 & 243 & 50 & 945 & 75 & 1706 \\ \hline
 0.5 & 2 & 25.5 & 337 & 50.5 & 945 & 75.5 & 1706 \\ \hline
 1 & 2 & 26 & 337 & 51 & 945 & 76 & 1739 \\ \hline
 1.5 & 3 & 26.5 & 337 & 51.5 & 945 & 76.5 & 1739 \\ \hline
 2 & 7 & 27 & 337 & 52 & 945 & 77 & 1739 \\ \hline
 2.5 & 10 & 27.5 & 337 & 52.5 & 945 & 77.5 & 1739 \\ \hline
 3 & 10 & 28 & 337 & 53 & 945 & 78 & 1739 \\ \hline
 3.5 & 10 & 28.5 & 337 & 53.5 & 945 & 78.5 & 1739 \\ \hline
 4 & 12 & 29 & 337 & 54 & 945 & 79 & 1739 \\ \hline
 4.5 & 24 & 29.5 & 337 & 54.5 & 945 & 79.5 & 1739 \\ \hline
 5 & 24 & 30 & 337 & 55 & 945 & 80 & 1739 \\ \hline
 5.5 & 24 & 30.5 & 337 & 55.5 & 945 & 80.5 & 1739 \\ \hline
 6 & 24 & 31 & 337 & 56 & 945 & 81 & 1739 \\ \hline
 6.5 & 32 & 31.5 & 337 & 56.5 & 945 & 81.5 & 1739 \\ \hline
 7 & 32 & 32 & 384 & 57 & 945 & 82 & 1739 \\ \hline
 7.5 & 32 & 32.5 & 384 & 57.5 & 1168 & 82.5 & 1739 \\ \hline
 8 & 32 & 33 & 384 & 58 & 1168 & 83 & 2064 \\ \hline
 8.5 & 33 & 33.5 & 384 & 58.5 & 1168 & 83.5 & 2064 \\ \hline
 9 & 72 & 34 & 384 & 59 & 1168 & 84 & 2064 \\ \hline
 9.5 & 72 & 34.5 & 709 & 59.5 & 1168 & 84.5 & 2064 \\ \hline
 10 & 72 & 35 & 709 & 60 & 1285 & 85 & 2064 \\ \hline
 10.5 & 72 & 35.5 & 709 & 60.5 & 1285 & 85.5 & 2064 \\ \hline
 11 & 72 & 36 & 709 & 61 & 1285 & 86 & 2064 \\ \hline
 11.5 & 98 & 36.5 & 709 & 61.5 & 1285 & 86.5 & 2064 \\ \hline
 12 & 98 & 37 & 709 & 62 & 1285 & 87 & 2064 \\ \hline
 12.5 & 98 & 37.5 & 709 & 62.5 & 1285 & 87.5 & 2064 \\ \hline
 13 & 98 & 38 & 709 & 63 & 1285 & 88 & 2064 \\ \hline
 13.5 & 98 & 38.5 & 709 & 63.5 & 1285 & 88.5 & 2064 \\ \hline
 14 & 148 & 39 & 709 & 64 & 1285 & 89 & 2064 \\ \hline
 14.5 & 148 & 39.5 & 709 & 64.5 & 1285 & 89.5 & 2064 \\ \hline
 15 & 148 & 40 & 709 & 65 & 1285 & 90 & 2064 \\ \hline
 15.5 & 148 & 40.5 & 709 & 65.5 & 1285 & 90.5 & 2064 \\ \hline
 16 & 148 & 41 & 709 & 66 & 1285 & 91 & 2064 \\ \hline
 16.5 & 148 & 41.5 & 709 & 66.5 & 1285 & 91.5 & 2064 \\ \hline
 17 & 148 & 42 & 709 & 67 & 1285 & 92 & 2064 \\ \hline
 17.5 & 148 & 42.5 & 709 & 67.5 & 1285 & 92.5 & 2064 \\ \hline
 18 & 148 & 43 & 709 & 68 & 1285 & 93 & 2064 \\ \hline
 18.5 & 148 & 43.5 & 709 & 68.5 & 1285 & 93.5 & 2064 \\ \hline
 19 & 148 & 44 & 709 & 69 & 1285 & 94 & 2064 \\ \hline
 19.5 & 148 & 44.5 & 709 & 69.5 & 1285 & 94.5 & 2590 \\ \hline
 20 & 148 & 45 & 709 & 70 & 1285 & 95 & 2590 \\ \hline
 20.5 & 233 & 45.5 & 709 & 70.5 & 1285 & 95.5 & 2590 \\ \hline
 21 & 233 & 46 & 709 & 71 & 1285 & 96 & 2590 \\ \hline
 21.5 & 233 & 46.5 & 709 & 71.5 & 1706 & 96.5 & 2590 \\ \hline
 22 & 233 & 47 & 709 & 72 & 1706 & 97 & 2590 \\ \hline
 22.5 & 233 & 47.5 & 709 & 72.5 & 1706 & 97.5 & 2590 \\ \hline
 23 & 243 & 48 & 709 & 73 & 1706 & 98 & 2590 \\ \hline
 23.5 & 243 & 48.5 & 945 & 73.5 & 1706 & 98.5 & 2590 \\ \hline
 24 & 243 & 49 & 945 & 74 & 1706 & 99 & 2780 \\ \hline
 24.5 & 243 & 49.5 & 945 & 74.5 & 1706 & 99.5 & 2780 \\
 \hline
\end{tabularx}
\renewcommand{\arraystretch}{1}
\caption{The required $N$\textsubscript{Trott} for a $\epsilon_{{\rm Trott}} < 0.1$ in the trivial vacuum-to-$d_r \overline{d}_r$ probability.}
\label{tab:drdrbNTrottData}
\end{table}
The decomposition of the energy, starting from trivial vacuum at $t=0$, is given in Table~\ref{tab:vacAnimData}.
\begin{table}[!ht]
\renewcommand{\arraystretch}{1.2}
\begin{tabularx}{.6\textwidth}{||c | Y  | Y  | Y ||}
\hline
\multicolumn{4}{||c||}{Decomposition of the energy starting from the trivial vacuum} \\
 \hline
 $t$ & $\langle H_m \rangle$ 
 & $\langle H_{kin} \rangle$ 
 & $\langle H_{el} \rangle$ 
 \\
 \hline\hline
  \ \ 0 \ \  & 0 & 0 & 0 \\ \hline
 0.3 & 0.254 & -0.3369 & 0.08281 \\ \hline
 0.6 & 0.8455 & -1.104 & 0.2584 \\ \hline
 0.9 & 1.393 & -1.781 & 0.3879 \\ \hline
 1.2 & 1.575 & -1.974 & 0.3997 \\ \hline
 1.5 & 1.31 & -1.625 & 0.3151 \\ \hline
 1.8 & 0.769 & -0.9532 & 0.1842 \\ \hline
 2.1 & 0.2684 & -0.3282 & 0.05983 \\ \hline
 2.4 & 0.1081 & -0.1256 & 0.01747 \\ \hline
 2.7 & 0.3939 & -0.5129 & 0.119 \\ \hline
 3.0 & 0.9349 & -1.245 & 0.31 \\ \hline
 3.3 & 1.359 & -1.788 & 0.4285 \\ \hline
 3.6 & 1.401 & -1.788 & 0.3873 \\ \hline
 3.9 & 1.079 & -1.335 & 0.2565 \\ \hline
 4.2 & 0.6237 & -0.7542 & 0.1305 \\ \hline
 4.5 & 0.3094 & -0.3586 & 0.0492 \\
 \hline
\end{tabularx}
\renewcommand{\arraystretch}{1}
\caption{Decomposition of the energy starting from the trivial vacuum at $t=0$.}
\label{tab:vacAnimData}
\end{table}
The trivial vacuum to $B \overline{B}$ probabilities are given in Table~\ref{tab:BBarprobData}.
\begin{table}[!ht]
\renewcommand{\arraystretch}{1.08}
\begin{tabularx}{.4\textwidth}{||c | Y  | Y  | Y | Y||}
\hline
\multicolumn{5}{||c||}{Trivial vacuum to $B \overline{B}$ probability} \\
 \hline
 $t$ & 1 Step
 & 2 Steps
 & 3 Steps
 & Exact
 \\
 \hline\hline
\ \ 0 \ \  & 0 & 0 & 0 & 0 \\ \hline
 0.1 & 0.0000 & 0.0000 & 0.0000 & 0.0000 \\ \hline
 0.2 & 0.0000 & 0.0000 & 0.0000 & 0.0000 \\ \hline
 0.3 & 0.0000 & 0.0000 & 0.0000 & 0.0000 \\ \hline
 0.4 & 0.0001 & 0.0001 & 0.0001 & 0.0001 \\ \hline
 0.5 & 0.0002 & 0.0002 & 0.0002 & 0.0002 \\ \hline
 0.6 & 0.0007 & 0.0005 & 0.0005 & 0.0005 \\ \hline
 0.7 & 0.0016 & 0.0011 & 0.001 & 0.001 \\ \hline
 0.8 & 0.0035 & 0.0021 & 0.0019 & 0.0018 \\ \hline
 0.9 & 0.0068 & 0.0036 & 0.0032 & 0.0029 \\ \hline
 1.0 & 0.0121 & 0.0055 & 0.0047 & 0.0042 \\ \hline
 1.1 & 0.0204 & 0.0079 & 0.0064 & 0.0055 \\ \hline
 1.2 & 0.0324 & 0.0104 & 0.0081 & 0.0067 \\ \hline
 1.3 & 0.0491 & 0.013 & 0.0095 & 0.0075 \\ \hline
 1.4 & 0.0715 & 0.0154 & 0.0105 & 0.0079 \\ \hline
 1.5 & 0.1003 & 0.0175 & 0.0109 & 0.0078 \\ \hline
 1.6 & 0.1363 & 0.0192 & 0.0108 & 0.0072 \\ \hline
 1.7 & 0.1798 & 0.0206 & 0.0101 & 0.0062 \\ \hline
 1.8 & 0.231 & 0.0218 & 0.0092 & 0.0052 \\ \hline
 1.9 & 0.2897 & 0.023 & 0.008 & 0.0041 \\ \hline
 2.0 & 0.355 & 0.0241 & 0.0069 & 0.0033 \\ \hline
 2.1 & 0.426 & 0.0252 & 0.0059 & 0.0025 \\ \hline
 2.2 & 0.501 & 0.026 & 0.005 & 0.002 \\ \hline
 2.3 & 0.5783 & 0.0263 & 0.0042 & 0.0015 \\ \hline
 2.4 & 0.6555 & 0.0257 & 0.0036 & 0.0011 \\ \hline
 2.5 & 0.7304 & 0.0239 & 0.003 & 0.0007 \\ \hline
 2.6 & 0.8003 & 0.0208 & 0.0026 & 0.0004 \\ \hline
 2.7 & 0.8629 & 0.0166 & 0.0024 & 0.0001 \\ \hline
 2.8 & 0.9158 & 0.0116 & 0.0025 & 0.0000 \\ \hline
 2.9 & 0.9571 & 0.0066 & 0.0029 & 0.0000 \\ \hline
 3.0 & 0.9851 & 0.0025 & 0.0036 & 0.0002 \\ \hline
 3.1 & 0.9987 & 0.0002 & 0.0045 & 0.0005 \\ \hline
 3.2 & 0.9974 & 0.0005 & 0.0057 & 0.0008 \\ \hline
 3.3 & 0.9813 & 0.0037 & 0.007 & 0.0011 \\ \hline
 3.4 & 0.951 & 0.01 & 0.0087 & 0.0014 \\ \hline
 3.5 & 0.9077 & 0.0189 & 0.0111 & 0.0016 \\ \hline
 3.6 & 0.853 & 0.0295 & 0.0148 & 0.0019 \\ \hline
 3.7 & 0.789 & 0.0409 & 0.0205 & 0.0023 \\ \hline
 3.8 & 0.7181 & 0.0517 & 0.0293 & 0.0027 \\ \hline
 3.9 & 0.6427 & 0.061 & 0.042 & 0.0032 \\ \hline
 4.0 & 0.5652 & 0.0677 & 0.0592 & 0.0036 \\ \hline
 4.1 & 0.4882 & 0.0714 & 0.0814 & 0.004 \\ \hline
 4.2 & 0.4137 & 0.0719 & 0.1085 & 0.0043 \\ \hline
 4.3 & 0.3436 & 0.0693 & 0.14 & 0.0045 \\ \hline
 4.4 & 0.2793 & 0.0641 & 0.1749 & 0.0046 \\ \hline
 4.5 & 0.2219 & 0.057 & 0.2117 & 0.0047 \\ \hline
 4.6 & 0.172 & 0.0488 & 0.2489 & 0.0049 \\ \hline
 4.7 & 0.1297 & 0.0401 & 0.2848 & 0.005 \\ \hline
 4.8 & 0.095 & 0.0317 & 0.3178 & 0.0051 \\ \hline
 4.9 & 0.0673 & 0.024 & 0.3464 & 0.0051 \\ \hline
 5.0 & 0.0459 & 0.0174 & 0.3695 & 0.0048 \\
 \hline
\end{tabularx}
\renewcommand{\arraystretch}{1}
\caption{Trivial vacuum to $B \overline{B}$ probability for $m=g=L=N_f=1$.}
\label{tab:BBarprobData}
\end{table}
%


\clearpage

\bibliographystyle{JHEP}
\bibliography{bibi,biblioKRS}

\end{document}